\documentclass[aps,pra,twocolumn,superscriptaddress,longbibliography]{revtex4-2}

\usepackage{amssymb,amsmath,amsthm,bbm,bbold}
\usepackage{amsfonts}
\usepackage{graphicx}
\usepackage{mathtools}
\usepackage[breaklinks,colorlinks,allcolors=blue]{hyperref}
\usepackage[normalem]{ulem}
\usepackage[T1]{fontenc}
\usepackage{color}
\usepackage{xcolor}
\usepackage{tcolorbox}

\usepackage{pifont}
\usepackage{amssymb}
\usepackage{tikz}
\usepackage{makecell}

\def\RR{\mathbbm{R}}
\def\CC{\mathbbm{C}}

\newcommand{\bd}[1]{\boldsymbol{#1}}

\newcommand{\Tr}{\mathrm{Tr}}
\usepackage{accents}

\newcommand{\bra}[1]{\mbox{$\langle #1 |$}}
\newcommand{\ket}[1]{\mbox{$| #1 \rangle$}}

\newcommand*\xbar[1]{\hbox{\vbox{
       \hrule height 0.6pt 
       \kern0.3ex
       \hbox{%
         \kern-0.2em
         \ensuremath{#1}%
         \kern 0.0em
         }}}}

\newcommand*\xxbar[1]{\hbox{\vbox{
       \hrule height 0.6pt 
       \kern0.3ex
       \hbox{%
         \kern-0.0em
         \ensuremath{#1}%
         \kern 0.0em
         }}}}

\newtheorem{prop}{Proposition}

\begin{document}
\title{A Witness for Quantum Correlations Using Nonlinear Spectroscopy}
\author{Julia Liebert}
\affiliation{Department of Chemistry, Princeton University, Princeton, New Jersey 08544, United States}
\author{Gregory D. Scholes}
\email{gscholes@princeton.edu}
\affiliation{Department of Chemistry, Princeton University, Princeton, New Jersey 08544, United States}

\begin{abstract}
The ability to detect and quantify entanglement in molecular systems is essential for understanding their quantum correlations and to evaluating molecules as platforms for quantum technologies. Here we show how signals from existing spectroscopic techniques can provide sufficient information to witness entanglement. The main ingredient is a spectroscopic map from admissible quantum states to a finite-dimensional space of signal coefficients, which are expectation values of state-independent effective observables. The molecular response model and measurement protocol determine the effective observables for a fixed system bipartition,  and the spectroscopic map defines sets of joint expectation values compatible with all admissible quantum states and their separable subset. We illustrate the construction of the effective observables and their expectation value sets using rotational invariants and the spectral structure of optical signals. Fluorescence dynamics and photon coincidences provide a complementary example. Our findings provide a theoretical basis for entanglement certification with established spectroscopic techniques.
\end{abstract}
\date{\today}

\maketitle

\section{Introduction \label{sec:intro}}

Molecular systems offer a huge potential for quantum technologies \cite{WFMKSYZB20, YKLF21, S23, ZOHYSW24, SOCMNHF25, NG25}. Not only are molecules highly tunable, but understanding how quantum correlations influence their behaviour may yield new insight into chemical mechanisms and design principles \cite{SFCAGABC17, faraday24}. To study quantum effects in molecular systems and explore how to leverage them, we must first be able to detect them. Quantum information theory provides formal criteria for entanglement \cite{HHHH09, GT09}. Translating these conditions into experimentally accessible tests is often difficult. Spectroscopy is a central tool for studying molecular structure and dynamics \cite{BKZS22, GS26, SLWF26}, but a given measurement provides access only to specific properties of the quantum state. This raises the question: Can the information contained in signals from established spectroscopic techniques suffice to certify molecular entanglement?

A natural starting point is quantum state tomography \cite{JKMW01, LR09}. For small systems, the density matrix can be reconstructed from sufficiently many informative measurements. Extensions to pump--probe and multidimensional spectroscopy have been proposed for excitonic dimers and single excitation manifolds \cite{HW13, YZKMAG11, YZAG11}. These methods require a model of the spectroscopic response and become increasingly demanding for larger Hilbert spaces \cite{HW13}. Even with the full density matrix, deciding if a general mixed state is separable remains computationally hard. The standard weak membership problem is NP-hard \cite{Gurvits03, Gharibian10}. Universal entanglement detection requires tomographic-order information in the worst case \cite{ZTE10}.
Entanglement witnesses avoid full state tomography \cite{HHH96, LKCH00, SBL01, LKHC01, BEKGWGH04}. Their experimental use depends on which observables can be measured. Molecular spectroscopy cannot generally measure arbitrary observables. Answering the question raised above requires determining which entanglement tests the available spectroscopic signals are able to support.

Previous work investigated how spectroscopic signals can be used to construct quantum witnesses.
This includes tests that distinguish electronic coherences from purely vibronic oscillations \cite{YZKAG12, JYZAGK14}. Recent proposals for witnessing multipartite entanglement focused on the quantum Fisher information (QFI) \cite{HHTZ16, SFS21, WFSHS25, SDZEW26}. Extracting QFI from spectroscopic signals is only possible under specific assumptions about the state and its response. For example, the standard relation based on the susceptibility assumes a thermal equilibrium and does not directly apply to photogenerated molecular states. In addition, the entanglement bounds depend on the chosen generator and its local structure. Here we take a complementary approach by testing if a single admissible separable state can explain all retained signal coefficients.

We address this compatibility problem by constructing a spectroscopic map, which assigns to each quantum state a vector of coefficients extracted from the measured signal. Specifically, we consider coefficients that can be expressed as expectation values of state-independent effective observables. The effective observables are determined by the molecular response model and measurement protocol. The image of all admissible states under the spectroscopic map forms the full reachable set. Restricting the domain to states separable across the chosen bipartition gives the separable reachable set. Mathematically, these sets are the full and separable joint numerical ranges of the effective observables \cite{toeplitz1918, LP99, GPJZ10, SCSZ21, PSW21}. A vector of signal coefficients outside the separable reachable set certifies entanglement, while a vector inside is inconclusive.

The paper is structured as follows. Section~\ref{sec:Fragestellung} introduces the scientific problem and compares entanglement witnesses in state space with tests based on joint numerical ranges. In Sec.~\ref{sec:main}, we develop the framework for constructing entanglement tests from spectroscopic signals. Outer approximations of the full and separable reachable sets are discussed in Sec.~\ref{sec:witness}. Section~\ref{sec:exp-workflow} discusses conceptual aspects of the witness construction and outlines a joint theoretical and experimental workflow for implementing the entanglement test.
Sections~\ref{sec:examples-excited} and \ref{sec:examples-initial} illustrate its applications to excited and initial molecular states. We conclude in Sec.~\ref{sec:concl} with a summary and outlook.

\section{Key concepts and scientific problem \label{sec:Fragestellung}}

\begin{figure*}[t]
\centering
\includegraphics[width=\linewidth]{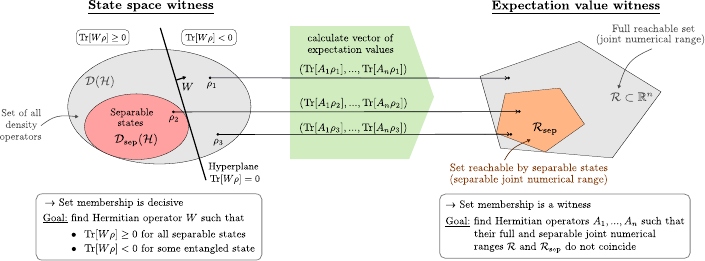}
\caption{Schematic comparison of entanglement witnessing via witness operators on state space (left) and joint numerical ranges which describe admissible expectation value tuples (right), and their relation through the trace map. 
The witness operator $W$ (left) detects the two entangled states $\rho_1$ and $\rho_3$, whereas the selected observables map $\rho_1$ outside the separable numerical range and $\rho_3$ inside it. This illustrates that set membership on the level of separable numerical ranges is a witness itself.}
\label{fig:comparison}
\end{figure*}

Entanglement can be certified without reconstructing the full density operator. Figure~\ref{fig:comparison} compares two established formulations: an operator witness on the state space (left) and a compatibility test in space of joint expectation values (right) \cite{HHH96, IT06, Spedalieri12, SCSZ21}.
Both approaches are based on convex separation, but organize the available information differently. In the following, we recall each formulation and explain the restrictions on its use in molecular spectroscopy. This leads us to the scientific problem addressed in this paper. 

The left panel of Fig.~\ref{fig:comparison} illustrates the standard operator formulation, which builds on the separability criterion of Horodecki et al.~\cite{HHH96} and subsequent theoretical and experimental developments \cite{Terhal00, LKCH00, GHBELMS03, BEKGWGH04}.
Let $\mathcal{D}(\mathcal{H})$ denote the density operators on a finite-dimensional Hilbert space $\mathcal{H}$ and $\mathcal{D}_{\mathrm{sep}}(\mathcal{H})$ the subset of separable states for a fixed physical bipartition \cite{HHHH09}. A state is entangled exactly when it lies outside this subset. An entanglement witness offers a more limited test than this membership question.
As illustrated in the left panel of Fig.~\ref{fig:comparison}, the key idea is to chooses a Hermitian operator $W$ with $\Tr[W\sigma] \geq 0$ for every separable $\sigma$ and $\Tr[W\rho] < 0$ for at least one state $\rho$.
The hyperplane $\Tr[W\rho] = 0$ separates the detected states from the separable set. A negative expectation value certifies entanglement, whereas a nonnegative value is generally inconclusive.

This raises the question: Can the expectation of a chosen witness be measured through nonlinear spectroscopy?
A common experimental strategy decomposes $W$ into products of local observables. In qubit models, these can be Pauli strings, $W = \sum_i w_i P_i$, and evaluating the corresponding correlators usually requires several
local measurement bases and independent control of the subsystems \cite{GHBELMS03}.
On the contrary, molecular spectroscopy offers different controls such as optical fields that couple through molecular transition operators, while the subsequent dynamics and detection determine the measured signal \cite{YFM89,Mukamel1995-book,HW13}. Currently, there is no practical scheme for arbitrary local basis measurements in those settings. In addition, a mathematically valid witness may lie outside the observable span accessible to the experiment.

A complementary approach starts from the observables available to the experiment and tests their joint expectation values \cite{GPJZ10, SCSZ21}. For Hermitian operators $A_1, \ldots, A_n$, the map $F(\rho) = (\Tr[\rho A_1], \ldots, \Tr[\rho A_n])$ assigns each state a vector in $\mathbbm{R}^n$.
The images of $\mathcal{D}(\mathcal{H})$ and $\mathcal{D}_{\mathrm{sep}}(\mathcal{H})$ define the full
and separable joint numerical ranges, $\mathcal{R}$ and $\mathcal{R}_{\mathrm{sep}}$. 
This transition from density operators to expectation values is illustrated in Fig.~\ref{fig:comparison}.
A vector in $\mathcal{R}\setminus\mathcal{R}_{\mathrm{sep}}$ certifies entanglement since no separable state reproduces all its components \cite{Spedalieri12}. 
Since the map $F$ is in general not injective, different states, including pairs of separable and entangled states, can yield the same vector of expectation values. 
We illustrate this in Fig.~\ref{fig:comparison}, where both the separable state $\rho_2$ and the entangled state $\rho_3$ map into $\mathcal{R}_{\mathrm{sep}}$, while $\rho_1$ maps outside of it.
Membership in the separable numerical range therefore implies a separable explanation of the data, but it does not mean that the actual state needs to be separable.

The numerical range witness construction is formulated for a chosen family of observables $\{A_i\}$. In molecular spectroscopy the experimental protocol constrains that choice.
The molecular response and detection protocol determine which effective observables are accessible. In particular, a response model needs to be able to describe the measured signals for all admissible states \cite{Mukamel1995-book, HW13}.
To certify entanglement it is not sufficient to reproduce a spectrum with an entangled state model, as a separable state may yield the same signal. Nevertheless, different components of a spectrum can provide complementary constraints. Together, they may exclude every admissible separable explanation.

What the above discussion shows is that theoretical entanglement criteria do not automatically provide practical spectroscopic tests.
Their experimental use requires a clear connection to the information contained in the measured signals. Furthermore, one would ideally like to avoid abstract new measurement proposals and see if entanglement can be specified from existing techniques. This leads us to the following scientific problem:

\begin{tcolorbox}[colframe=blue!50!black,  colback=blue!5!white, title={Scientific problem}, fonttitle=\bfseries ]
\textbf{How can existing spectroscopic techniques certify molecular entanglement?}

\smallskip 
We seek tests that require neither full state tomography nor access to arbitrary observables. For a specified physical bipartition, the task is to identify the effective observables associated with the measured signal coefficients. Their joint expectation values over all admissible separable states determine which vectors of signal coefficients are not compatible with any separable state and, thus, must come from an entangled state. Certification requires the experimental confidence region to remain disjoint from this set after uncertainties in the response model are taken into account.
\end{tcolorbox}

In Sec.~\ref{sec:main}, we construct entanglement tests from selected coefficients of optical signals. We then illustrate the framework for excited and initial molecular states in Secs.~\ref{sec:examples-initial} and \ref{sec:examples-excited}. The examples show how polarization and spectral information, fluorescence dynamics, and photon coincidences can provide constraints on separability. Notably, even after rotational averaging, the remaining polarization dependence may suffice to certify entanglement.

\section{Spectroscopic entanglement certification\label{sec:main}}

\begin{figure*}[tb]
\includegraphics[width=\linewidth]{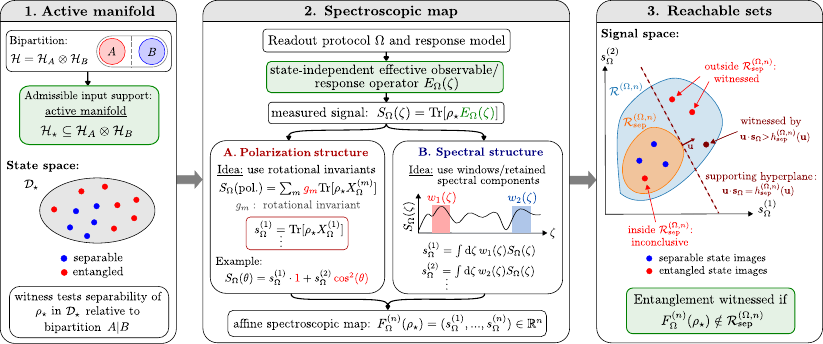}
\caption{Workflow for witnessing entanglement directly from spectroscopic data by mapping the high-dimensional state space (left) to a low-dimensional space of signal coefficients (right).
After choosing an active manifold $\mathcal{H}_\star$ and bipartition $A|B$, the protocol $\Omega$ and response model determine an effective response observable $E_\Omega(\zeta)$.
Signal coefficients, such as prefactors of rotational invariants or linearly post-processed spectral components, define the operators $X_\Omega^{(j)}$ and hence the affine spectroscopic map $F_\Omega^{(n)}(\rho_\star)$. The images of all states and their separable subset under this map give the reachable sets $\mathcal{R}^{(\Omega,n)}$ (blue) and $\mathcal{R}_{\mathrm{sep}}^{(\Omega,n)}$ (orange) (see Eq.~\eqref{eq:reachable-sets}) in the right panel. A measured point outside $\mathcal{R}_{\mathrm{sep}}^{(\Omega,n)}$ certifies entanglement, while points inside are inconclusive. A direction $\mathbf{u}$ defines a supporting hyperplane through the set of points $\mathbf s_\Omega = F_\Omega^{(n)}(\rho_\star)$ satisfying $\mathbf{u}\cdot\mathbf{s}_\Omega=h_{\mathrm{sep}}^{(\Omega,n)}(\mathbf{u})$ (c.f.~Eq.~\eqref{eq:witness}). If a measured point lies in the half-space described by $\mathbf{u}\cdot\mathbf{s}_\Omega >h_{\mathrm{sep}}^{(\Omega,n)}(\mathbf{u})$ is its entanglement is witnessed, as illustrated by the brown point. See text for more information. \label{fig:overview}}
\end{figure*}

Figure~\ref{fig:overview} summarizes the construction developed in this section. We follow its panels from left to right, starting with the admissible molecular states and their physical bipartition.
The subsequent subsections construct the spectroscopic map and establish which signal vectors exclude an explanation of the signal coefficients through any separable state.

\subsection{Active state manifold \label{subsec:manifold}}

We consider a finite-dimensional effective Hilbert space $\mathcal{H} = \mathcal{H}_A\otimes\mathcal{H}_B$, whose tensor product structure encodes the physical bipartition $A|B$ of interest. In turn the bipartition fixes the commuting local observable algebras associated with $A$ and $B$, and entanglement is defined relative to this choice \cite{Zanardi01,ZLL04, Landsman}. The set of density operators on $\mathcal{H}$ is denoted by $\mathcal{D}(\mathcal{H})$, and $\mathcal{D}_\mathrm{sep}(\mathcal{H})$ is the set of separable states. Recall that a state $\sigma\in\mathcal{D}(\mathcal{H})$ is called separable if it can be written as a convex combination of product states, that is $\sigma = \sum_i p_i \rho_A^{(i)}\otimes \rho_B^{(i)}$ with $\rho_A\in \mathcal{D}(\mathcal{H}_{A})$, $\rho_B\in \mathcal{D}(\mathcal{H}_{B})$, $p_i\geq 0$ and $\sum_ip_i=1$. The set of entangled states $\mathcal{D}_\mathrm{ent}(\mathcal{H}):=\mathcal{D}(\mathcal{H})/\mathcal{D}_\mathrm{sep}(\mathcal{H})$ consists of all density operators that are not separable. 

Moreover, preparation and subsequent dynamics may further restrict admissible states to a physically motivated subspace $\mathcal{H}_\star\subseteq\mathcal{H}$. In the following, we refer to $\mathcal{H}_\star$ as the active manifold. The orthogonal projector onto the subspace $\mathcal{H}_\star$ is denoted by $P_\star$. At the time when entanglement is tested, we assume that the state has no population outside the active manifold. The relevant space of physically admissible density operators is defined as 
\begin{equation}\label{eq:D-star}
\mathcal{D}_\star := \left\{\rho\in\mathcal{D}(\mathcal{H}): \rho =P_\star\rho P_\star \right\} \cong \mathcal{D}(\mathcal{H}_\star)\,, 
\end{equation}
which is a proper subset of $\mathcal{D}(\mathcal{H})$ unless $\mathcal{H} = \mathcal{H}_\star$. 
Every density operator on the active manifold $\mathcal{H}_\star$ extends uniquely to a density operator in $\mathcal{D}_\star$ by acting as zero on the orthogonal complement $\mathcal{H}_\star^\perp$. The inverse statement follows from restricting a state in $\mathcal{D}_\star$ to $\mathcal{H}_\star$.

Restricting the admissible states to $\mathcal{D}_\star$ leaves the physical bipartition $A|B$ unchanged. What is important is that entanglement has to be certified on the level of the full Hilbert space $\mathcal{H} = \mathcal{H}_A\otimes \mathcal{H}_B$, even when $\mathcal{H}_\star$ is \textit{not} a tensor product of local subspaces of $\mathcal{H}_A$ and $\mathcal{H}_B$.
The admissible separable states are obtained by imposing the support restriction on $\mathcal{D}_{\mathrm{sep}}(\mathcal{H})$,
\begin{equation}
\mathcal{D}_{\star, \mathrm{sep}} := \mathcal{D}_\star \cap  \mathcal{D}_{\mathrm{sep}}(\mathcal{H})\,.
\label{eq:D-star-sep}
\end{equation}
Membership in $\mathcal{D}_{\star,\mathrm{sep}}$ means that the state is separable across the physical partition $A|B$ in $\mathcal{H}$. It does not require that the active manifold $\mathcal{H}_\star$ itself factorizes into local subspaces. 
The sets $\mathcal{D}_\star$ and $\mathcal{D}_{\star,\mathrm{sep}}$ in Eqs.~\eqref{eq:D-star} and \eqref{eq:D-star-sep} are both compact and convex. We illustrate the state space description in the left panel of Fig.~\ref{fig:overview}. The blue and red points represent admissible states that are separable and entangled across the physical bipartition $A|B$, respectively. 

As an example from physical chemistry, consider an energy transfer model restricted to at most one electronic excitation with an active manifold $\mathcal{H}_\star =  \mathrm{span}\{|g\rangle\}\oplus\mathcal{H}_e$. 
Here $|g\rangle$ denotes the overall ground state and $\mathcal{H}_e$ is the single excitation manifold.
If all higher lying excited states can be neglected, including $|g\rangle$ ensures that the state on $\mathcal{H}_\star$ remains normalized to unity without conditioning on an excitation.
The exclusion of higher excited states is a physical model assumption and its validity must be assessed for the considered system parameters and state preparation.

The active manifold $\mathcal{H}_\star$ introduced in this section specifies the support of the state whose entanglement is tested. 
Spectroscopy accesses information about this state through a detected optical signal. Depending on the specifics of the protocol, this signal may depend only on the restriction of the state to a smaller subspace of $\mathcal{H}_\star$.
We describe its dependence on the input state through the spectroscopic map which will be introduced in Sec.~\ref{subsec:spectroscopic-map} below.
For instance, under suitable response assumptions, a nonlinear spectroscopic signal in the energy transfer example above can depend only on the subnormalized block $P_e \rho P_e$, where $P_e$ projects onto the single excitation manifold $\mathcal{H}_e \subseteq \mathcal{H}_\star$.
This means that the full map from the state to the signal consists of extracting this block and applying the reduced spectroscopic map. The admissible state space remains $\mathcal{D}_\star$, and the separable comparison is obtained by applying the same procedure to all states in $\mathcal{D}_{\star,\mathrm{sep}}$.
Moreover, the optical fields may also induce transitions to levels outside the active manifold during the measurement. These transitions must be included when constructing the effective response observables (see Sec.~\ref{subsec:spectroscopic-map}), which relate the input state to the measured signal.
At the same time, their occurrence does, depending on the context, not necessarily require enlarging the active manifold. The latter is only needed if the state whose entanglement is tested has support on those levels.

Equipped with the admissible state space, we now turn to the information accessible through spectroscopy. This step follows the transition from the left to the middle panel of Fig.~\ref{fig:overview}. In Sec.~\ref{subsec:spectroscopic-map}, we will introduce the effective response observables associated with the experimental protocol and use them to construct the spectroscopic map from $\mathcal{D}_\star$ to the retained signal coefficients.

\subsection{Spectroscopic map \label{subsec:spectroscopic-map}}

The middle panel of Fig.~\ref{fig:overview} summarizes how a spectroscopic protocol maps the input state to a finite vector of retained signal coefficients.
The key step in translating the framework of joint numerical ranges \cite{GPJZ10, SCSZ21} into spectroscopic entanglement tests is to determine the effective observables associated with these coefficients using currently available spectroscopic techniques, rather than relying on access to arbitrarily chosen observables.

We first consider a single input state $\rho\in\mathcal{D}_\star$, defined at the time at which entanglement is to be tested. Depending on the experiment, this may be before optical excitation or at a chosen stage after preparation. The subsequent molecular evolution and signal detection constitute the readout. The readout possibly involves a probe interaction or detection of subsequent molecular emission alone. Whenever the measurement dynamics is described by a fixed linear map on the chosen input states, detecting expectation values preserves this linearity. Therefore, a real signal component can be written as
\begin{equation}\label{eq:S-linear}
S_\Omega(\zeta) = \Tr[\rho E_\Omega(\zeta)]\,. 
\end{equation}
Here, $\Omega$ specifies the readout protocol and the fixed parameters of its response model. Within that protocol, $\zeta$ labels a detected coordinate, such as frequency or time, or a readout setting. The label can also distinguish the type of signal. For example, Sec.~\ref{subsec:DBT} uses both fluorescence intensity and an unnormalized photon coincidence density.
Each real signal component has a corresponding effective observable $E_\Omega(\zeta)$ through Eq.~\eqref{eq:S-linear}. 
The effective observable $E_\Omega(\zeta)$ describes how the molecular evolution and detection relate the input $\rho$ to the measured signal. However, it is independent of the unknown state. Since the effective observable $E_\Omega(\zeta)$ provides the map from density operators to signal coefficients, it determines the signal for all states $\rho\in\mathcal{D}_\star$ simultaneously. 
Different readout settings can be applied to repeated preparations of the same state. The corresponding effective observables incorporate the subsequent dynamics, and the state may evolve during detection. By contrast, changing the preparation during a scan may produce different target states. Then, the resulting signals cannot be treated as joint expectation values of a single state.

We now recall the construction of $E_\Omega(\zeta)$ for common nonlinear spectroscopy protocols. For a given molecular system and protocol, this observable incorporates both the dynamics after the chosen input time and the detection procedure. 
For a signal of the form $\Tr[M_\zeta\Lambda_{\Omega,\zeta}(\rho)]$, the adjoint representation
gives $E_\Omega(\zeta)=\Lambda_{\Omega,\zeta}^*(M_\zeta)$. Here, $\Lambda_{\Omega,\zeta}$ is a dynamical map which describes the system's dynamics and $M_\zeta$ is an observable.
The label on $\Lambda_{\Omega,\zeta}$ allows the propagation to depend on the detection time or chosen readout setting. This allows us to separate the time at which entanglement is tested from the time at which the signal is detected. In particular, the input time can be chosen to investigate a particular stage of the molecular dynamics as long as $\Lambda_{\Omega, \zeta}$ describes the remaining protocol for all admissible states at that time.
The effective observable can be calculated using either perturbative response theory \cite{YFM89, Mukamel1995-book} or non-perturbative propagation of the light-matter dynamics \cite{SSD95}. For temporally separated strong pulses, the latter also admits a doorway-window formulation \cite{GED11}. 

A fixed background, such as the reference field intensity in heterodyne detection, contributes an offset independent of $\rho$ \cite{HW13}. It can be subtracted or, for normalized input states, absorbed into $E_\Omega(\zeta)$ as a multiple of the identity. This is possible as long as the measured data and the separable benchmark use the same convention.
Moreover, for complex signals linear in $\rho$, experimentally accessible real and imaginary parts of the signal are treated as separate components, each represented by a Hermitian response observable.

A single real component of Eq.~\eqref{eq:S-linear} yields one coordinate for the entanglement test. Several coefficients have the ability to impose a stronger constraint on the signal because one separable state must reproduce all their values. Indeed, a joint compatibility test might be able to certify entanglement in cases where each coefficient alone is inconclusive. The scientific task is to determine such coefficients within the available signals and derive their effective observables.

Let $\mathcal{F}_j$ denote a fixed real linear functional that extracts the $j$th retained coefficient from the signals. The same operation determines its effective observable. Equation~\eqref{eq:S-linear} gives
\begin{equation}\label{eq:signal-selec}
\begin{split}
s_\Omega^{(j)}(\rho) &= \mathcal{F}_j[S_\Omega(\cdot; \rho)]  = \Tr[\rho X_\Omega^{(j)}]\,,\\
X_\Omega^{(j)} &= \mathcal{F}_j[E_\Omega(\cdot)]\,.
\end{split}
\end{equation}
Here, the dependence of the signal on $\rho$ is explicit.
For $E_\Omega$, the functionals $\mathcal{F}_j(\cdot)$ act on the real and imaginary parts of each matrix element. Since $E_\Omega(\zeta)$ is Hermitian, Eq.~\eqref{eq:signal-selec} yields a set of Hermitian operators $X_\Omega^{(j)}$. The functional is allowed to depend on fixed model and calibration parameters, but not on the unknown input state (similarly to the effective observable).
The middle panel of Fig.~\ref{fig:overview} illustrates this idea exemplary through coefficients of rotational invariants and spectral integrals.
If both constructions refer to the same input, their coefficients may also be combined in one signal vector.

The polarization dependence provides one example for identifying the operators $X_\Omega^{(j)}$. For an isotropic ensemble whose input state is independent of molecular orientation, a real third-order signal in the electric dipole approximation has the form (see also middle panel of Fig.~\ref{fig:overview})
\begin{equation}\label{eq:rot-invariants}
S_\Omega(\mathbf{e}_1, \mathbf{e}_2, \mathbf{e}_3, \mathbf{e}_4) = \sum_{m = 1}^{3} g_m (\mathbf{e}_1, \mathbf{e}_2, \mathbf{e}_3, \mathbf{e}_4)\, \Tr\left[\rho X_\Omega^{(m)}\right]
\end{equation}
in the perturbative regime. Here, $\mathbf{e}_i$ are real unit polarization vectors associated with the optical interactions and detection. The functions $g_m$ are the three fourth-rank rotational invariants, formed from the three pairings of polarization scalar products \cite{AT77}.
Moreover, the introduced operators $X_\Omega^{(m)}$ depend on the particular third-order process.
Thus the retained signal coefficients are the prefactors $s_\Omega^{(m)}(\rho_\star) = \Tr[\rho_\star X_\Omega^{(m)}]$ and can be extracted from sufficiently many independent polarization configurations.

The isotropic fourth-rank response simplifies when the four polarization vectors reduce to two equal pairs,
\begin{align}\label{eq:S-TPA}
S_\Omega(\theta) &= s_\Omega^{(1)} + s_\Omega^{(2)}\cos^2\theta \nonumber\\
&=\Tr \left[\rho\left(X_\Omega^{(1)} + X_\Omega^{(2)}\cos^2\theta\right)\right]\,, 
\end{align}
where $\theta$ is the angle between the two polarization vectors. With the input preparation and all other pulse
parameters fixed, the coefficients follow from a fit to this angular dependence. Rotational symmetry fixes the form of the signal in Eq.~\eqref{eq:S-TPA}, while the effective observables depend on the measured quantity. For a two-photon absorption (TPA) example, we show the explicit derivation of the operators $X_\Omega^{(1)}, X_\Omega^{(2)}$ in Appendix~\ref{app:Ax-Bx}.

Spectral selection provides another choice of retained signal coefficients. At fixed preparation and delay time, different spectral regions can supply complementary constraints on the same input state.
The excitonic dimer in Sec.~\ref{subsec:excitonic-dimer} illustrates this choice. For fixed real window functions $w_j$, Eq.~\eqref{eq:S-linear} gives 
\begin{equation} \label{eq:wind-op-gen}
s_\Omega^{(j)}(\rho) = \int \mathrm{d}\zeta\, w_j(\zeta) S_\Omega(\zeta) = \Tr[\rho X_\Omega^{(j)}]\,,
\end{equation}
where
\begin{equation}
X_\Omega^{(j)} =  \int\mathrm{d}\zeta\, w_j(\zeta) E_\Omega(\zeta)\,.
\end{equation}
The weights $w_j$ determine how each spectral region contributes to a retained coefficient. These integrals are examples of the application of the linear functional in Eq.~\eqref{eq:signal-selec}.
At the same time, the selected coefficients need not be spectral.
Section~\ref{subsec:DBT} illustrates this point with fluorescence after removal of the preparation drive.

For molecular ensembles, the reduction to one averaged input state requires a careful analysis. A polarized pump can prepare different states at different molecular orientations, while the probe sensitivity also depends on orientation \cite{HW13}. In that case, the averaged signal need not take the form of Eq.~\eqref{eq:S-linear} with the averaged molecular state.
Measurements on oriented samples or individual molecules offer promising alternatives to isotropic averaging.

If orientation-dependent preparation or static disorder prevents a signal representation in terms of a single averaged state, one may keep the classical label $\lambda$ explicitly.
Let $p(\lambda)$ be its fixed, normalized distribution and $\rho_\lambda\in\mathcal D_\star$ the state of the subensemble. Define $\Gamma(\lambda) = p(\lambda)\rho_\lambda$.
Provided the response is linear in $\rho_\lambda$ at each label, the retained coefficients are
\begin{equation}
s_\Omega^{(j)} = \int\mathrm{d}\lambda\, \Tr[\Gamma(\lambda)X_{\Omega, \lambda}^{(j)}]\,.
\end{equation}
where $X_{\Omega,\lambda}^{(j)}$ is the effective observable for coefficient $j$ at label $\lambda$. The signal map remains affine in $\Gamma$ but the averaged molecular state alone need not determine the signal.
The separable comparison keeps $p(\lambda)$ fixed but allows an admissible separable state
$\sigma_\lambda\in\mathcal{D}_{\star,\mathrm{sep}}$ to vary with $\lambda$. Further details with specifications for orientational and static disorder are provide in Appendix~\ref{app:ensemble-res}.  

We return to measurements with the single-state representation in Eq.~\eqref{eq:S-linear}.
Each retained coefficient has a Hermitian observable $X_\Omega^{(j)}$ as defined in Eq.~\eqref{eq:signal-selec}, irrespective of the signal feature used to obtain it.
The observables define the affine spectroscopic map shown at the bottom of the middle panel in Fig.~\ref{fig:overview},
\begin{equation}\label{eq:spectral-map}
\begin{split}
F_\Omega^{(n)}&: \mathcal{D}_\star\to \mathbbm{R}^n\,,\\
F_\Omega^{(n)}(\rho) &= \left( \Tr\left[\rho X_\Omega^{(1)}\right], \ldots, \Tr\left[\rho X_\Omega^{(n)}\right] \right)\,.
\end{split}
\end{equation}
The domain of $F_\Omega^{(n)}$ is the admissible state space $\mathcal{D}_\star$ in Eq.~\eqref{eq:D-star}.
We now revisit the energy transfer example from Sec.~\ref{subsec:manifold}. There, a coefficient may depend linearly on the subnormalized excited state block $P_e \rho P_e$ through an observable $X_{\Omega,e}^{(j)}$ on $\mathcal{H}_e$. Extending this observable by zero outside $\mathcal{H}_e$ gives $X_\Omega^{(j)}$ and preserves the signal,
\begin{equation}
\Tr_{\mathcal{H}_e}\left[(P_e\rho P_e)X_{\Omega,e}^{(j)}\right] = \Tr_{\mathcal{H}}\!\left[\rho X_\Omega^{(j)}\right]\,.
\end{equation}
Thus, the spectroscopic map defines a map on $\mathcal{D}_\star$ through extraction of the relevant block.
The state whose entanglement is tested is a density operator $\rho\in \mathcal{D}_\star$, and the comparison to admissible vectors of signal coefficients from separable states included all states in $\mathcal{D}_{\star,\mathrm{sep}}$.

The spectroscopic map $F_\Omega^{(n)}$ in Eq.~\eqref{eq:spectral-map} being affine means that convex mixtures of input states produce the same mixtures of signal vectors.
Each coefficient alone takes values within an interval. For several signal coefficients, a separable explanation must reproduce all of them with the same admissible separable state. This joint requirement can exclude separability in cases where each coefficient lies within its individual separable range. We formulate this compatibility test using the reachable sets shown in the right panel of Fig.~\ref{fig:overview} and developed in Sec.~\ref{subsec:reachable-sets}.

\subsection{Reachable sets and entanglement certification \label{subsec:reachable-sets}}

The spectroscopic map $F_\Omega^{(n)}$ in Eq.~\eqref{eq:spectral-map} determines the geometry of the signal space as illustrated in the right panel of Fig.~\ref{fig:overview}. The images of the admissible states and their separable subset under $F_\Omega^{(n)}$ are the full and separable reachable sets,
\begin{align}\label{eq:reachable-sets}
\mathcal{R}^{(\Omega, n)} :=  F_\Omega^{(n)}(\mathcal{D}_\star)\,,\,\,\,\,
\mathcal{R}_{\mathrm{sep}}^{(\Omega, n)} := F_\Omega^{(n)}(\mathcal{D}_{\star, \mathrm{sep}})\,.
\end{align}
We illustrate $\mathcal{R}^{(\Omega, n)}$ and $\mathcal{R}_{\mathrm{sep}}^{(\Omega, n)}$ by the overlapping blue and orange regions in the right panel of Fig.~\ref{fig:overview}. 
Mathematically, they are the joint numerical ranges of the effective observables restricted to $\mathcal{D}_\star$ and $\mathcal{D}_{\star,\mathrm{sep}}$, respectively, with mixed states included \cite{GPJZ10, SCSZ21}.
We refer to them as spectroscopic reachable sets to emphasize which signal vectors the fixed protocol can produce from the admissible input states. At the same time, this does not imply that every such input can be prepared by a particular control sequence. Both sets in Eq.~\eqref{eq:reachable-sets} are compact and convex, since they are images of convex compact sets under an affine map. 

There are several relevant differences between the state space and signal space picture, as anticipated in Sec.~\ref{sec:Fragestellung}. 
For a vector $\mathbf{s}\in\mathcal{R}^{(\Omega, n)}$ of signal coefficients,
\begin{equation}
\mathbf{s}\notin \mathcal{R}_{\mathrm{sep}}^{(\Omega, n)}
\end{equation}
means that \textit{no} admissible separable state reproduces the measured coefficients.
In other words, any admissible state that is consistent with this signal is entangled across the physical partition $A|B$. It is the existence of a strict inclusion relation between the full and separable sets, i.e., $\mathcal{R}_{\mathrm{sep}}^{(\Omega, n)} \subsetneq\mathcal{R}^{(\Omega, n)}$, which allows for entanglement certification in case a signal vectors lies outside of $\mathcal{R}_{\mathrm{sep}}^{(\Omega, n)}$. The explicit properties of $\mathcal{R}_{\mathrm{sep}}^{(\Omega, n)}, \mathcal{R}^{(\Omega, n)}$ that matter for entanglement witnessing, such as a comparison of their volumes, are properties of the spectroscopic map $F_\Omega^{(n)}$.   

The distinction between the membership problems on the state space and signal space is crucial. 
As explained in Sec.~\ref{sec:Fragestellung}, $\rho\in\mathcal{D}_{\star,\mathrm{sep}}$ ensures separability of that state. 
By contrast, what $F_\Omega^{(n)}(\rho)\in \mathcal{R}_{\mathrm{sep}}^{(\Omega, n)}$ implies is only the existence of some separable state leading to the same observed signal. 
A non-injective spectroscopic map cannot characterize the signal of every admissible state uniquely. It can assign the same signal to both separable and entangled states.
Membership in the separable reachable set neither certifies entanglement, nor does it prove separability of the input state $\rho$ in Eq.~\eqref{eq:S-linear}. This is the key difference to entanglement certification on the level of the full density operator $\rho$. 
At the same time, exclusion from the separable reachable set $\mathcal{R}_{\mathrm{sep}}^{(\Omega, n)}$ constitutes the strongest possible entanglement test based on the given set of experimentally accessible signal coefficients.

What is more, witnessing entanglement through membership on the signal space level is complementary to the more well-known operator witness description recalled in Sec.~\ref{sec:Fragestellung}. 
A single operator witness on the state space specifies one separating half-space, whereas the separable reachable set encodes all compatibility constraints among the retained coefficients.
Since $\mathcal{R}_{\mathrm{sep}}^{(\Omega, n)}$ is compact and convex, for any signal outside $\mathcal{R}_{\mathrm{sep}}^{(\Omega, n)}$ there exists a hyperplane which separates it from the set. 
An inequality that describes this hyperplane corresponds to a linear combination of the effective observables and the identity. 

An experimental test requires a robust separation between signals and the separable reachable set. 
Therefore, one needs both an informative protocol and a calibrated response model relating the measured data to Eq.~\eqref{eq:spectral-map}. The latter is likely the most challenging part of the experimental design. 
Two potential options for calibration are reference measurements or using additional normalization channels that are insensitive to the particular entanglement distinction. For instance, this could be spin-blind signals when testing a state for entanglement between spin degrees of freedom.
To preserve the affine map \eqref{eq:spectral-map}, a state-dependent normalization signal can be included as an additional coordinate instead of being used as a denominator.
Entanglement is certified if the experimental confidence region does not overlap the separable reachable set. 

Finally, we note that the set membership criterion for witnessing entanglement applies to nonlinear maps as well. Indeed, for any well-defined map $G$ on $\mathcal{D}_\star$, the condition $G(\rho) \notin G(\mathcal{D}_{\star, \mathrm{sep}})$ excludes every admissible separable explanation of the signal.
The affine representation used here yields compact, convex reachable sets. By the hyperplane separation theorem \cite{rockafellar2015}, the separable reachable set is fully determined by (a possibly infinite number of) linear witness inequalities. This is in analogy to Refs.~\cite{LKCH00, LKHC01, GPJZ10, SCSZ21}, which use the same standard result from convex analysis.

\section{Separable bounds and witnesses \label{sec:witness}}

The separable reachable set in Eq.~\eqref{eq:reachable-sets} yields a compatibility test for the vector of signal coefficients. However, its full geometry need not be known to certify entanglement in a particular setting. It can suffice to derive bounds on selected linear combinations of the coefficients over all admissible separable states. Each bound defines a half-space containing the separable reachable set. Together, several such bounds provide an outer approximation, which is particularly useful when calculating the exact set is computationally demanding. The following construction applies well-known witness and outer approximation tools \cite{IT06,Spedalieri12,SCSZ21} to the spectroscopic observables derived in Sec.~\ref{subsec:spectroscopic-map}.
In the following, we show how entanglement tests based on witness operators translate into inequalities restricting the signal coefficients. This allows us to derive rigorous bounds on experimentally accessible signal combinations. Each bound must hold for every separable state supported on the active manifold $\mathcal{H}_\star$.

We consider a fixed protocol $\Omega$. A direction $\mathbf{u}\in\mathbbm{R}^n$ specifies a weighted combination of signal coefficients. Its operator representation is the combined response observable
\begin{equation}\label{eq:W-h}
W_\Omega(\mathbf{u}) := \sum_{j=1}^{n} u_j X_\Omega^{(j)}\,. 
\end{equation}
As in Sec.~\ref{sec:main}, the operators are understood through their zero extensions to $\mathcal{H}$ on which the bipartition is defined.
Assuming $\mathcal{D}_{\star, \mathrm{sep}}\neq\emptyset$, the largest value attainable by an admissible separable state is the support function of $\mathcal{D}_{\star, \mathrm{sep}}$
\begin{equation}\label{eq:witness}
h_{\mathrm{sep}}^{(\Omega, n)}(\mathbf{u}) := \max_{\sigma \in \mathcal{D}_{\star, \mathrm{sep}}} \Tr[\sigma W_\Omega(\mathbf{u})]\,,
\end{equation}
which is also a support function of the separable reachable set $\mathcal{R}_{\mathrm{sep}}^{(\Omega, n)}$. Therefore, every admissible separable state $\sigma\in \mathcal{D}_{\star, \mathrm{sep}}$ satisfies
\begin{equation}\label{eq:signal-witness-inequality}
\mathbf{u}\cdot F_\Omega^{(n)}(\sigma) = \Tr[\sigma W_\Omega(\mathbf{u})] \leq h_{\mathrm{sep}}^{(\Omega, n)}(\mathbf{u})\,.
\end{equation}
The first equality in Eq.~\eqref{eq:signal-witness-inequality} holds for every admissible state, not only separable ones. This is a consequence of the Hilbert--Schmidt pairing $\Tr[\rho W_\Omega(\mathbf{u})]$ in operator space with the ordinary dot product $\mathbf{u}\cdot\mathbf{s}_\Omega$ in signal space, where $\mathbf{s}_\Omega = F_\Omega^{(n)}(\rho)$.
Thus, an expectation value bound on the state can be evaluated directly from the measured coefficients.
For directions $\mathbf{u}\neq 0$, the equation $\mathbf{u}\cdot\mathbf{s}_\Omega = h_{\mathrm{sep}}^{(\Omega, n)}(\mathbf{u})$ defines a supporting hyperplane of the separable reachable set, as illustrated in the right panel of Fig.~\ref{fig:overview}.
A signal on the side $\mathbf{u}\cdot\mathbf{s}_\Omega > h_{\mathrm{sep}}^{(\Omega, n)}(\mathbf{u})$ excludes every explanation of the signal in terms of admissible separable states, and thereby certifies entanglement.

The combined observable $W_\Omega(\mathbf{u})$ allows for witnessing entanglement through the separable bound in Eq.~\eqref{eq:signal-witness-inequality}. It is not the same as the witness operator described in Sec.~\ref{sec:Fragestellung} despite it contains the same information. 
Indeed, the same entanglement test can be equivalently expressed in terms of a negative expectation value criterion. Taking the active manifold and its orthogonal projection $P_\star$ introduced in Sec.~\ref{subsec:manifold} into account, we define
\begin{equation}
\widetilde{W}_\Omega(\mathbf{u}) :=  h_{\mathrm{sep}}^{(\Omega, n)}(\mathbf{u})P_\star - W_\Omega(\mathbf{u}).
\end{equation}
Since $\Tr[\rho P_\star] = 1$ on $\mathcal{D}_\star$, it follows that
\begin{equation}
\Tr[\rho\widetilde W_\Omega(\mathbf{u})] = h_{\mathrm{sep}}^{(\Omega, n)}(\mathbf{u}) - \mathbf{u}\cdot F_\Omega^{(n)}(\rho)\,.
\end{equation}
By construction, the operator $\widetilde{W}_\Omega(\mathbf{u})$ is nonnegative on $\mathcal{D}_{\star, \mathrm{sep}}$.
Thus, a negative expectation certifies entanglement of the admissible input state.
However, the signal inequality in Eq.~\eqref{eq:signal-witness-inequality} and the operator test are two representations of the same entanglement certificate.
Both test respect the support assumption of Sec.~\ref{subsec:manifold}, since $\widetilde{W}_\Omega(\mathbf{u})$ need not be nonnegative on separable states outside that domain.

Furthermore, a direction $\mathbf{u}$ provides a nontrivial witness if
\begin{equation}\label{eq:signal-witness-gap}
h_{\mathrm{sep}}^{(\Omega, n)}(\mathbf{u}) < \lambda_{\max} \left(W_\Omega(\mathbf{u})|_{\mathcal H_\star}\right)\,,
\end{equation}
where the right-hand side is the maximum expectation over all admissible states in $\mathcal{D}_\star$.
Therefore, the eigenvalue must be evaluated on the active manifold $\mathcal{H}_\star$, and not on the zero extension outside it. Such a direction exists if and only if $\mathcal{R}_{\mathrm{sep}}^{(\Omega, n)} \subsetneq \mathcal{R}^{(\Omega, n)}$. Equivalently, the real span of $P_\star, X_\Omega^{(1)},\ldots, X_\Omega^{(n)}$ contains a witness relative to $\mathcal{D}_{\star, \mathrm{sep}}$. A short proof and the precise condition for an operator expectation to be determined by the retained coefficients are given in Appendix~\ref{app:witness}. 

The spectroscopic aspect of the witness criterion lies in determining which observables are accessible through the measurement. In Sec.~\ref{subsec:spectroscopic-map} accessible observables from the implemented readout have been derived. Their span determines which witness tests can be evaluated from the retained signal coefficients. In particular, no separate measurement of $W_\Omega(\mathbf{u})$ in Eq.~\eqref{eq:W-h} is needed.
The expectation value of $W_\Omega(\mathbf{u})$ follows directly from the retained coefficients, which implies that different directions can be tested using the same data.
The gap condition in Eq.~\eqref{eq:signal-witness-gap} determines whether the protocol can detect at least one admissible entangled state. To certify entanglement of a particular state, its signal needs to violate Eq.~\eqref{eq:signal-witness-inequality}.

In many cases, the separable reachable set may not be known explicitly. Then, several witness inequalities provide an outer approximation that can detect signals missed by a single fixed inequality.
We apply the polyhedral witness interpretation \cite{SCSZ21} and the half-space approximation approach for joint numerical ranges \cite{SDW17} to the spectroscopic signal space.
Each inequality uses a certified upper bound on a signal combination over all admissible separable states.
Consider a set of $L$ directions $\mathbf{u}_1, \ldots, \mathbf{u}_L$ and bounds $\widehat{h}_\ell \geq h_{\mathrm{sep}}^{(\Omega, n)}(\mathbf{u}_\ell)$, and define
\begin{equation}
\mathcal{O}_L := \bigcap_{\ell = 1}^{L}\left\{\mathbf{s} \in\mathbbm{R}^n: \mathbf{u}_\ell \cdot \mathbf{s}\leq \widehat{h}_\ell \right\}\,.
\end{equation}
Clearly, the polyhedral set $\mathcal{O}_L$ contains the separable reachable set $\mathcal R_{\mathrm{sep}}^{(\Omega,n)}$.
Hence, every signal in $\mathcal R^{(\Omega,n)}\setminus\mathcal{O}_L$ certifies entanglement. 
Finally, adding valid inequalities cannot enlarge $\mathcal{O}_L$ and does not require further measurements in case the retained coefficients remain unchanged.

\section{Interpretation and implementation \label{sec:exp-workflow}}

Before we illustrate the witness construction in Secs.~\ref{sec:examples-excited} and Sec.~\ref{sec:examples-initial}, we discuss how the framework of Sec.~\ref{sec:main} connects a response model to an experimental entanglement test. The discussion is organized around three questions, concerning the role of the experiment, the response model, and how the entanglement test can be applied in practise. 

The first question addresses the role of the experiment: \textit{If the response model is known, what does the measurement add?} A forward simulation requires both a response model and a specified input state.
Then, the response model describes how the input state leads to the signal. At the same time, a prediction of the state requires further information regarding its preparation and any evolution before the chosen reference time at which entanglement of the evolved state is tested.
In contrast, the entanglement test proposed in Sec.~\ref{sec:main} starts from the experimental data and asks if it can be explained by any admissible separable state. For this purpose, the response model determines the signals that separable states can produce, and the experiment supplies the vector of signal coefficients to be tested. If the signal vector lies outside the separable reachable set, entanglement is certified provided the response model correctly describes the measurement. What is important here is that this conclusion does not require a prediction of the particular state produced in the experiment. So even in cases where a simulation predicts entanglement, the witness tells us whether the measured data supports that prediction.

The state tomography of Ref.~\cite{HW13} likewise uses a model of the probe response to infer an unknown prepared state. Here, selected signal coefficients can suffice to certify entanglement without full state reconstruction. To compare the measured coefficients with the separable reachable set, experiment and theory must describe the same measured quantity. Then, the role of calibration is to relate the experimental signal scale to the theoretical one.  In addition, the molecular response model has to describe the measurement for all admissible input states. Therefore, it is important that neither the response model nor its calibration rely on the assumption that the target state is entangled. As discussed in Sec.~\ref{subsec:reachable-sets}, entanglement can be certified only if no admissible separable state can explain the measured coefficients within the allowed model and calibration uncertainties.

This raises a related question: \textit{Can the same measurement data be tested against different response models?} The same recorded signals can be analysed with different response models of the implemented protocolwhich is particularly useful when system--environment couplings are uncertain.
For each model, one first revisits the calibration against the available reference data and, where needed, obtains additional reference measurements.
In case the coefficient definitions and calibration remain unchanged, the measured coefficient vector can be reused. Otherwise, the coefficients and their uncertainties are recalculated from the stored signals. We then derive the effective observables and separable benchmark for each model, using the same coefficient definitions and calibration as in the experimental analysis.
A new measurement of the target signal is not required if the stored record contains all necessary information. Among the response models which are compatible with the measured data, the results show whether entanglement certification persists or depends on particular response assumptions.

\begin{figure}[t]
\centering
\includegraphics[width=\linewidth]{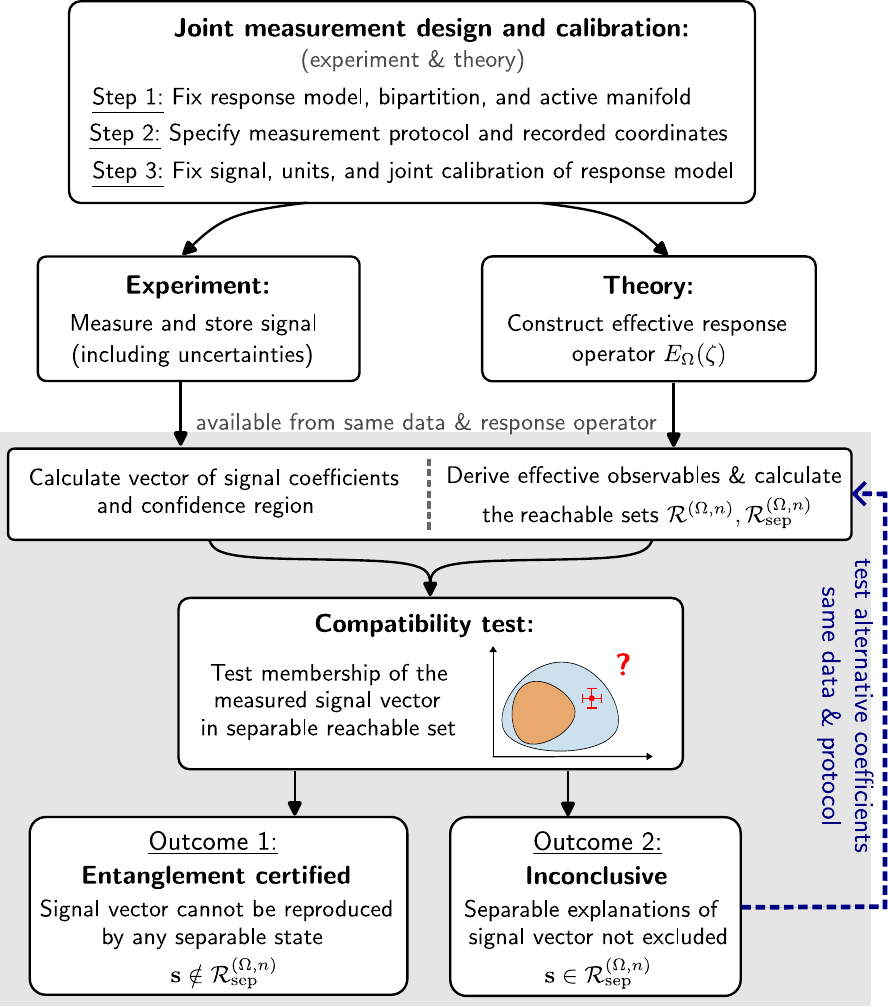}
\caption{Workflow for spectroscopic entanglement certification. A common protocol and calibration connect measured coefficients and separable benchmark. The feedback loop revises coefficient selection from the same data. (See text for more explanations.) \label{fig:workflow-exp}}
\end{figure}

\textit{How would experimentalists and theorists apply the test?}
Figure~\ref{fig:workflow-exp} explains the required steps for one response model. The joint setup which requires both theory and experiment comes first. It specifies the physical bipartition, the active manifold. In addition, additional measurement related decisions such as the reference time of the target state and the measurement protocol are fixed, including the chosen detected coordinates, and readout channels (c.f., Sec.~\ref{subsec:spectroscopic-map}). The explicit details depend on the experiment and vary for different experimental techniques. 
For instance, in the pump--probe example that will be discussed in Sec.~\ref{subsec:excitonic-dimer}, the probe frequency is resolved at fixed preparation and delay.
The fluorescence example in Sec.~\ref{subsec:DBT} instead records a time-resolved intensity and photon coincidences, while the two two-photon absorption examples in Sec.~\ref{sec:examples-initial} require suitable polarization settings.
Together, these decisions fix the recorded variable(s) $\zeta$, which enters the experimental design and the derivation of the effective response operator in Eq.~\eqref{eq:S-linear}. Thus, they must be made before any data is acquired.
Furthermore, theory and experiment need to agree on the signal definition, the chosen units, and calibration as discussed in more detail in the previous paragraphs. 
From there the workflow splits. The experiment records signals from repeated preparations of the same target state, while the theoretical analysis constructs the effective response operator $E_\Omega(\zeta)$. From there onwards all steps can be performed without the need of repeating the measurement and the same operator $E_\Omega(\zeta)$, as indicated by the gray box in Fig.~\ref{fig:workflow-exp}.
The two branches then share one choice of retained coefficients, defined via the linear map in Eq.~\eqref{eq:signal-selec}.
Experimentally, this choice yields the vector $\mathbf{s} = (s_\Omega^{(1)},\ldots, s_\Omega^{(n)})$ and its confidence region.
On the theoretical side, it determines the effective observables $X_\Omega^{(j)}$ and hence the spectroscopic map. After constructing the spectroscopic map, the full and separable reachable sets $\mathcal{R}^{(\Omega, n)}$ and $\mathcal{R}_\mathrm{sep}^{(\Omega, n)}$ are calculated, where $\mathcal{R}_\mathrm{sep}^{(\Omega, n)}$ may be replaced by an outer approximation as discussed in Sec.~\ref{sec:witness}. 

The decisive step is then to test membership of the measured signal vector $\mathbf{s}$ in $\mathcal{R}_\mathrm{sep}^{(\Omega, n)}$ while taking the confidence region into account. 
This is the compatibility test described in Sec.~\ref{subsec:reachable-sets}. There are two possible outcomes as illustrated in Fig.~\ref{fig:workflow-exp}.
The first option is that the experimental confidence region excludes every separable explanation within the allowed model uncertainties. In that case, entanglement is certified under the response model and confidence region.
If the confidence region of the measured signal vector $\mathbf{s}$ overlaps $\mathcal{R}_{\mathrm{sep}}^{(\Omega,n)}$, the test is inconclusive.
There is a third option, which we do not include in Fig.~\ref{fig:workflow-exp}, and that is the case where the full reachable set is not compatible with the data, meaning that $\mathbf{s}$ lies outside of $\mathcal{R}^{(\Omega, n)}$. Since this is not possible in a physically consistent setup, in that case the response model and calibrations need to be revisited. 

Finally, we note that an inconclusive result for one choice of retained signal coefficients does not mean no other spectroscopic map can witness the states' entanglement if present. It is possible to choose a different set of signal coefficients within the same response model and measurement protocol such that same recorded signals can be evaluated with additional or alternative coefficients, as indicated by the blue feedback arrow in Fig.~\ref{fig:workflow-exp}. The analysis can then be repeated for a different spectroscopic map. While not applicable to all spectroscopic techniques, testing different constructions of the signal vector might be particularly helpful in case $s_\Omega^{(j)}$ are constructed from spectral windows (recall Eq.~\eqref{eq:wind-op-gen}). This setup will  be used in the pump-probe example for an excitonic dimer in Sec.~\ref{subsec:excitonic-dimer}.

\section{Examples for witnessing entanglement in the excited state \label{sec:examples-excited}}

We apply the entanglement witness construction of Sec.~\ref{sec:main} to an excitonic dimer and a pair of coupled fluorescent molecules. The first example uses spectral windows from a pump--probe measurement. In the second example we derive suitable effective observables from fluorescence dynamics and photon coincidences and illustrate their ability to witness entanglement. 

\subsection{Excitonic entanglement from pump-probe spectra \label{subsec:excitonic-dimer}}

Entanglement between protein pigments has been investigated in models of photosynthetic light harvesting \cite{CCDHP09, SIFW10, IF10, CCDHP10, BWVU10, FOC10, SMWMB11, TPB12}.
For a state confined to the ground and single excitation sectors, coherence between local excitations implies entanglement across the chromophore partition \cite{SIFW10}. However, this correspondence is dependent on the actual state support, which is not established by a weak illumination alone \cite{TPB12}.
This raises a separate question: Can a pump-probe spectrum exclude every separable explanation of the molecular state?
The answer to this measurement question is yes, as we illustrate here for an excitonic dimer. As a minimal model, we assume that the active manifold comprises only the ground and single excitation manifold. 
The example follows the construction of Sec.~\ref{sec:main}, from the admissible states to the effective response observables and their reachable sets. Our aim is to establish what the accessible optical information can certify. We clearly do not aim to infer a functional advantage of entanglement for energy transfer. 

The two chromophores define the physical subsystems, $\mathcal{H} = \mathbbm{C}_A^2 \otimes \mathbbm{C}_B^2$.
At the chosen pump-probe delay time $T$, we assume that the state has no population in $|ee\rangle$. This implies that the active manifold (c.f.~\ref{subsec:manifold}) is the subspace
\begin{equation}\label{eq:dimer-Hstar}
\mathcal{H}_\star = \mathrm{span}\{\ket{gg}, \ket{eg}, \ket{ge} \} \subsetneq \mathcal{H}\,.
\end{equation}
This realizes the example anticipated in Sec.~\ref{subsec:manifold}.
Moreover, the ground state contains the unexcited population, and all states in $\mathcal{D}_\star \cong \mathcal{D}(\mathcal{H}_\star)$ are normalized. 
Separability across the chromophore partition is assessed across $A|B$ in the four-dimensional
space $\mathcal{H}$, with comparison set $\mathcal{D}_{\star, \mathrm{sep}}$ as defined in Eq.~\eqref{eq:D-star-sep}.
It is important to keep in mind that the absence of double excitation constitutes an input assumption that has to be justified by the preparation an model properties. However, it does not exclude transitions to
$\ket{ee}$ during the probe, which contribute to excited state absorption (see also the corresponding remark in Sec.~\ref{subsec:manifold}).

We remark that the general construction of the full and separable reachable sets does not require the restriction to $\mathcal{H}_\star$. One may instead work with $\mathcal{H}_\star = \mathcal{H}$, which admits double excitation in the dimer input and includes all separable states of the four-dimensional model in the comparison.
This avoids the inference from a truncated state questioned in Ref.~\cite{TPB12}. The general construction of the entanglement witness through the spectral windows remains the same. What needs to be recalculated are the response observables and the separable reachable set.

The next step is to construct the spectroscopic map of Sec.~\ref{subsec:spectroscopic-map}.
The readout is the difference in probe absorption with and without the pump, evaluated at fixed preparation and delay $T$. Moreover, the response follows the weak-probe assumption of Ref.~\cite{HW13} within the rotating wave approximation, with nonoverlapping pulses and a probe short compared with the evolution of the input state. The angle between the polarization vectors of the two pulses is taken as a magic angle with $\cos^2\theta = 1/3$ for which the rotational average of the response factorizes as discussed in Ref.~\cite{HW13}. Then, the isotropic response has the operator representation
\begin{equation}\label{eq:Rpp-main}
R_{\mathrm{PP}}^{\mathrm{iso}}(\omega, \Delta\rho) = \Tr[\mathcal{P}_{\mathrm{iso}}(\omega)\Delta\rho]\,.
\end{equation}
Here, $\omega$ is the detected angular frequency, and $\Delta\rho$ describes the change from the unpumped reference in the sectors to which the response is sensitive. The input state and $\Delta\rho$ denote the corresponding averages over molecular orientations.
The complex operator $\mathcal{P}_{\mathrm{iso}}$ includes the dipole interactions and the subsequent propagation. We show the derivation of the matrix form of $\mathcal{P}_{\mathrm{iso}}$ for the model considered here and its connection to the detected signal in Appendix~\ref{app:dimer-response}. 

Self-heterodyne detection selects the absorptive part of Eq.~\eqref{eq:Rpp-main}, which is represented by the Hermitian operator (see Appendix~\ref{app:dimer-response} for more details)
\begin{equation}\label{eq:dimer-Aabs}
A_{\mathrm{abs}}(\nu) = \frac{\mathcal{P}_{\mathrm{iso}}(2\pi c\nu) - \mathcal{P}_{\mathrm{iso}}(2\pi c\nu)^\dagger}{2i}\,.
\end{equation}
For the illustration, the probe intensity and detection factors are assumed to contribute a constant positive scale and absorbed into a fixed signal unit. Subtraction of the unpumped response then gives
\begin{align}\label{eq:dimer-Enu}
E_\Omega(\nu) &= P_\star A_{\mathrm{abs}}(\nu) P_\star - A_g(\nu) P_\star\,,\\
A_g(\nu) &= \Tr[\rho_g A_{\mathrm{abs}}(\nu)]\,,\\
S_\Omega(\nu) &= \Tr[\rho E_\Omega(\nu)]  = \Tr[(\rho - \rho_g) A_{\mathrm{abs}}(\nu)]\,.
\end{align}
Equation~\eqref{eq:dimer-Enu} expresses the difference from the unpumped reference as an expectation value of the normalized molecular state $\rho\in\mathcal{D}_\star$, as required by Sec.~\ref{subsec:manifold}.
The effective observable $E_\Omega(\nu)$ accounts for ground state bleach, stimulated emission, and excited state absorption into the double excited state $\ket{ee}$, even though it acts on the active manifold $\mathcal{H}\star$. 

From $E_\Omega(\nu)$ in Eq.~\eqref{eq:dimer-Enu}, two spectral coefficients are obtained from the same spectrum using fixed Gaussian weights,
\begin{align}
w_j(\nu) &= \frac{\exp[-(\nu - \nu_j)^2/(2\sigma^2)]}{\sqrt{2\pi}\sigma}\,, \label{eq:dimer-Gaussian} \\
X_\Omega^{(j)} &= \int\mathrm{d}\nu\,w_j(\nu)E_\Omega(\nu)\,,\quad\quad  j=1,2\,.\label{eq:dimber-obs}
\end{align}

\begin{figure}[t]
\centering
\includegraphics[width=\linewidth]{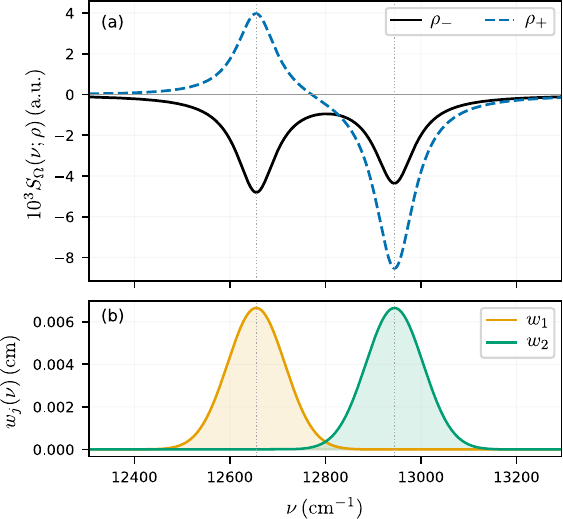}
\caption{Spectral selection for the excitonic dimer. (a) Differential pump-probe spectra of the states $\rho_-$ and $\rho_+$ in Eq.~\eqref{eq:dimer-states-ex}, both with excitation probability $p = 0.2$. (b) Gaussian weights of unit area define the two retained coefficients. Dotted lines mark the exciton transition wavenumbers. All spectra use the same fixed signal unit which is specified in Appendix~\ref{app:dimer-response}. \label{fig:dimer-spectra}}
\end{figure}

To illustrate these concepts, we use the electronic parameters and relative dipoles of the dimer used in Refs.~\cite{HW13, YZKMAG11}.
In the site basis $(|eg\rangle,|ge\rangle)$, the corresponding Hamiltonian block describing the excited state is 
\begin{equation}\label{eq:dimer-H}
\frac{H_e}{hc} =\begin{pmatrix} 12881 & 120\\
120 & 12719 \end{pmatrix} \,\mathrm{cm}^{-1}\,. 
\end{equation}
The local transition dipoles have magnitude ratio $|\boldsymbol d_B|/|\boldsymbol d_A| = 2$ and relative angle
$0.3$ radians. The ground energy is set to zero. Following Refs.~\cite{HW13,YZKMAG11}, we neglect any additional interaction energy when both chromophores are excited.
Then, the energy of $|ee\rangle$ is the sum of the local excitation energies, $E_{ee} = \Tr[H_e]$.
Furthermore, we replace the explicit phonon baths of Ref.~\cite{HW13} by dephasing that gives every optical transition a Lorentzian half-width $\gamma = 50\,\mathrm{cm}^{-1}$.
This propagator retains coherent evolution under $H$ and preserves energy populations. Relaxation of those populations and static disorder are omitted as they are not required to illustrate the witness construction. For comparison to experiment or specific molecular systems, these additional factors would need to be included accordingly.

We pick two example states. Let $|-\rangle$ and $|+\rangle$ be the lower and upper eigenstates of $H_e$. To compare entanglement detection with an inconclusive outcome, we consider as two example target states $\rho_{\pm}$,
\begin{equation}\label{eq:dimer-states-ex}
\rho_\pm = (1 - p)\rho_g + p \ket{\pm}\!\bra{\pm}\,, \quad  \rho_g = \ket{gg}\!\bra{gg}
\end{equation}
with $p = 0.2$. Both states $\rho_{\pm}$ are entangled across the chromophore partition $A|B$, since the nonzero coupling produces coherence between $|eg\rangle$ and $|ge\rangle$ while the double excitation population is assumed to vanish. We choose the states $\rho_{\pm}$ to illustrate different outcomes of the entanglement test in the following. It is important to recall that the full and separable reachable sets are constructed from all admissible states and all admissible separable states, respectively.
An experimentally prepared state need not coincide with either example, and can map to any signal vector in the full and separable reachable sets. 

Figure~\ref{fig:dimer-spectra} displays the two example spectra and the selected regions. 
The centres of the two Gaussian in Eq.~\eqref{eq:dimer-Gaussian} are chosen as the lower and upper exciton wavenumbers, $\nu_1 \approx 12655.2\,\mathrm{cm}^{-1}$ and $\nu_2\approx 12944.8\,\mathrm{cm}^{-1}$, with $\sigma = 60\,\mathrm{cm}^{-1}$.
The windows retain different combinations of the population and coherence contributions to the response.
The spectral windows chosen here illustrate the construction and are not assumed to be optimal. For a given molecular system and experimental setup, alternative weights can be assessed by the separation they provide between the target signal and the separable reachable set. The relevant goal is a separation that remains conclusive after experimental and response model uncertainties are included, as further explained in Sec.~\ref{subsec:reachable-sets}.

To witness entanglement of states, we need to determine the full and separable reachable sets. 
The spectrosopic map $F_\Omega^{(2)}(\rho) = (\Tr[\rho X_\Omega^{(1)}], \Tr[\rho X_\Omega^{(2)}])$ build from the operators in Eq.~\eqref{eq:dimber-obs} determines the full and separable reachable sets of Sec.~\ref{subsec:reachable-sets}.
On the active manifold $\mathcal{H}_\star$, the response operators have the block form $X_\Omega^{(j)} = 0\oplus X_{\Omega, e}^{(j)}$.
The zero block follows from subtraction of the ground reference and coherences between excitation sectors are not detected within the given model
At the same time, the variable excited population remains part of the normalized input domain.
To derive the full and separable reachable sets, we write $X_{\Omega,e}^{(j)} = a_j\mathbbm{1}_e + \mathbf{b}_j \cdot \boldsymbol{\sigma}$, where $\boldsymbol{\sigma}$ are the Pauli matrices in the site basis, and define $\mathbf{a} = (a_1,a_2)$ and $B_{jk} = (\mathbf{b}_j)_k$.
The normalized excited states have Bloch vectors $\mathbf{r}\in\mathbbm{R}^3$ with $\|\mathbf{r}\|\leq 1$. Therefore, the full reachable set $\mathcal{R}^{(\Omega,2)}$ follows as
\begin{align}\label{eq:dimer-full-range}
\mathcal{R}^{(\Omega,2)} &= \mathrm{conv}(\{\boldsymbol{0}\}\cup\mathcal{E}_\Omega)\,,\\
\mathcal{E}_\Omega &= \{\mathbf{a} + B\mathbf{r}:\|\mathbf{r}\|\leq 1\}\,.
\end{align}
For the chosen parameters, $B$ has rank two and $\mathcal{E}_\Omega$ is a filled ellipse.
Thus, the two coefficients $s_\Omega^{(1/2)}$ give a genuinely two-dimensional signal set. 

For any admissible separable state, positivity of its partial transpose and the absence of double excitation imply $\bra{eg}\rho\ket{ge}=0$ and thus its signal depends only on the site populations. Conversely, every allowed population distribution is realized by a separable diagonal state, which implies that
\begin{align}\label{eq:dimer-Rsep}
\mathcal{R}_{\mathrm{sep}}^{(\Omega, 2)} &= \mathrm{conv}\{\boldsymbol{0}, \mathbf{v}_A, \mathbf{v}_B\}\,,\\
\mathbf{v}_A &= F_\Omega^{(2)}(\ket{eg}\!\bra{eg})\,,\\
\mathbf{v}_B &= F_\Omega^{(2)}(\ket{ge}\!\bra{ge})\,.
\end{align}
In particular, for any direction $\mathbf{u}$, the separable support function of Sec.~\ref{sec:witness} takes the simple form $h_{\mathrm{sep}}^{(\Omega, 2)}(\mathbf{u}) = \max\{0, \mathbf{u}\cdot\mathbf{v}_A, \mathbf{u}\cdot\mathbf{v}_B\}$.

\begin{figure}[t]
\centering
\includegraphics[width=\linewidth]{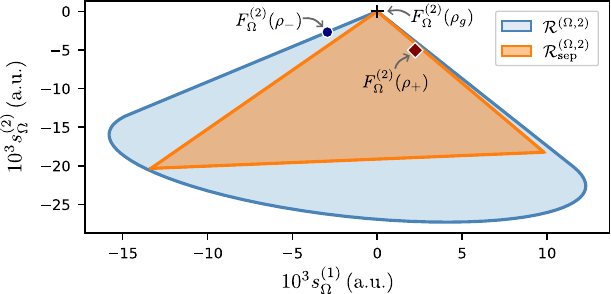}
\caption{Full (blue) and separable (orange) reachable sets for the two spectral coefficients defined through the spectral windows in Fig.~\ref{fig:dimer-spectra}. The set $\mathcal{R}^{(\Omega, 2)}$ is the convex hull of the origin and the excited state ellipse (c.f., Eq.~\eqref{eq:dimer-full-range}, and $\mathcal{R}^{(\Omega, 2)}_\mathrm{sep}$ is the triangle with vertices given by the ground state and the two local excitations.
The lower exciton state $\rho_-$ (blue circle) as defined in Eq.~\eqref{eq:dimer-states-ex} is detected, whereas the upper exciton example state $\rho_+$ (red diamond) also admits a separable explanation.
The cross marks the separable ground state $\rho_g$.\label{fig:dimer-reachable-sets}}
\end{figure}

We compare in Fig.~\ref{fig:dimer-reachable-sets} the full and separable reachable sets $\mathcal{R}^{(\Omega,2)}$ (blue) and $\mathcal{R}_{\mathrm{sep}}^{(\Omega,2)}$ (orange). Their analytic expressions are given in Eqs.~\eqref{eq:dimer-full-range} and \eqref{eq:dimer-Rsep}. 
The three markers depict the signals of the states $\rho_\pm$ in Eq.~\eqref{eq:dimer-states-ex} (blue and red) and the ground state reference $\rho_g$, which maps to the origin (black cross).
The signal of $\rho_-$ (blue circle) lies outside the separable reachable set. At the same time, each individual coefficients $s_\Omega^{(1/2)}(\rho_-)$ is contained in its corresponding separable interval.
Thus, while neither of the two coefficients alone certifies entanglement, their joint values cannot be explained by any admissible separable state, which certifies its entanglement.
The signal of the entangled state $\rho_+$ (red diamond) is contained in the separable reachable set. Therefore, there exists a separable state in $\mathcal{D}_{\star, \mathrm{sep}}$ which reproduces both coefficients. It follows that our entanglement test is inconclusive for $\rho_+$ despite $\rho_+$ in Eq.~\eqref{eq:dimer-states-ex} is entangled. 
The two different outcomes of the membership test illustrate at the same time the advantage of joint spectral constraints and the limits of the information they can provide.

To understand this further, we note that in the present model, an admissible separable state diagonal in the site basis reproduces the entire absorptive spectrum of $\rho_+$. The spectra of these two states are combinations of the same two Lorentzian profiles, whose amplitudes are determined uniquely by the two Gaussian integrals. Therefore, they cannot be distinguished by any spectral windows of this spectrum. What this means is that the inconclusive result of the entanglement test reflects information absent from the full spectrum and not information lost through the chosen windows.
Furthermore, the matching separable state has a different excitation probability from $\rho_+$. It remains an admissible separable explanation of the observed signal because the comparison of membership in $\mathcal{R}^{(\Omega, 2)}$, $\mathcal{R}_\mathrm{sep}^{(\Omega, 2)}$ allows for any excitation probability, even though the plotted state has $p = 0.2$.
At the same time, the same observables in Eq.~\eqref{eq:dimber-obs} detect the pure upper exciton $|+\rangle\langle+|$. This illustrates that the ability to witness entanglement in this example depends on the full input state rather than only on which exciton is excited. 
In summary, we have seen that two spectral integrals suffice to certify entanglement in the above dimer model. With suitable state and response models, this approach could test for entanglement between pigments during energy transfer in light-harvesting complexes without full tomography. 

A further direction is to explore how temperature affects entanglement certification from pump-probe spectra. In a thermal phonon bath model, the relaxation and dephasing rates depend on the temperature, which enters both the system evolution up to the chosen delay $T$ and the subsequent optical response. As a consequence, the effective response operators that define the spectroscopic map become temperature dependent as well.

\subsection{Entanglement tests from fluorescence \label{subsec:DBT}}

The effective observables available for entanglement certification depend on the measurement protocol. In the pump-probe example of Sec.~\ref{subsec:excitonic-dimer}, different spectral windows provide coefficients for the same molecular state at a fixed preparation and delay. Fluorescence measurements require a different approach when the excitation laser frequency is varied. Changing the laser detuning in general leads to a different prepared state. This has the consequence that fluorescence intensities do not represent joint expectation values of a single state. Instead, we construct entanglement tests from fluorescence dynamics and photon coincidences measured for the same preparation.

\subsubsection{Molecular system}

As an experimentally motivated illustration, we use a pair of coupled dibenzoterrylene (DBT) molecules as studied in Ref.~\cite{LDWHH24}, and derive a set of effective observables and corresponding full and separable reachable sets in this section. The experiment resolves collective resonances and measures fluorescence lifetimes and photon correlations. However, agreement with a model that predicts entanglement as in  Ref.~\cite{LDWHH24} does not exclude the possibility that separable states may reproduce the measured information. This raises the following question: Can the fluorescence record also certify entanglement of the prepared molecular pair?

To answer this question, we derive in the following effective observables from the fluorescence intensity and photon coincidences, with the molecular evolution included as in Sec.~\ref{subsec:spectroscopic-map}.
The drive is assumed to vanish completely and instantaneously after preparation of the steady state.
Each molecule is represented by its ground and excited electronic levels. We are interested in bipartite entanglement across the partition $A|B$. The effective Hilbert space of the molecule pair is $\mathcal{H}_\star = \mathcal{H} = \mathbbm{C}_A^2 \otimes \mathbbm{C}_B^2$ such that active manifold and full Hilbert space coincide. Thus, we include the doubly excited state $|ee\rangle$, in contrast to the preceding excitonic example of Sec.~\ref{subsec:excitonic-dimer}. 
Let $\omega_L$ denote the laser frequency and $\omega_i, i=A, B$ the transition frequencies of the individual molecules $A, B$. 
At each fixed detuning $\delta = \omega_{\mathrm{L}} - (\omega_A + \omega_B)/2$, the target state $\rho_\delta$ is the stationary state of the continuously driven model immediately before the laser is removed. We provide the details of the master equation modelling the time evolution in Appendix~\ref{app:DBT}. 
Specifically, each detuning $\delta$ corresponds to a separate entanglement test, with the same subsequent readout. Moreover, the stationary solutions of the Lindblad master equation determine only the example
inputs. As in Sec.~\ref{subsec:manifold}, the sets $\mathcal{D}_\star$ and $\mathcal{D}_{\star, \mathrm{sep}}$ consist of all density operators on the Hilbert space $\mathcal{H}$ and their separable subset.

\subsubsection{Proposed joint theoretical and experimental workflow}

\begin{figure}[tb]
\includegraphics[width=\linewidth]{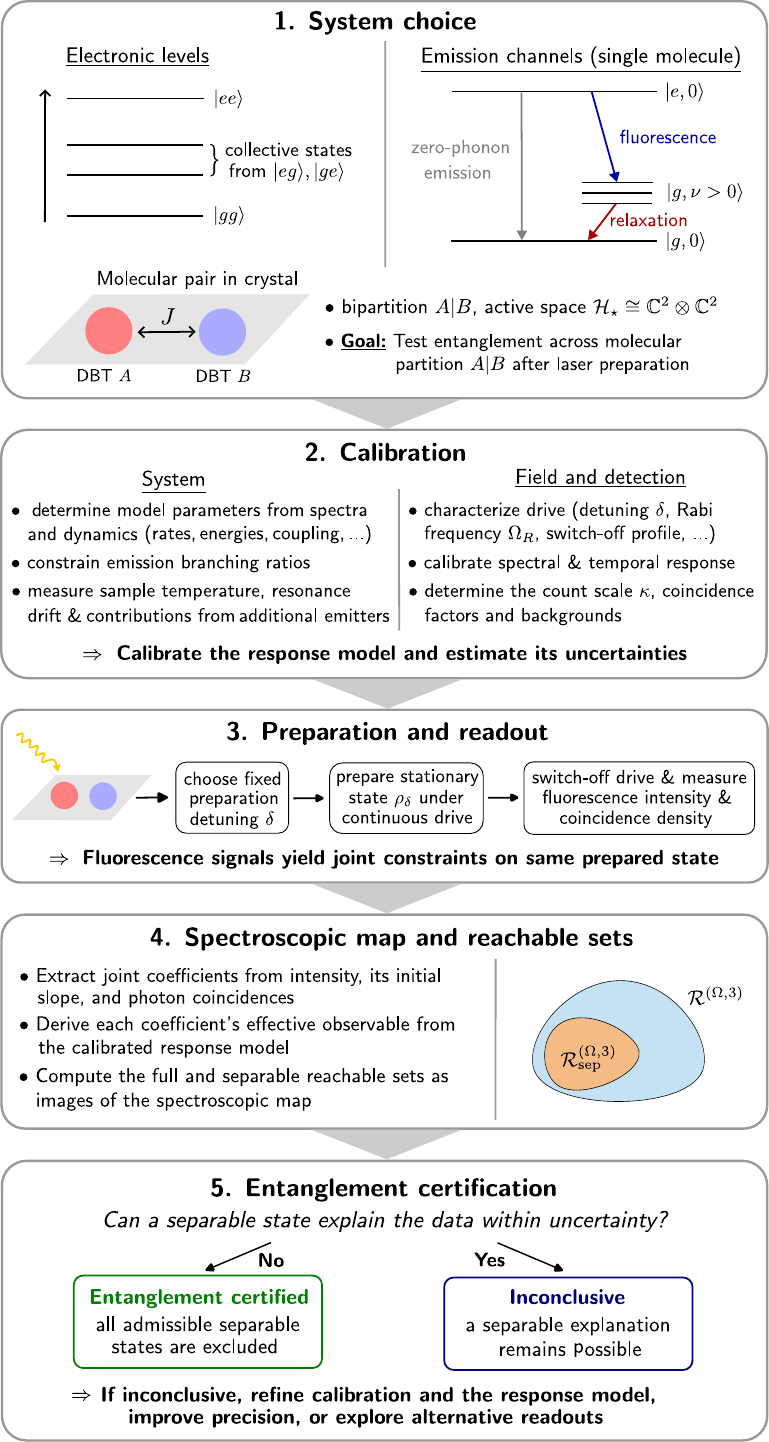}
\caption{Proposed workflow for fluorescence-based entanglement certification
of a DBT pair. Fluorescence and photon coincidences at fixed preparation provide
the measured signal coefficients. The corresponding effective observables are derived from the calibrated response model and define the full and separable reachable sets. Entanglement certification requires the joint confidence region of the measured signal coefficients (i.e., expectation values of suitable effective observables) to remain disjoint from all separable reachable sets consistent with the calibration and model uncertainties. \label{fig:DBT-workflow}}
\end{figure}

In Fig.~\ref{fig:DBT-workflow} we propose a joint experimental and theoretical workflow for witnessing entanglement in a pair of DBT molecules in an anthracene nanocrystal. The active manifold comprises the ground and excited electronic states of each molecule. Entanglement is tested with respect to the molecular partition $A|B$. The singly excited states $\ket{eg}$ and $\ket{ge}$ are mixed by the dipole--dipole coupling $J$. Here the doubly excited state $\ket{ee}$ is included since its population contributes to photon coincidences used for constructing the spectroscopic map below. 
We follow Ref.~\cite{LDWHH24} for the detection scheme, which records Stokes-shifted fluorescence and filters out emission on the zero-phonon line. Here zero-phonon decay from either $\ket{eg}$ or $\ket{ge}$ results in $\ket{gg}$, with both molecules in their vibrational ground states.
Suppose that the emitted photon occupies the same optical mode in both cases.
Then the two paths end in the same final state of the molecule pair and field, and their amplitudes interfere. Since the detected fluorescence intensity is proportional to the remaining excitation number, this interference enters its initial slope. We use this property to construct the spectroscopic map below.

Calibration requires reference measurements of both the molecular pair and the optical setup.
The individual transition frequencies and the coupling $J$ follow from weak-drive excitation spectra recorded at several detunings between the molecules. The model parameters needed to match theory and experiment such as the coupling $J$, the total decay rate $\gamma$, and the zero-phonon branching parameter $\alpha$ may be estimated by fitting a model jointly to fluorescence decay curves and second-order photon correlations as in Ref.~\cite{LDWHH24}.
Moreover, when the detuning between the molecular transition frequencies greatly exceeds the coupling $J$, each molecule behaves effectively as an independent emitter. Its measured excited state lifetime determines the linewidth which originates from population decay alone. Comparing this value with the homogeneous linewidth of the zero-phonon line yields an approximate pure dephasing rate. 
Furthermore, the fitted $\alpha$ can be verified by measuring the zero-phonon branching ratio of a single molecule at large intermolecular detuning. 

The spectroscopic reference measurements and the fluorescence readout should be performed at the same temperature, since phonon interactions affect both the homogeneous linewidth of the zero-phonon line and the distribution of emission between that line and the sidebands \cite{CSMISCMC20}. Furthermore, a change in the local environment can shift the molecular resonance frequencies. At fixed laser frequency, a shift in their mean changes the preparation detuning $\delta$. In the model described below and in Appendix~\ref{app:DBT}, the detuning $\delta$ enters through the driven steady state $\rho_\delta$ whose entanglement is tested. Under the stated model assumptions, this has the consequence that the spectroscopic map and, hence, the full and separable reachable sets are independent of $\delta$. However, an error in the estimated detuning $\delta$ can lead to a difference between a forward simulated simulated and measured vector of joint signal coefficients by changing the steady state used in the simulation.
Furthermore, additional emitters may be detected through further peaks in the frequency scan. They can alter the photon coincidences used to determine the double excitation population of the molecule pair, and thus the spectroscopic map specified below. This contribution and its uncertainty need to be considered when testing membership of a signal vector in the full and separable reachable sets.
Furthermore, preparation and readout are repeated to accumulate photon statistics. Even if the local charge environment remains stable during an individual readout, it may vary between repetitions. 
If the readout is linear in the input density operator for each configuration, Appendix~\ref{app:ensemble-res} applies. Then, the measured signal is modelled as an average of the corresponding signals and weighted by the calibrated configuration probabilities. 
For a fixed distribution of configurations, the full reachable set consists of the signal vectors obtained via the spectroscopic map by allowing any admissible input state for each configuration. The separable reachable set follows similarly. In case there are several distributions that are consistent with the calibration, the union over those distributions is taken separately for the full and separable reachable sets. 

Following our discussion in Sec.~\ref{sec:exp-workflow}, the measured signal can be compared with the signal predicted for the steady state $\rho_\delta$ from a forward simulation at the calibrated laser settings. The forward simulation requires knowledge of the detuning $\delta$ and the Rabi frequency $\Omega_R$. In the model below we assume an instantaneous switch-off of the laser field. However, this idealization may not hold in practise leading to a change in the fluorescence intensity at early times and the photon coincidences used in the witness construction. This means that the measured time dependence of the field needs to be included in the model whenever this effect is non-negligible. 

As illustrated in the second panel in Fig.~\ref{fig:DBT-workflow}, the response model includes information about the fluorescence detection itself. 
The detected count rate per excitation $\kappa$ relates the initial fluorescence rate to the number $n$ of electronic excitations. 
A fixed timing offset shifts all recorded arrival times by the same amount relative to the start of laser switch-off. Random timing errors, by contrast, can record photons that arrive at the same delay at different times. It is important to include both effects in the calculation of the signal coefficients defined in Eqs.~\eqref{eq:DBT-n-main}-\eqref{eq:DBT-d-main} below. A photon pair contributes to the coincidence count when one photon is registered by each detector and the associated detection probability depends on the efficiency of the detectors. Finally, we account for background contributions to the fluorescence and coincidence counts in the response model used to compare the measured signal coefficients with the full and separable reachable sets.

For each test, the workflow is then to repeat the preparation at fixed drive settings and record fluorescence and the unnormalized photon coincidences after the laser drive has been turned off, as illustrated in the third panel in Fig.~\ref{fig:DBT-workflow}. Below we further explain the spectroscopic map and how the joint signal coefficients are derived from these measured data. Then, the next step is to calculate the full and separable reachable sets as shown in the fourth panel. Here it is important that the experimental comparison accounts for uncertainty in the calibration discussed above. The final step is to test membership of the signal vector in the full and separable reachable sets (see panel five). As explained in Sec.~\ref{subsec:reachable-sets}, entanglement is certified if the confidence region for the signal vector is disjoint from the separable reachable set.

\subsubsection{Effective observables and spectroscopic map}

The experiment measures fluorescence from transitions to vibrational levels of the electronic ground state \cite{LDWHH24}. For the illustrations of the full and separable reachable sets, we describe emission into these channels before collection and detection losses.
As explained in more detail in Appendix~\ref{app:DBT}, the operator associated with the photon detection count is $C_{\mathrm{det}} = \kappa N$, with $\kappa = \eta(1-\alpha)\gamma$.
Here, $\eta$ is the collection and detection efficiency, which is assumed equal for both molecules, and an experimental use of this relation requires calibration of $\kappa$. Moreover, $\alpha$ is the combined Debye--Waller and Franck--Condon factor, and $\gamma$ denotes the total decay rate. 

Let $\sigma_A =\ket{g_A}\!\bra{e_A}\otimes \mathbbm{1}_B$ and $\sigma_B = \mathbbm{1}_A\otimes\ket{g_B}\!\bra{e_B}$ denote the local lowering operators of molecules $A, B$. Moreover, the total excitation number operator is $N=n_A+n_B$ with $n_i = \sigma_i^\dagger\sigma_i$. Then, the observable for the detected photon rate follows as $C_{\mathrm{det}} = \kappa N$.
The readout protocol $\Omega$ fixes the evolution after removal of the drive, generated by the Lindblad generator $\mathcal{L}_{\mathrm{off}}$ specified in Appendix~\ref{app:DBT}. Following the adjoint construction of Sec.~\ref{subsec:spectroscopic-map}, the operator $C_{\mathrm{det}}$ allows us to express the fluorescence intensity $I_\Omega(t;\rho)$ as the expectation value
\begin{equation}\label{eq:DBT-I}
I_\Omega(t;\rho) = \Tr[\rho E_\Omega(t)]\,, \quad E_\Omega(t) = e^{t\mathcal L_{\mathrm{off}}^*}(\kappa N). 
\end{equation}
Here, $\rho$ is the input at $t=0$, immediately after the ideal the driving laser is switched off.
The fluorescence signal is resolved in time after a fixed preparation, and the subsequent molecular evolution is included in $E_\Omega(t)$. Therefore, we can use Eq.~\eqref{eq:DBT-I} in the following to construct signal coefficients which refer to the input state $\rho$ at the moment the drive is removed.

\begin{figure*}[tb]
\includegraphics[width=\linewidth]{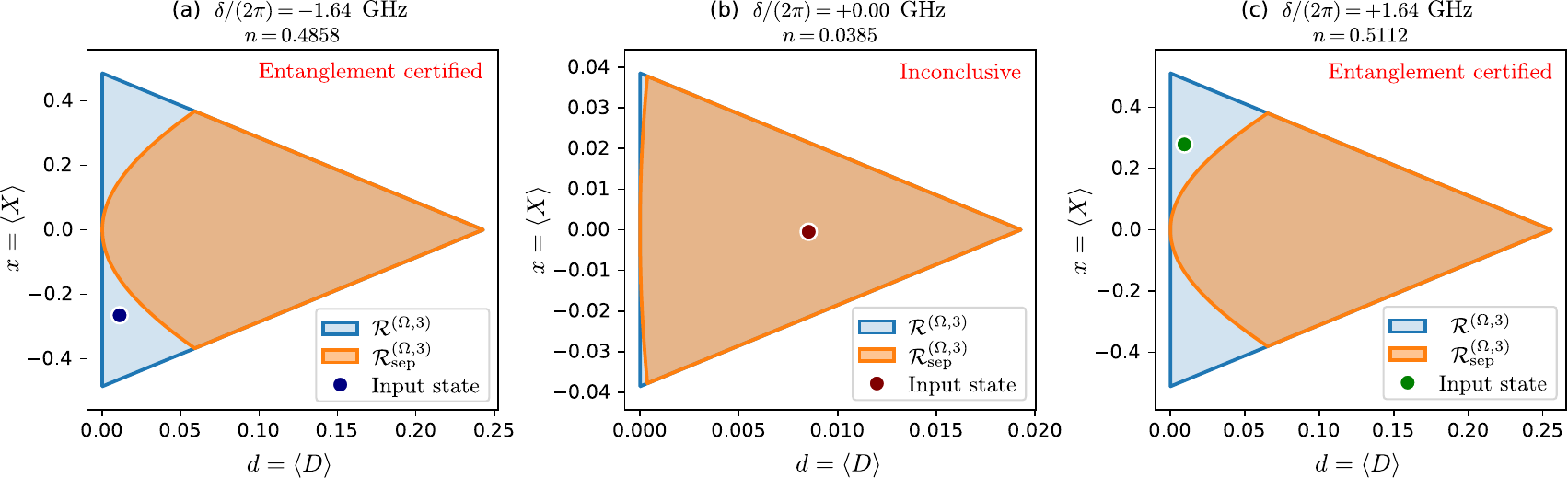}
\caption{Cross sections of the full (blue) and separable (orange) reachable sets in Eqs.~\eqref{eq:DBT-R} and \eqref{eq:DBT-Rsep} at $n = n_\delta$ of the steady-state whose entanglement is tested. 
The markers correspond to the input signals for the lower-resonance (a), central (b), and upper-resonance (c) preparations. The two resonance signals certify entanglement under the readout assumptions, while the central signal remains inconclusive (see text for more information).\label{fig:DBT-slices}}
\end{figure*}

The initial fluorescence intensity determines the average number of electronic excitations, $n = \Tr[\rho N]$. The full and separable ranges of $n$ both equal $[0,2]$. Thus, $n$ alone cannot certify entanglement.
However, the subsequent fluorescence can provide further information because collective decay through the undetected zero-phonon channel affects the molecular evolution. To describe this sensitivity and the information supplied by photon coincidences, we introduce the operators
\begin{equation}
X = \sigma_A^\dagger\sigma_B + \sigma_B^\dagger\sigma_A\,,\quad  D = n_A n_B = \ket{ee}\!\bra{ee}\,.
\end{equation}
The expectation value $x=\Tr[\rho X] = 2\mathrm{Re}\bra{ge}\rho\ket{eg}$ is twice the real part of the exchange coherence, while $d = \Tr[\rho D]$ describes the probability that both molecules are excited.

Furthermore, the three expectation values $n, x$ and $d$ can be determined from the fluorescence intensity $I_\Omega(t;\rho)$ and the unnormalized photon-coincidence density $G_\Omega^{(2)}(0, 0; \rho)$ specified in Appendix~\ref{app:DBT}.
For fixed model parameters and detection scale, with $\alpha>0$ and $\kappa>0$, we have
\begin{align}
n &= \Tr[\rho N] = \frac{I_\Omega(0; \rho)}{\kappa}\,,\label{eq:DBT-n-main}\\
x &= \Tr[\rho X]  = -\frac{\dot{I}_\Omega(0^+; \rho) + \gamma I_\Omega(0; \rho)}{\alpha\gamma\kappa}\,,\label{eq:DBT-x-main}\\
d &= \Tr[\rho D] = \frac{G_\Omega^{(2)}(0, 0; \rho)}{2\kappa^2}\,. \label{eq:DBT-d-main}
\end{align}
Here, $0^+$ indicates that the derivative $\dot{I}_\Omega(0^+;\rho)$ is evaluated from the right. We derive these relations from the Lindblad generator of the molecular system in Appendix~\ref{app:DBT}. 
Equations~\eqref{eq:DBT-n-main}-\eqref{eq:DBT-d-main} determine the three effective observables $(X_\Omega^{(1)},X_\Omega^{(2)},X_\Omega^{(3)}) = (N,X,D)$ used for the definition of the spectroscopic map which gives the full and separable reachable sets,
\begin{equation}\label{eq:DBT-map}
F_\Omega^{(3)}(\rho) = (n, x, d)\,.
\end{equation}
Similar to the spectral integrals used for the excitonic dimer in Sec.~\ref{subsec:excitonic-dimer}, the
expectation values of $N, X, D$ follow from fixed linear operations on the obtained signals.

\subsubsection{Reachable sets and entanglement certification}

Moreover, the calculation of $n, x, d$ via the expectation values in Eqs.~\eqref{eq:DBT-n-main}-\eqref{eq:DBT-d-main} are valid for every admissible input state $\rho\in \mathcal{D}_\star$ and not only for the stationary states of the assumed Lindblad model.
The images of $\mathcal{D}_\star$ and $\mathcal{D}_{\star, \mathrm{sep}}$ under the spectroscopic map in Eq.~\eqref{eq:DBT-map} can be calculated analytically, which leads to the full and separable reachable sets
\begin{align}
\mathcal{R}^{(\Omega, 3)} = \left\{(n, x,  d) : d\geq 0, 1-n+d \geq 0, |x| \leq n - 2d\right\}\,,\label{eq:DBT-R}\\
\mathcal{R}_{\mathrm{sep}}^{(\Omega, 3)} = \left\{(n, x, d)\in \mathcal{R}^{(\Omega, 3)}:  |x|\leq 2 \sqrt{d(1-n+d)}\right\}\,.\label{eq:DBT-Rsep}
\end{align}
We provide the derivation of Eqs.~\eqref{eq:DBT-R} and \eqref{eq:DBT-Rsep} in Appendix~\ref{app:DBT}. 

With the sets $\mathcal{R}^{(\Omega, 3)}, \,\mathcal{R}_{\mathrm{sep}}^{(\Omega, 3)}$ we are now able to test entanglement of particular states. Based on Ref.~\cite{LDWHH24} we choose the steady states of the Lindblad master equation as specified in Appendix~\ref{app:DBT}. Different detunings $\delta$ give different steady states, which we denote by $\rho_\delta$. We then calculate their mean electronic excitation number $n \equiv n_\rho = \Tr[N\rho_\delta]$ and plot in Fig.~\ref{fig:DBT-slices} the corresponding slices of the full and separable reachable sets for fixed $n_\delta$ and three different values of the detuning. The system parameters are chosen in accordance with Ref.~\cite{LDWHH24} and are $\Delta/(2\pi) = (\omega_A-\omega_B)/(2\pi)= 2.6\,\mathrm{GHz}$, the exchange coupling $J/(2\pi)=1.0\,\mathrm{GHz}$, $\gamma/(2\pi) = 33\,\mathrm{MHz}$, the local pure dephasing coefficient $\gamma_\phi/(2\pi)=10\,\mathrm{MHz}$ and $\alpha=0.11$. Both molecules experience the same drive phase and Rabi frequency $\Omega_R/(2\pi)\approx 138.4\,\mathrm{MHz}$ (see Appendix~\ref{app:DBT} for more details on the used rate conventions and conversions). 

We compare in Fig.~\ref{fig:DBT-slices} the ability of the reachable sets to witness entanglement for the three detunings $\delta=-\widetilde{\omega}, 0, +\widetilde{\omega}$, where $\widetilde\omega = \sqrt{J^2 + (\Delta/2)^2}$ and $\widetilde\omega/(2\pi)\approx 1.640\,\mathrm{GHz}$. The detunings $\pm\widetilde\omega$ locate the single excitation resonances of the undriven Hamiltonian relative to the mean molecular transition frequency. The choice $\delta=0$ corresponds to the two-photon resonance. Each panel displays the allowed pairs $(d,x)$ at fixed $n = n_\delta$. This is a valid choice as any admissible separable explanation of the fluorescence signal must reproduce all three coefficients, including the fixed value of $n$. The geometry of the full sets $\mathcal{R}^{(\Omega, 3)}, \,\mathcal{R}_{\mathrm{sep}}^{(\Omega, 3)}$ is independent of the detuning $\delta$, and only their slices for different $n=n_\delta$ differ. In particular, we find that while entanglement can be witnessed for $\delta = \pm\widetilde{\omega}$, the test is inconclusive for $\delta=0$.

\begin{figure}[tb]
\includegraphics[width=\linewidth]{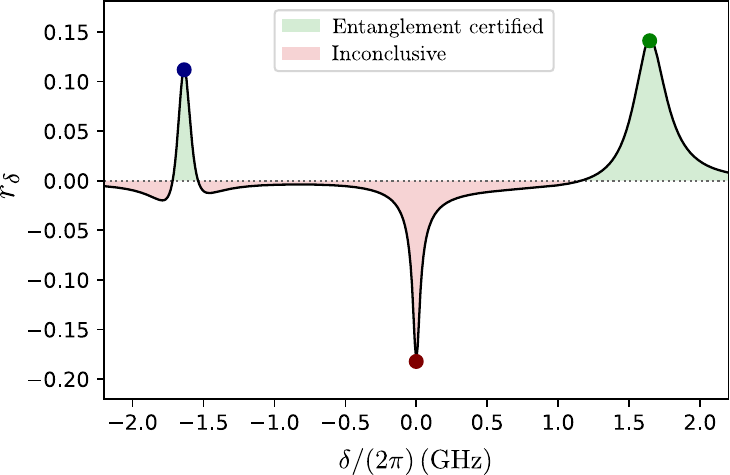}
\caption{Residual $r_\delta$ in Eq.~\eqref{eq:DBT-margin} versus preparation detuning for the fixed fluorescence readout. Green shaded areas below the residual curve mark positive residuals that certify entanglement, while red shading indicates inconclusive results. Circles indicate the three preparations in Fig.~\ref{fig:DBT-slices}. \label{fig:DBT-residual}}
\end{figure}

To elaborate on this detuning dependence of the witnessing power via the selected coefficients, we have a closer look at the separable reachable set in Eq.~\eqref{eq:DBT-Rsep} and the inequality in its definition. For each target $\rho_\delta$, we define the signed residual of this inequality as
\begin{equation}
r_\delta = |x_\delta| - 2\sqrt{d_\delta(1-n_\delta+d_\delta)}\,,
\label{eq:DBT-margin}
\end{equation}
where $x_\delta = \Tr[\rho_\delta X], \, d_\delta = \Tr[\rho_\delta D]$ in analogy to $n_\delta$. 
For a physical signal vector $F_\Omega^{(3)}(\rho_\delta)$, membership in $\mathcal{R}_{\mathrm{sep}}^{(\Omega, 3)}$ is equivalent to $r_\delta\leq 0$. A positive residual therefore excludes every admissible separable explanation of the three retained coefficients. For $r_\delta\leq 0$, the test is inconclusive.
The residual summarizes the joint test of $n_\delta,x_\delta,d_\delta$ and displays its outcome across the detuning scan in Fig.~\ref{fig:DBT-residual}. Green shaded areas mark positive residuals and hence entanglement certification under the stated model. In contrast, red shaded areas illustrate negative residuals, for which the retained coefficients admit a separable explanation. The circles mark the three preparations in Fig.~\ref{fig:DBT-slices}, with the same colours. What is important here is that a nonpositive residual does not imply that the state $\rho_\delta$ is separable. In fact, the steady states have positive negativity at every sampled detuning as shown in Fig.~\ref{fig:DBT-negativity} in Appendix~\ref{app:DBT}. 
Therefore, the states corresponding to the red regions are entangled, but their signal coefficients are also compatible with some separable state in $\mathcal{D}_{\star, \mathrm{sep}}$. 

Furthermore, Appendix~\ref{app:DBT} shows that $d$ is essential for certification of the two resonance inputs at these parameters. In fact, if $d$ is omitted and only $n, x$ are retained, also the entanglement test for $\delta = \pm \widetilde{\omega}$ becomes inconclusive since a separable projection permits $|x|\leq n(2-n)/2$, and this condition is satisfied by both resonance input states.
The independent constraint on double excitation therefore removes separable explanations compatible with $n$ and $x$ alone and allows us to witness their entanglement according to Figs.~\ref{fig:DBT-slices} and \ref{fig:DBT-residual}.

In summary, this example shows how fluorescence intensity and photon coincidence can jointly certify molecular entanglement without full state reconstruction. The conclusion depends on the response model, whose effective observables define the separable reachable set in signal space for the chosen molecular bipartition and admissible state space. Therefore, an inaccurate response model can result in an incorrect entanglement claim. The present illustration uses ideal signals and does not yet certify entanglement from the published data of Ref.~\cite{LDWHH24}. An application of the witness construction to similar experiment requires a response model that includes the detector response and the actual removal of the laser drive, which depend on the specific setting. 
In addition, a calibration of fluorescence intensity and coincidence records is required as explained in Sec.~\ref{subsec:reachable-sets} all inferred signal coefficients need to refer to the same state preparation. Entanglement is certified only when no admissible separable state reproduces all retained coefficients within the stated experimental and model uncertainties.

\section{Examples for witnessing entanglement in the initial state\label{sec:examples-initial}}

In this section, we use the polarization dependence of orientationally averaged two-photon signals to test entanglement in the initial molecular state.
The first example illustrates the construction of the full and separable reachable sets for orbital entanglement in an open-shell system in Sec.~\ref{subsec:RP-TPA}. As a second example we discuss vibrational entanglement in a molecular dimer in Sec.~\ref{subsec:vibronic-dimer}.

\subsection{Orbital entanglement from two-photon absorption \label{subsec:RP-TPA}}

We first illustrate how the coefficients of rotational invariants in Sec.~\ref{subsec:spectroscopic-map} can support entanglement certification after averaging over molecular orientations.
The symmetric two-orbital model of open shell molecules provides a chemically motivated example whose two-photon response depends on the contributions of neutral and ionic configurations~\cite{Nakano09, Nakano17}.
Here, the entanglement certification concerns orbital entanglement across the local Fock spaces associated with the two localized orbitals. Both spin and occupation are included as local degrees of freedom.
For the resolved two-photon transition considered below, the transition dipoles lie along a common molecular axis. Thus, the resulting effective observables are proportional. This results in the two polarization coefficients constrain the same expectation value leading to one-dimensional full and separable reachable sets. 

In this example, we consider two electrons in the localized spatial orbitals $A$ and $B$ with spins $\sigma=\uparrow, \downarrow$.
Each orbital has four local states, $\ket{0},\ket{\uparrow}, \ket{\downarrow}, \ket{\uparrow\downarrow}$. 
Their local Fock spaces define the orbital partition $\mathcal{H}\cong\mathbbm{C}_A^4\otimes\mathbbm{C}_B^4$. For the two electrons, the active manifold $\mathcal{H}_\star$ has dimension six and contains both singly occupied and ionic configurations. The domains $\mathcal{D}_\star$ and $\mathcal{D}_{\star,\mathrm{sep}}$ contain all density operators on this manifold and those separable across $A|B$, respectively, as explained in Sec.~\ref{subsec:manifold}. 
The Hamiltonian and dipole operator specified in Appendix~\ref{app:two-orbital-TPA} determine the molecular
eigenstates and optical couplings. We consider the regime with a singlet ground state $\ket{g}$ and use $P_g = \ket{g}\!\bra{g}$ as the example input for the two-photon transition tensor derived in Appendix~\ref{app:Ax-Bx}.
For the two-photon transition to the higher singlet $\ket{f}$, only the odd singlet $\ket{o}$ contributes to the sum over intermediate states in the two-photon transition tensor.
Both transition dipoles lie along the molecular axis. From the resulting matrix elements, we derive in Appendix~\ref{app:two-orbital-TPA} the explicit form of the transition tensor to obtain the effective response observable.

For the optical readout, the two pulse frequencies satisfy $\omega_1 + \omega_2 = (E_f - E_g)/\hbar$, while each pulse remains far detuned from the allowed one-photon transitions.
We neglect relaxation during the pulses and average their optical phases, which does by construction not affect the input state whose entanglement is to be tested. Under the assumptions specified in Appendix~\ref{app:two-orbital-TPA}, the contribution proportional to the product of the pulse intensities describes the net energy transfer through the $g\leftrightarrow f$ transition. We denote this contribution per molecule, divided by the energy difference $\Lambda = E_f-E_g$, by $S_\Omega(\theta)$, which is the signal in Eq.~\eqref{eq:S-linear}. 

To obtain the angular dependence, we consider an isotropic ensemble with a common input state in the molecular frame. The response derived in Appendix~\ref{app:two-orbital-TPA}, combined with the rotational contractions of Appendix~\ref{app:Ax-Bx}, yields the effective observable
\begin{align}
E_\Omega(\theta) &= C_\Omega (1 + 2\cos^2\theta) (P_g-P_f)\,,\nonumber\\
X_\Omega^{(1)} &= C_\Omega (P_g-P_f), \quad X_\Omega^{(2)} = 2C_\Omega (P_g-P_f)\,.
\label{eq:OS-response}
\end{align}
Here $\theta$ is the angle between the polarization vectors, $P_f$ the projector onto the final state $\ket{f}$, and $C_\Omega > 0$ defined in Eq.~\eqref{eq:OS-pulse-scale} is determined by the transition strength and pulse envelopes. The two operators $X_\Omega^{(1)}$ and $X_\Omega^{(2)}$ represent the angular coefficients of Eq.~\eqref{eq:S-TPA}. Therefore, they define the spectroscopy map in Eq.~\eqref{eq:spectral-map}. Absorption contributes through $P_g$, whereas stimulated two-photon emission leads to the term $-P_f$. Both processes must enter the comparison because admissible separable input states can populate $|f\rangle$ despite the illustrative target $P_g$ has no population in that level.

The signal point $F_\Omega^{(2)}(\rho) = C_\Omega \Tr[\rho (P_g-P_f)] (1, 2)$ follows from fitting the signal to Eq.~\eqref{eq:S-TPA}. 
Thus, the full and separable reachable sets, $\mathcal{R}^{(\Omega,2)}$ and $\mathcal{R}_{\mathrm{sep}}^{(\Omega,2)}$, are line segments since  both signal coefficients $s_\Omega^{(1)}$ and $s_\Omega^{(2)}$ constrain the same quantity. 
The explicit interval of the full and separable reachable sets depends on the difference $U > 0$ between on-site and intersite Coulomb repulsion and the effective transfer magnitude $\tau > 0$ (see Appendix~\ref{app:two-orbital-TPA} for more details).
With $\Lambda = \sqrt{U^2 + 16\tau^2}$, we obtain
\begin{align}\label{eq:OS-ranges}
\mathcal{R}^{(\Omega, 2)} &=\{C_\Omega z(1,2): -1 \leq z\leq 1\}\,, \nonumber\\
\mathcal{R}_{\mathrm{sep}}^{(\Omega, 2)} &= \left\{C_\Omega z(1,2) : |z|\leq\frac{U}{2\Lambda}\right\}\,.
\end{align}
where $z  =\Tr[\rho (P_g-P_f)]$. The derivation of Eq.~\eqref{eq:OS-ranges} is provided in Appendix~\ref{app:two-orbital-TPA}. 

The separable reachable set $\mathcal{R}_{\mathrm{sep}}^{(\Omega, 2)}$ allows for witnessing entanglement. 
For instance, the state $\rho = P_g$ has $z=1$ and thus its orbital entanglement is witnessed. 
In general, any admissible pure or mixed state with $|z|>U/(2\Lambda)$ is certified entangled across the orbital partition. Any value within the interval admits a separable state that
reproduces both coefficients. Then, the entanglement test remains inconclusive, without implying that the input state is separable.

What this open shell example shows is that one independent expectation value can suffice for certification after rotational averaging.
Although the angular fit yields two signal coefficients, their fixed relation $s_\Omega^{(2)} = 2s_\Omega^{(1)}$ means that the second signal coefficient adds no new constraint on the state. The relevant information is contained in the signal amplitude relative to the calibrated scale $C_\Omega$.
The following vibronic example illustrates how two independent polarization coefficients provide additional constraints.

\subsection{Vibrational entanglement in a molecular dimer \label{subsec:vibronic-dimer}}

As a fourth illustration, we use the spectroscopic map based on rotational invariants to witness vibrational entanglement in a molecular dimer.
This example distinguishes two aspects of the measurement: its ability to certify entanglement and the dimension of its reachable sets. We consider two identical fragments $A$ and $B$. Each fragment has one vibrational mode, described by harmonic potentials with different equilibrium positions in the ground and excited electronic states \cite{PKP12}.
The preparation is confined to the two lowest vibrational levels of each fragment in its electronic ground state. These levels define the active input manifold
\begin{equation}\label{eq:vd-HG}
\mathcal{H}_\star = \mathcal{H}_G = \mathrm{span} \{\ket{00}, \ket{01}, \ket{10}, \ket{11}\}
\end{equation}
with $\ket{ij} :=\ket{g_A,i}\otimes\ket{g_B,j}$ for $i,j\in\{0,1\}$. 
Here $\ket{g_Z,i}$ denotes vibrational level $i$ in the electronic ground state of fragment $Z=A, B$.
The set of admissible density operators on that active manifold is $\mathcal{D}_\star = \mathcal{D}(\mathcal{H}_G)$. The goal is to witness vibrational entanglement across the physical fragment partition $A|B$.

The doubly excited electronic manifold is shifted by an exciton--exciton interaction energy $U$ relative to the sum of the local excitation energies \cite{BLPSKS20}. We neglect electronic excitation transfer between the fragments. As in Sec.~\ref{subsec:RP-TPA}, the input state is probed through two-photon absorption.
Two weak Gaussian pulses drive transitions to the selected final vibronic level $\ket{F} = \ket{e_A,0; e_B,0}$. The chosen readout is assumed to resolve the population of $\ket{F}$ before it is redistributed by relaxation processes. The retained signal $S_\Omega(\theta)$ is the population of $\ket{F}$ associated with absorption of one photon from each pulse.
Here $\theta$ denotes the angle between the linear polarizations of the two laser pulses as in Sec.~\ref{subsec:spectroscopic-map}.

The derivation of the effective observable $E_\Omega(\theta)$ for this measurement example is provided in
Appendix~\ref{app:vibronic-complete}. To this end, we evaluate the molecular transition tensor
of Appendix~\ref{app:Ax-Bx} for the dimer and combine it with the finite pulse spectra.
This results in a second-order transition amplitude which includes both chronological orders of absorption of the two pulses. Finally, we perform an average over molecular orientations to obtain an explicit form for $E_\Omega(\theta)$. 
In addition, Appendix~\ref{app:vibronic-complete} explains how the signal $S_\Omega(\theta)$ at leading order in the fields is extracted. For this, the measured populations of $\ket{F}$ are averaged over the relative pulse phase, and the populations measured with each pulse alone are subtracted.
For an isotropic distribution of molecular orientations, the signal takes the form
\begin{equation}\label{eq:vd-Stheta}
S_\Omega(\theta) = \Tr[\rho_G E_\Omega(\theta)] = s_\Omega^{(1)} + s_\Omega^{(2)}\cos^2\theta\,.
\end{equation}
A signal point $\mathbf{s} = (s_\Omega^{(1)}, s_\Omega^{(2)})$ can be extracted through an angular fit of its $\theta$ dependence. 
In the following, we derive the full and separable reachable sets of these signal coefficients for parallel and nonparallel local transition dipoles. This leads to line segments as reachable sets in the first case, and genuinely two-dimensional sets in the second one.

For parallel local transition dipoles, the extracted signal in Eq.~\eqref{eq:vd-Stheta} depends on a single channel vector $\ket{\chi_\parallel}$ on $\mathcal{H}_G$.
The state $\ket{\chi_\parallel}$ contains the information about the complete two-photon transition amplitudes from the four input basis states in Eq.~\eqref{eq:vd-HG} to the selected final level $\ket{F}$.
With $p = \bra{\chi_\parallel}\rho_G\ket{\chi_\parallel}$, the signal takes the form $S_\Omega(\theta) = p(1 + 2\cos^2\theta)/15$.
For a nonzero channel, let $\Lambda_{\max}$ denote
the largest Schmidt coefficient of the normalized
vector $\ket{\chi_\parallel}/\|\chi_\parallel\|$.
Then, the full and separable reachable sets are the line segments
\begin{align}
\mathcal{R}^{(\Omega, 2)} &= \left\{(p/15, 2p/15): 0\leq p\leq \|\chi_\parallel\|^2 \right\}\,,\label{eq:vd-main-R}\\
\mathcal{R}_{\mathrm{sep}}^{(\Omega, 2)} &=  \left\{(p/15, 2p/15): 0\leq p\leq \|\chi_\parallel\|^2 \Lambda_{\max}^2\right\}\,.\label{eq:vd-main-Rsep}
\end{align}
The separable line segment defining $\mathcal{R}_{\mathrm{sep}}^{(\Omega, 2)}$ is strictly shorter than than the line segment admissible for all $\rho_G\in \mathcal{D}_\star$ whenever $\Lambda_{\max} < 1$, that is when the normalized channel vector is entangled.
For the channel vector $\ket{\chi_\parallel}$ to be entangled, both retained vibrational levels of each fragment have to contribute to the transition to $\ket{F}$. 
Specifically,  the complete two-photon transition amplitudes are need to be not factorizable across $A|B$.
Membership of $\mathbf{s}$ in $\mathcal{R}^{(\Omega, 2)}/\mathcal{R}^{(\Omega, 2)}_\mathrm{sep}$ certifies vibrational entanglement of the input state $\rho_G$. 
We refer to Appendix~\ref{app:parallel-vd} for a more detailed derivation of the sets in Eqs.~\eqref{eq:vd-main-R} and \eqref{eq:vd-main-Rsep} and the equivalent criterion for an entangled channel that is stated directly in terms of the transition amplitudes. 

For nonparallel local transition dipoles, the polarization factors for the two possible assignments of the pulses to the fragments can be different. 
The channel vector $\ket{\chi_1}$ contains the information about the transition amplitudes for excitation of fragment $A$ by pulse 1 and $B$ by pulse 2. Conversely, in $\ket{\chi_2}$, pulse 1 excites $B$ and pulse 2 excites $A$.
For the identical fragments and a symmetric final state $\ket{F}$ considered here, the two linear combinations
\begin{equation}
\ket{\chi_s} = \frac{\ket{\chi_1} + \ket{\chi_2}}{2}\,,\quad  \ket{\chi_a} = \frac{\ket{\chi_1} - \ket{\chi_2}}{2}
\end{equation}
are symmetric and antisymmetric under exchange of $A$ and $B$, respectively. 
Let $N_k = \|\chi_k\|^2$ denote their squared norms for $k = s,a$.
The dipole geometry is specified through the parameter $\eta = \mathbf{u}_A\cdot\mathbf{u}_B$, where $\mathbf{u}_A$ and $\mathbf{u}_B$ are unit vectors of the local transition dipoles.
We show in Appendix~\ref{app:vibronic-dimer-2D} that the full and separable reachable sets are both two-dimensional in case the dipoles are non-parallel and both channels are nonzero, i.e., for $|\eta|<1$, $N_s > 0$, and $N_a > 0$. For the simultaneous Gaussian pulses of equal width considered here, different carrier frequencies are required for a second independent signal coefficient requires different carrier frequencies.
The signal takes the form in Eq.~\eqref{eq:vd-Stheta} as for parallel dipoles. 
For nonparallel dipoles, the overlaps of $\rho_G$ with the normalized channel vectors, $p_k = \bra{\chi_k}\rho_G \ket{\chi_k}/N_k$ for $k=s,a$, are related to the fitted coefficients $s_\Omega^{(1)}, s_\Omega^{(2)}$ through
\begin{equation}\label{eq:vd-main-slsla}
\begin{split}
\mathbf{s} &= p_s\bd\lambda_s + p_a\bd\lambda_a\,,\\
\bd\lambda_s &= \frac{N_s}{15}(3 + \eta^2, \, 1 + 7\eta^2)\,,\\
\bd\lambda_a &= \frac{N_a(1 - \eta^2)}{3}(1,\,-1)\,.
\end{split}
\end{equation}
Here, $\bd\lambda_k$ is the image of the normalized channel state $\ket{\chi_k}/\sqrt{N_k}$ under the spectroscopic map. The explicit form of the spectroscopic map is given in Appendix~\ref{app:vibronic-dimer-2D}. 
Equation~\eqref{eq:vd-main-slsla} uniquely determines $p_s$ and $p_a$ from the signal coefficients $s_\Omega^{(1)}, s_\Omega^{(2)}$, provided that the vectors $\bd\lambda_s$ and $\bd\lambda_a$ are linearly independent. Input states $\rho_G\in \mathcal{D}_\star$ that have zero overlap with both channels $\ket{\chi_a}, \ket{\chi_s}$, map to the origin $\mathbf{s} = \mathbf{0}$.

Both the full and separable reachable sets can be calculated analytically as shown in Appendix~\ref{app:vibronic-dimer-2D}. The full reachable set is a triangle with vertices $\bd\lambda_s, \bd\lambda_a$ and $\mathbf{0}$. 
To define the boundary of the separable reachable set, let $c\in[0,1]$ denote the concurrence of the normalized symmetric channel $\ket{\chi_s}/\sqrt{N_s}$. The concurrence $c$ is fixed by the transition amplitudes. Moreover, $c=0$ for a product channel and $c=1$ for a maximally entangled channel. 
Then, the full and separable reachable sets follow as 
\begin{align}
\mathcal{R}^{(\Omega, 2)} &=\mathrm{conv}\{\mathbf{0}, \bd\lambda_s, \bd\lambda_a\}\,, \label{eq:vd-main-nonparallel-R}\\
\mathcal{R}_{\mathrm{sep}}^{(\Omega, 2)} &= \left\{p_s\bd\lambda_s + p_a\bd\lambda_a: 0\leq  p_a \leq \tfrac{1}{2},\, 0 \leq p_s \leq f_c(p_a)\right\}\label{eq:vd-main-nonparallel-Rsep}
\end{align}
where
\begin{equation}\label{eq:vd-main-fc}
f_c(p_a) = \frac{1 - (1 - c)p_a + \sqrt{(1 - c^2)(1 - 2p_a)}}{2}\,.
\end{equation}
For a fixed antisymmetric overlap $p_a$, the value of the function $f_c(p_a)$ is the largest symmetric overlap $p_s$ that can be attained by a separable input state $\rho_G\in \mathcal{D}_{\star, \mathrm{sep}}$.

For identical fragments and symmetric final state $\ket{F}$, the exchange $\ket{ij} \mapsto \ket{ji}$ maps $\ket{\chi_1}$ to $\ket{\chi_2}$. The channel vector $\ket{\chi_a}$ changes sign under this exchange.
The one-dimensional antisymmetric subspace of $\mathcal{H}_G$ in Eq.~\eqref{eq:vd-HG} is spanned by the vibrational singlet $\ket{\psi_-} = (\ket{10} - \ket{01})/\sqrt2$. Therefore, $\ket{\chi_a}/\sqrt{N_a}$ equals the Bell state $\ket{\psi_-}$ up to an overall phase. The corresponding vertex $\lambda_a$ lies outside the separable reachable set. 
In parameter regime of two-dimensional reachable sets, Eq.~\eqref{eq:vd-main-slsla} determines $p_a$ from $s_\Omega^{(1)}$ and $s_\Omega^{(2)}$. Admissible separable states $\rho_G\in \mathcal{D}_{\star, \mathrm{sep}}$ satisfy $p_a\leq 1/2$. Hence, a value $p_a > 1/2$ certifies vibrational entanglement of $\rho_G$. In particular, the singlet input state has $p_a = 1$ and produces the signal point $\bd\lambda_a$.
For parallel local transition dipoles,
$\bd\lambda_a=\mathbf0$.
The signal then depends only on the symmetric channel overlap $p_s$ and vanishes for the singlet input.
Then, the full and separable reachable sets reduce to the line segments in Eqs.~\eqref{eq:vd-main-R} and \eqref{eq:vd-main-Rsep}.

\begin{figure}[tb]
\includegraphics[width=\linewidth]{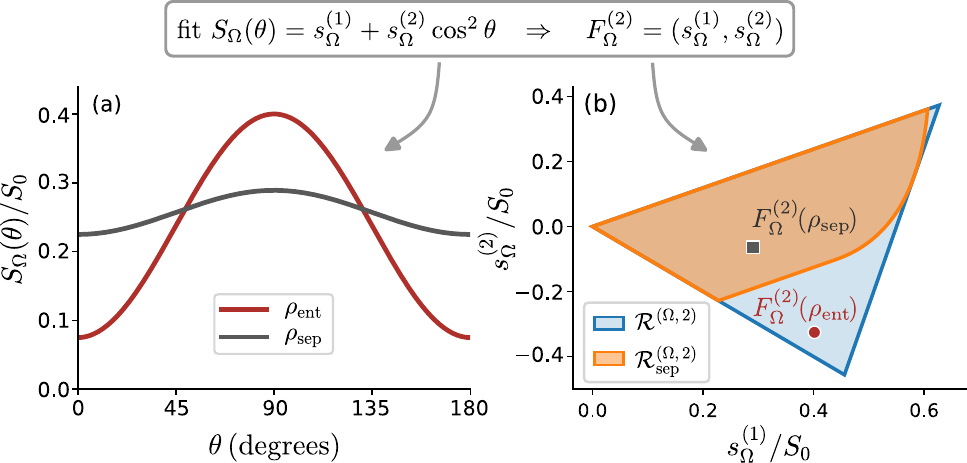}
\caption{Angular two-photon signals and reachable sets
for the molecular dimer. (a) Signal $S_\Omega(\theta)/S_0$ for the separable input state $\rho_{\mathrm{sep}}$ (gray) and entangled input state $\rho_{\mathrm{ent}}$ (red).
The angle $\theta$ defines the relative orientation of the pulse polarizations. (b) Fitted coefficient pairs, together with the full reachable set $\mathcal{R}^{(\Omega, 2)}$ (blue) and the separable reachable set $\mathcal{R}_{\mathrm{sep}}^{(\Omega, 2)}$ (orange).
The signal vector $(s_\Omega^{(1)}, s_\Omega^{(2)})$ of $\rho_{\mathrm{ent}}$ lies outside the separable set and certifies its entanglement. (See text for more explanations.) \label{fig:TPA-dimer}}
\end{figure}

We illustrate in Fig.~\eqref{fig:TPA-dimer} how the signals from both a separable and entangled example map map into the respective full and reachable sets in Eqs.~\eqref{eq:vd-main-nonparallel-R} and \eqref{eq:vd-main-nonparallel-Rsep}. The two example input states are $\rho_\epsilon = (1 - \epsilon)\ket{\psi_-}\!\bra{\psi_-} + \epsilon\mathbbm{1}_4/4$, with $\rho_{\mathrm{ent}} = \rho_{0.3}$ (labelled by $\rho_\mathrm{ent}$) and $\rho_{\mathrm{sep}} = \rho_{0.9}$ (labelled by $\rho_\mathrm{sep}$).
The used model parameters are $\varepsilon_{00} = 2.10\,\mathrm{eV}$, $\varepsilon_v = 0.180\,\mathrm{eV}$, $U = 0.100\,\mathrm{eV}$, a Huang--Rhys factor $S = 0.4$, and the dipole geometry $\eta = 0.35$ (see also Eq.~\eqref{eq:vd-identical}).
The carrier energies of the Gaussian pulses are $\varepsilon_1 = 2.04\,\mathrm{eV}$ and $\varepsilon_2 = 2.16\,\mathrm{eV}$, a common amplitude width $\tau = 12.5\,\mathrm{fs}$, and positive frequency dipole coupling energies $\mu\mathcal{E}_{01} = \mu\mathcal{E}_{02} =1\,\mathrm{meV}$ (see Eq.~\eqref{eq:vd-pulses} for definitions).
The pulse parameters and nonparallel dipole geometry leads to two nonzero channels in the regime of weak excitation. The signal and its coefficients in panels (a) and (b) of Fig.~\ref{fig:TPA-dimer} are rescaled by a common parameter $S_0 = 4N_s(1 + 2\eta^2)/15$. 

Panels (a) and (b) illustrate how signals with the same qualitative angular dependence can yield different conclusions about entanglement. The fitted coefficient pair for $\rho_{\mathrm{ent}}$ lies outside the separable reachable set and therefore certifies its entanglement.
For the chosen parameters, a separable input can reproduce the signal $S_\Omega(\theta)$ of $\rho_{\mathrm{ent}}$ at any fixed polarization angle $\theta$.
The required separable state may depend on $\theta$.
However, the fitted coefficient $s_\Omega^{(2)} = S_\Omega(0) - S_\Omega(\pi/2)$ lies below the minimum attainable by separable inputs.
This implies that the difference between the signals for parallel and perpendicular pulse polarizations certifies entanglement, even though each signal $S_\Omega(\theta)$ alone is compatible with a separable state. The separable state $\rho_{\mathrm{sep}}$ maps into the separable reachable set as required.

\section{Conclusions and outlook \label{sec:concl}}

We have developed a framework for certifying entanglement directly from spectroscopic signals.
The key step is to use the structure of these signals to select informative coefficients. The corresponding effective observables are independent of the state whose entanglement properties are tested and are determined by a model of the molecular response. The expectation values of the observables yield the signal coefficients for any admissible input state. For a chosen physical bipartition, these observables define the full and separable reachable sets in signal space. Entanglement is certified if the vector of signal coefficients lies outside the separable reachable set, since in that case no admissible separable state can reproduce all its components. At the same time, the dependence of the effective observables on the response model presents a fundamental difficulty inherent to any spectroscopic entanglement test. Its validity requires an accurate description of the experimental response. 

The examples in Secs.~\ref{sec:examples-excited} and \ref{sec:examples-initial} demonstrate how distinct features of spectroscopic signals can be used to exclude separable explanations. In the excitonic dimer, two spectral windows jointly exclude separability, although neither coefficient alone is conclusive. Fluorescence measurements require an alternative construction. A scan of the excitation laser detuning changes the prepared states. Thus, the resulting signals cannot be treated as joint expectation values of a single state. Instead, fluorescence measurements require a complementary approach for constructing a suitable spectroscopic map for a fixed preparation. 
Moreover, many molecular spectroscopy experiments are performed in solution, where signals are averaged over molecular orientations. However, rotational averaging does leave sufficient structure in the signal, which can be exploited to construct entanglement tests. We have illustrated this with polarization-dependent two-photon absorption in an open-shell system and a vibrational molecular dimer. In sum, the examples show how the information accessible through a spectroscopic measurement determines what can be certified about molecular entanglement.

Turning these theoretical tests into experimental certificates requires validated models of the spectroscopic response. Reproducing the measured signals alone is not sufficient. The model must describe the response of every admissible comparison state, including the detection process. In addition, entanglement certification requires the experimental confidence region to remain disjoint from all separable reachable sets allowed by the uncertainties of the response model. These requirements can guide the choice of signal coefficients and measurement settings for robust entanglement tests.

The idea of witnessing entanglement through joint expectation value regions is not restricted to the spectroscopic techniques considered here. Other protocols may provide additional observables and improve the ability to distinguish signals from separable and entangled states. Two-dimensional electronic and double quantum coherence spectroscopy are promising examples \cite{KMS09, KHCS09, BKZS22}. This raises a question for future work: How can we identify the most informative spectroscopic measurements while keeping the calculation of their separable bounds tractable?
Addressing this question could guide the selection of signal coefficients and experimental protocols for robust entanglement certification. A further extension concerns nonlinear signal maps, whose separable images still define valid compatibility tests but need not be convex.
A long-term goal is to establish a calibrated framework that uses spectroscopic measurements to certify molecular entanglement and analyse its evolution during photochemical reactions.

\begin{acknowledgments}
This research was funded by the National Science Foundation under Grant No.~CHE-2537080. 
\end{acknowledgments}

\appendix

\section{Measurement operators for two-photon absorption\label{app:Ax-Bx}}

In this appendix we derive the effective observables $X_\Omega^{(1)}$ and $X_\Omega^{(2)}$ for a selected two-photon excitation yield in the weak field electric dipole regime. Its orientationally averaged polarization dependence has the form
\begin{equation}\label{eq:SM-EAB}
E_\Omega(\theta) = X^{(1)}_\Omega +  X^{(2)}_\Omega \cos^2\theta\,,
\end{equation}
For this selected signal,
\begin{equation}
S_\Omega(\theta) = \Tr\left[\rho_G\left(X^{(1)}_\Omega + X^{(2)}_\Omega \cos^2\theta \right)\right]\,.
\end{equation}
Here $\mathcal{H}_\star \subseteq \mathcal{H}_A\otimes\mathcal{H}_B$ is the active input manifold introduced in the main text, $\rho_\star \in\mathcal{D}_\star$ is the corresponding input state, and $\Omega$ denotes the fixed experimental control parameters. Since we consider explicitly states on the ground state manifold we write in the following $\mathcal{H}_G\equiv \mathcal{H}_\star$. Detailed examples with extensions to finite pulses are discussed in Secs.~\ref{subsec:excitonic-dimer} and \ref{subsec:vibronic-dimer} and their respective Appendix~\ref{app:two-orbital-TPA} and \ref{app:vibronic-complete}. We consider a dimer of two weakly interacting molecules labelled by $A$ and $B$ and work in the electric-dipole approximation and in the perturbative regime. In addition, we assume an isotropic distribution of molecular orientations. The derivation below is formulated for a general active input manifold $\mathcal{H}_G$. 

Let $\mathcal{H}_E$ denote the manifold of one-photon accessible intermediate states and $\mathcal{H}_F$ the detected final manifold, with projectors $P_E$ and $P_F$, respectively. 
We consider the population transferred into $\mathcal{H}_F$ by absorption of one photon from each field. Before the fields act, this manifold is assumed to be unoccupied, i.e., $P_FP_\star = 0$.
Within the response model, there are no transition amplitudes connecting the input manifold to $\mathcal{H}_F$ at first order in the fields. Other processes at second order must be absent or excluded from the detected signal by the protocol.
Moreover, these assumptions apply not only to the target state but to every $\rho\in\mathcal{D}_\star$.
Then, for energy eigenstates $\ket{i}\in\mathcal{H}_G$ and $\ket{f}\in\mathcal{H}_F$, the two-photon transition tensor is \cite{MMC70, Mukamel1995-book, FBR14, Beerepoot15}
\begin{align}\label{eq:Tfi_ab}
T^{fi}_{\alpha\beta}(\omega_1, \omega_2) &= \frac{1}{\hbar}\sum_{n\in\mathcal{H}_E} \left( \frac{\langle f|\mu_\alpha|n\rangle\langle n|\mu_\beta|i\rangle} {\omega_{ni} - \omega_1-i\gamma_n} \right. \nonumber\\
&\qquad \left. + \frac{\langle f|\mu_\beta|n\rangle\langle n|\mu_\alpha|i\rangle} {\omega_{ni}-\omega_2-i\gamma_n} \right)\,,
\end{align}
where $\alpha,\beta\in\{x,y,z\}$, $\omega_{ni}=(E_n-E_i)/\hbar$, and $\gamma_n$ is a phenomenological damping rate. In the non-degenerate case $T^{fi}_{\alpha\beta}(\omega_1, \omega_2)$
is in general not symmetric under $\alpha \leftrightarrow \beta$, whereas in the degenerate case $\omega_1=\omega_2=\omega$ one has
\begin{equation}
T^{fi}_{\alpha\beta}(\omega, \omega) = T^{fi}_{\beta\alpha}(\omega, \omega)\,.
\end{equation}
The tensor above describes the electronic part of the two-photon transition amplitude. 
For pulsed excitation, the transferred population also depends on the complex spectra of the applied fields and the transition frequency \cite{DDGFS01}. 
We denote by $T^{(\Omega)}_{\alpha\beta}: \mathcal{H}_G \to \mathcal{H}_F$ the operator-valued tensor, defined by its matrix elements
\begin{equation}
\langle f|T^{(\Omega)}_{\alpha\beta}|i\rangle = T^{fi}_{\alpha\beta}(\omega_1, \omega_2)\,.
\end{equation}
In the simplified derivation, we assume that the field and pulse envelopes contribute a common factor to the amplitudes of the transitions. Here, the aim is to illustrate how the effective observables arise from the polarization structure of the signal. The derivations for two specific two-photon absorption examples which include further pulse properties explicitly are provided in Appendices~\ref{app:two-orbital-TPA} and \ref{app:vibronic-complete}. 
For a molecule at orientation $R\in SO(3)$, the laboratory frame tensor is given by
\begin{equation}
T^{(R, \Omega)}_{ij} = R_{i\alpha}R_{j\beta}\,  T^{(\Omega)}_{\alpha\beta}\,,
\end{equation}
where $i, j\in\{x, y, z\}$ are laboratory frame Cartesian indices. For two linearly polarized fields with real unit polarization vectors $\mathbf{e}_1, \mathbf{e}_2$, the orientation-dependent two-photon amplitude is defined as
\begin{equation}\label{M-Omega-R}
M_\Omega^{(R)}(\mathbf{e}_2, \mathbf{e}_1) := e_{2,i}e_{1,j}\, T^{(R, \Omega)}_{ij}\,.
\end{equation}
Furthermore, the relative polarization is parametrized by
\begin{equation}
c :=  \cos\theta = \mathbf{e}_1 \cdot \mathbf{e}_2\,.   
\end{equation}

Detection on the final manifold is described by a positive semidefinite operator $D_\Omega\succeq 0$ on $\mathcal{H}_F$. For a single detected final state $\ket{f}$, this reduces to $D_\Omega=\ket{f}\!\bra{f}$. Under these assumptions, the measured signal takes the form
\begin{align}
S_\Omega(\theta) &= \Big\langle \Tr \big[D_\Omega M_\Omega^{(R)}(\theta) \rho_G\,M_\Omega^{(R)}(\theta)^\dagger\big] \Big\rangle_R \nonumber\\
&= \Tr \big[\rho_G E_\Omega(\theta)\big]\,,
\end{align}
where $\langle\cdot\rangle_R$ denotes averaging over molecular orientations and
\begin{equation}\label{eq:E_Omega-app-R}
E_\Omega(\theta):=\Big\langle M_\Omega^{(R)}(\theta)^\dagger D_\Omega M_\Omega^{(R)}(\theta) \Big\rangle_R \in \mathcal{B}_{\mathrm{sa}}(\mathcal{H}_G)\,.
\end{equation}
Under the assumptions above, the positive operator $E_\Omega(\theta)$ in Eq.~\eqref{eq:E_Omega-app-R} represents the population transferred into the selected final manifold.
For absorption or transmission measurements, the effective observable must be derived from the complete response at the considered perturbative order. If admissible input states populate the final manifold $\mathcal{H}_F$, there can be also contributions from stimulated emission and interference between transition amplitudes at that order.
Expanding the laboratory frame tensor in Eq.~\eqref{eq:E_Omega-app-R} yields
\begin{align}
E_\Omega(\theta) &= e_{2,i}e_{1,j}e_{2,k}e_{1,l}\, \Big\langle R_{i\alpha}R_{j\beta}R_{k\gamma}R_{l\delta}\Big\rangle_R T_{\alpha\beta}^{(\Omega)\dagger}D_\Omega T_{\gamma\delta}^{(\Omega)}\,.
\end{align}
For an isotropic ensemble, the fourth-rank rotational average is given by \cite{AT77, FBR14}
\begin{widetext}
\begin{equation}\label{eq:rot-average}
\left\langle R_{i\alpha} R_{j\beta} R_{k\gamma} R_{l\delta} \right\rangle_R =
\begin{pmatrix}
\delta_{ij}\delta_{kl}\\
\delta_{ik}\delta_{jl}\\
\delta_{il}\delta_{jk}
\end{pmatrix}^{\!T}
\frac{1}{30}
\begin{pmatrix}
4 & -1 & -1\\
-1 & 4 & -1\\
-1 & -1 & 4
\end{pmatrix}
\begin{pmatrix}
\delta_{\alpha\beta}\delta_{\gamma\delta}\\
\delta_{\alpha\gamma}\delta_{\beta\delta}\\
\delta_{\alpha\delta}\delta_{\beta\gamma}
\end{pmatrix}\,.    
\end{equation}
\end{widetext}
For real linear polarizations one further has
\begin{align}\label{eq:contractions}
e_{2,i} e_{1,j} e_{2, k}  e_{1,l}\,\delta_{ij} \delta_{kl} &= e_{2, i} e_{1,j} e_{2, k}e_{1,l}\, \delta_{il}\delta_{jk} = c^2\,,\\
e_{2,i} e_{1,j} e_{2, k} e_{1,l}\,\delta_{ik} \delta_{jl} &= 1\,.
\end{align}
Therefore,
\begin{align}
E_\Omega(\theta) &= \frac{1}{30}\left[(3c^2 - 1)\mathcal{I}_{F, \Omega} + (4 - 2c^2) \mathcal{I}_{G,\Omega}\right.\nonumber\\
&\qquad\quad \left.  + (3c^2  - 1)\mathcal{I}_{H, \Omega} \right]\,,
\end{align}
where we introduced the three operator-valued contractions
\begin{align}
\mathcal{I}_{F,\Omega} &:= \left(\sum_{\alpha}T^{(\Omega)}_{\alpha\alpha} \right)^\dagger D_\Omega \left(\sum_{\beta}T^{(\Omega)}_{\beta\beta}\right),\\
\mathcal{I}_{G,\Omega} &:= \sum_{\alpha, \beta} T_{\alpha \beta}^{(\Omega)\dagger}D_\Omega T_{\alpha \beta}^{(\Omega)},\\ 
\mathcal{I}_{H,\Omega} &:= \sum_{\alpha, \beta} T_{\alpha\beta}^{(\Omega)\dagger} D_\Omega T_{\beta \alpha}^{(\Omega)}.
\end{align}
These are the operator analogues of the standard contractions in Ref.~\cite{MMC70}.

Since $c^2 = \cos^2\theta$, comparison with Eq.~\eqref{eq:SM-EAB} yields
\begin{align}
X^{(1)}_\Omega &=\frac{1}{30}\left( -\mathcal{I}_{F, \Omega} + 4\mathcal{I}_{G, \Omega} -\mathcal{I}_{H, \Omega} \right)\,, \label{eq:X-Omega} \\ 
X^{(2)}_\Omega &= \frac{1}{30}\left( 3\mathcal{I}_{F, \Omega} - 2\mathcal{I}_{G, \Omega} + 3\mathcal{I}_{H, \Omega} \right)\,.\label{eq:Y-Omega} 
\end{align}
Hence, it follows that
\begin{equation}
S_\Omega(\theta) = \Tr\left[\rho_G  \left(X^{(1)}_\Omega + X^{(2)}_\Omega\cos^2\theta \right) \right]\,.
\end{equation}
For parallel linear polarizations $(\theta = 0)$,
\begin{equation}
E_\Omega(0) = X^{(1)}_\Omega + X^{(2)}_\Omega = \frac{1}{15}\left(\mathcal{I}_{F, \Omega} + \mathcal{I}_{G, \Omega} + \mathcal{I}_{H, \Omega} \right)\,,
\end{equation}
while for perpendicular linear polarizations $(\theta = \pi/2)$,
\begin{equation}
E_\Omega(\pi/2) = X^{(1)}_\Omega = \frac{1}{30}\left( -\mathcal{I}_{F, \Omega} + 4\mathcal{I}_{G, \Omega} - \mathcal{I}_{H, \Omega}\right)\,.
\end{equation}
These are the standard linearly polarized isotropic combinations of the three contractions \cite{MMC70,AT77,FBR14}.

The degenerate case is recovered as a special case. If $\omega_1 = \omega_2$, then $T_{\alpha\beta}^{(\Omega)} = T_{\beta\alpha}^{(\Omega)}$ as shown above. Therefore, $\mathcal{I}_{H, \Omega} = \mathcal{I}_{G, \Omega}$. Then, the above formulas reduce to
\begin{equation}
X^{(1)}_\Omega = \frac{1}{30}\left(3\mathcal{I}_{G, \Omega} - \mathcal{I}_{F, \Omega}\right)\,,\,\,\,
X^{(2)}_\Omega  = \frac{1}{30}\left(\mathcal{I}_{G, \Omega} + 3\mathcal{I}_{F, \Omega}\right)\,,
\end{equation}    
and we recover the symmetric tensor result for degenerate TPA \cite{MMC70, FBR14, Beerepoot15}.

\section{Ensemble-resolved readout for rotational averaging and static disorder \label{app:ensemble-res}}

The main text and previous section assume that the measured signal can be represented by an affine map from the relevant state space to a finite-dimensional signal vector in $\RR^n$.
In many spectroscopic settings, the sample is an ensemble of molecules with different orientations or different static disorder realizations.
In such cases the measured signal need not factorize through a single averaged density matrix and a single averaged response observable.
This does not invalidate the construction of entanglement witnesses through testing membership in corresponding full and separable reachable sets (i.e., joint numerical ranges).
The correct state variable is then an ensemble resolved by the corresponding classical label, as detailed in the following.

Let $\lambda\in\Lambda$ denote a classical ensemble label, and let $p(\lambda)$ be its probability density. We refer the interested reader to the textbook \cite{Cohn2013} for an introduction to measure theory. 
For rotational averaging, $\Lambda = SO(3)$ and $\lambda = R$ is the molecular orientation.
For static disorder, $\Lambda$ is the space of disorder parameters, such as site-energy disorder, exchange couplings or other fixed Hamiltonian parameters.
For each fixed $\lambda$, assume that the reduced spectroscopic readout is linear on the active Hilbert space $\mathcal{H}_\star$.
Thus, for an observable $X_{\Omega, \lambda}^{(j)}$, the contribution from the subensemble $\lambda$ is
\begin{equation}
s_{\Omega,\lambda}^{(j)} = \Tr\left[\rho_{\star, \lambda}X_{\Omega,\lambda}^{(j)}\right]\,.
\end{equation}
Here $\rho_{\lambda}\in\mathcal{D}_\star$ is the state of the subensemble with label $\lambda$ and  $X_{\Omega,\lambda}^{(j)}$ is the effective observable after the chosen linear extraction of a signal coefficient, as in Sec.~\ref{subsec:spectroscopic-map}.
Throughout the section, the distribution $p(\lambda)$ is fixed and normalized, $\int\mathrm{d}\lambda\,p(\lambda) = 1$.
Moreover, it is assumed that
\begin{equation}
\int\mathrm{d}\lambda\, p(\lambda) \left\|X_{\Omega,\lambda}^{(j)}\right\|_\infty < \infty\,,
\quad j = 1, \ldots, n\,,
\end{equation}
where $\|\cdot\|_\infty$ denotes the operator norm.
These conditions ensure that the signal integrals are finite for every admissible ensemble.

The measured ensemble signal reads
\begin{equation}\label{eq:ensemble-windowed-signal}
s_{\Omega}^{(j)} = \int \mathrm{d}\lambda\, p(\lambda) \Tr\left[\rho_{\star, \lambda}X_{\Omega, \lambda}^{(j)}\right]\,. 
\end{equation}
Equivalently, define the unnormalized operator-valued ensemble
\begin{equation}
\Gamma(\lambda) = p(\lambda)\rho_{\star, \lambda}\,.
\end{equation}
Then Eq.~\eqref{eq:ensemble-windowed-signal} becomes
\begin{equation}
s_\Omega^{(j)} = \int \mathrm{d}\lambda\,\Tr\left[\Gamma(\lambda)X_{\Omega,\lambda}^{(j)}\right]\,.
\end{equation}
The integral is part of the readout map. The relevant domain is therefore \textit{not} the single-state set $\mathcal{D}_\star$, but instead the ensemble-resolved state set
\begin{align}\label{eq:ensemble-state-space}
\mathfrak{D}_p(\mathcal{H}_\star) &= \left \{\Gamma:\Lambda\to\mathrm{Herm}_+ (\mathcal{H}_\star): \Tr[\Gamma(\lambda)] = p(\lambda)\right. \nonumber\\
&\qquad\qquad  \left.  \mathrm{for\ almost\ every}\ \lambda\right\}\,.
\end{align}
Equivalently, $\Gamma\in\mathfrak{D}_p(\mathcal{H}_\star)$ if and only if $\Gamma(\lambda)=p(\lambda)\rho_{\star,\lambda}$ with $\rho_{\star, \lambda}\in\mathcal{D}(\mathcal{H}_\star)$ for almost every $\lambda$.
If the signal depends only on a subnormalized excited state
block $\tau_{\lambda} = P_e\rho_{\star, \lambda}P_e$ (recall Sec.~\ref{subsec:manifold}), the underlying physical state still belongs to $\mathcal{D}_\star$ and remains normalized.
The reduced ensemble block is $p(\lambda)\tau_\lambda$.
Its domain compatible with separable states on $\mathcal{H}= \mathcal{H}_A\otimes \mathcal{H}_B$ is obtained by applying the same compression to the admissible separable states, $\{P_e\sigma P_e: \sigma\in \mathcal{D}_{\star, \mathrm{sep}}\}$. 

For $n$ retained signal coefficients, i.e., $n$ operators $X_{\Omega, \lambda}^{(j)}$, define the ensemble-resolved signal map
\begin{align}\label{eq:ensemble-signal-map}
F_{\Omega,p}^{(n)}(\Gamma) &=\left(\int \mathrm{d}\lambda\,\Tr\left[\Gamma(\lambda)X_{\Omega,\lambda}^{(1)}\right],\int \mathrm{d}\lambda\,\Tr\left[\Gamma(\lambda)X_{\Omega,\lambda}^{(2)}\right], \right.\nonumber\\
&\qquad \qquad \left.\ldots,  \int \mathrm{d}\lambda\,\Tr\left[\Gamma(\lambda)X_{\Omega,\lambda}^{(n)}\right]\right)\,.
\end{align}
The map $F_{\Omega,p}^{(n)}(\cdot)$ is affine on $\mathfrak{D}_p(\mathcal{H}_\star)$. Indeed, for $\Gamma_1, \Gamma_2\in\mathfrak{D}_p(\mathcal{H}_\star)$ and $0\leq t\leq 1$, one has $\Gamma_t=t\Gamma_1+(1-t)\Gamma_2\in\mathfrak{D}_p(\mathcal{H}_\star)$.
Using the linearity of the trace and the integral gives
\begin{equation}
F_{\Omega, p}^{(n)}(\Gamma_t) = tF_{\Omega, p}^{(n)}(\Gamma_1) + (1 - t)F_{\Omega,p}^{(n)}(\Gamma_2)\,.
\end{equation}
Thus the ensemble average only changes the domain on which the affine map is defined but not the general convex structure of the reachable sets.

The ensemble-resolved separable set is
\begin{align}
\mathfrak{D}_{p,\mathrm{sep}}(\mathcal{H}_\star) &= \left\{\Gamma(\lambda) = p(\lambda)\sigma_{\star, \lambda}: \sigma_{\star, \lambda}\in\mathcal{D}_{\star, \mathrm{sep}}\right.\nonumber\\ 
&\qquad \left.  \mathrm{for\ almost\ every}\ \lambda\right\}\,.
\end{align}
The explicit $\lambda$-dependence of $\sigma_{\star, \lambda}$ is important.
It means that each orientation or disorder subensemble is allowed to carry its own separable state.
Restricting to a common state $\sigma_{\star,\lambda}=\sigma_\star$ for all $\lambda$ is a stronger model assumption and should only be used if justified by the preparation and averaging protocol.

The set $\mathfrak{D}_{p, \mathrm{sep}}(\mathcal{H}_\star)$ is convex.
If $\Gamma_1(\lambda) = p(\lambda)\sigma_{1, \star,\lambda}$ and $\Gamma_2(\lambda) = p(\lambda)\sigma_{2, \star,\lambda}$ are ensemble-resolved separable states, then
\begin{equation}
t\Gamma_1(\lambda) + (1 - t)\Gamma_2(\lambda) = p(\lambda)\left(t\sigma_{1, \star, \lambda} + (1 - t)\sigma_{2, \star, \lambda}\right)\,.
\end{equation}
Since $\mathcal{D}_{\star, \mathrm{sep}}$ is convex, the state $t\sigma_{1, \star,\lambda} + (1 - t)\sigma_{2,\star, \lambda}$ is separable for almost every $\lambda$.
Hence $t\Gamma_1 + (1 - t)\Gamma_2\in\mathfrak{D}_{p,\mathrm{sep}}(\mathcal{H}_\star)$.

The full and separable ensemble-resolved reachable sets are
\begin{equation}
\mathcal{R}_{p}^{(\Omega,n)} = F_{\Omega,p}^{(n)}\left(\mathfrak{D}_p(\mathcal{H}_\star)\right)
\end{equation}
and
\begin{equation}
\mathcal{R}_{p, \mathrm{sep}}^{(\Omega, n)} = F_{\Omega,p}^{(n)}\left(\mathfrak{D}_{p, \mathrm{sep}}(\mathcal{H}_\star)\right)\,.
\end{equation}
Because $F_{\Omega,p}^{(n)}$ is affine and both domains are convex, both reachable sets are convex.
Moreover, $\mathfrak{D}_{p,\mathrm{sep}}(\mathcal{H}_\star)\subseteq\mathfrak{D}_p(\mathcal{H}_\star)$ implies
\begin{equation}
\mathcal{R}_{p,\mathrm{sep}}^{(\Omega,n)}\subseteq \mathcal{R}_{p}^{(\Omega,n)}\,.
\end{equation}
A signal in $\mathcal{R}_{p}^{(\Omega,n)} \setminus \mathcal{R}_{p, \mathrm{sep}}^{(\Omega, n)}$ cannot be reproduced by an ensemble whose subensemble states are separable with probability one.

For a direction $\mathbf{u}\in\mathbbm{R}^n$, define the orientation- or disorder-resolved witness operator
\begin{equation}
W_{\Omega,\lambda}(\mathbf{u}) = \sum_{j=1}^n u_jX_{\Omega,\lambda}^{(j)}\,.
\end{equation}
The separable support function is
\begin{align}
h_{p,\mathrm{sep}}^{(\Omega,n)}(\mathbf{u}) &:= \sup_{\Gamma\in\mathfrak{D}_{p,\mathrm{sep}}(\mathcal{H}_\star)} \mathbf{u}\cdot F_{\Omega,p}^{(n)}(\Gamma) \nonumber\\
&= \sup_{\sigma_{\star,\lambda}\in\mathcal{D}_{\star, \mathrm{sep}}} \int \mathrm{d}\lambda\,p(\lambda)\Tr\left[\sigma_{\star,\lambda}W_{\Omega,\lambda}(\mathbf{u})\right]\,.
\end{align}
The ensemble domain permits independent choices of the separable state at different labels.
Assuming $\mathcal{D}_{\star, \mathrm{sep}}$ is nonempty, the objective is continuous in the state on this compact set and measurable in $\lambda$.
Thus, maximizing states can be determined label by label, which leads to
\begin{equation}\label{eq:ensemble-sep-app}
h_{p,\mathrm{sep}}^{(\Omega,n)}(\mathbf{u}) = \int\mathrm{d}\lambda\,p(\lambda) \max_{\sigma\in\mathcal{D}_{\star,\mathrm{sep}}} \Tr[\sigma W_{\Omega, \lambda}(\mathbf{u})]\,.
\end{equation}
A common preparation may impose additional relations between states at different labels.
If these relations are not included in the optimization, Eq.~\eqref{eq:ensemble-sep-app} reduces to a conservative upper bound on the corresponding separable signals, which need not be tight.
A valid ensemble-resolved witness condition is therefore
\begin{equation}
\mathbf{u} \cdot\mathbf{s} > h_{p, \mathrm{sep}}^{(\Omega,n)}(\mathbf{u})\,,
\end{equation}
where in practise statistical and model uncertainties should be included.
The statement to be tested is that the measured signal cannot be produced by any ensemble for which $\rho_{\lambda}$ is separable across the chosen bipartition for almost every $\lambda$.

If exact optimization over $\mathcal{D}_{\star, \mathrm{sep}}$ is not tractable, we use a compact convex outer relaxation
\begin{equation}
\mathcal{D}_{\star, \mathrm{sep}} \subseteq \mathcal{C}_k \subseteq \mathcal{D}_\star\,.
\end{equation}
The relaxation is defined with respect to the physical bipartition $A|B$. 
For example, the positive partial transpose relaxation \cite{Peres96,HHH96} yields
\begin{equation}
\mathcal{C}_{\mathrm{PPT}} = \left\{\sigma\in\mathcal{D}_\star: \sigma^{T_B}\succeq 0 \right\}\,,
\end{equation}
where the transposition $T_B$ acts on $\mathcal{H}_B$ in the full space $\mathcal{H}_A\otimes\mathcal{H}_B$.
Higher levels of the DPS hierarchy constrain symmetric extensions of the physical bipartite state \cite{DPS02,DPS04}. 
This gives outer approximations through supporting hyperplanes 
\begin{equation}
h_{p,k}^{(\Omega,n)}(\mathbf{u}) = \int \mathrm{d}\lambda\, p(\lambda)\max_{\rho\in\mathcal{C}_k} \Tr\left[\rho W_{\Omega,\lambda}(\mathbf{u})\right]\,.
\end{equation}
of the enlarged separable reachable set.
Since $\mathcal{D}_{\star, \mathrm{sep}}\subseteq \mathcal{C}_k$, one has $h_{p,\mathrm{sep}}^{(\Omega,n)}(\mathbf{u})\leq h_{p,k}^{(\Omega,n)}(\mathbf{u})$.
Thus $\mathbf{u}\cdot\mathbf{s} > h_{p,k}^{(\Omega,n)}(\mathbf{u})$ is a sufficient condition for entanglement.
Each inner objective is linear in the state. Thus, the maximization is a semidefinite program whenever $\mathcal{C}_k$ admits a semidefinite representation.

\subsection{Rotational averaging}

For rotational averaging, the classical label is the molecular orientation $R\in SO(3)$.
In this subsection, we emphasize the excited state case, where a pump can photoselect different molecular orientations and prepare orientation-dependent states $\rho_{\star, R}(T)$. For ground state spectroscopy without orientation-dependent preparation, the situation is simpler. For a fixed molecule, the initial ground state $\rho_0$ is the same for every orientation, while the response operator depends on $R$ through the laboratory frame polarization geometry, and the rotational average factorizes.

We therefore now focus on the excited state setting.  Then, for each fixed orientation $R$, the windowed signal is
\begin{equation}
s_{\Omega,R}^{(j)} = \Tr\left[\rho_{\star, R}X_{\Omega,R}^{(j)}\right]
\end{equation}
while the orientationally averaged signal is
\begin{equation}\label{eq:rotational-average-general}
s_\Omega^{(j)}=\int_{SO(3)} \mathrm{d}R\,p(R)\Tr\left[\rho_{\star, R}X_{\Omega,R}^{(j)}\right]\,,
\end{equation}
which corresponds to an affine map on the orientation-resolved ensemble $\Gamma(R) = p(R)\rho_{\star, R}$.
Its separable counterpart is the operator-valued ensemble $\Gamma(R) = p(R)\sigma_{\star, R}$ with $\sigma_{\star, R}\in\mathcal{D}_{\star, \mathrm{sep}}$ for almost every $R$.

In special cases, Eq.~\eqref{eq:rotational-average-general} can factorize through a single averaged state and a single averaged response observable.
For example, if the prepared state is independent of orientation, $\rho_{\star, R} = \rho_\star$, then
\begin{equation}
s_\Omega^{(j)}=\Tr\left[\rho_\star \bar{X}_\Omega^{(j)}\right]\,, \quad \bar{X}_\Omega^{(j)} =\int_{SO(3)} \mathrm{d}R\,p(R) X_{\Omega,R}^{(j)}\,.
\end{equation}
For some pump-probe polarization configurations, such as the magic-angle setting discussed in Ref.~\cite{HW13}, orientational averaging may also simplify the response.
When such a factorization is justified, the witness can be formulated on the usual single-state domain with the averaged operators $\bar{X}_\Omega^{(j)}$.
When it is not justified, the orientation-resolved formulation in Eq.~\eqref{eq:rotational-average-general} is the appropriate formulation of an entanglement witness on the convex set of retained signal coefficients.

\subsection{Static disorder}

For static disorder, the classical label $\lambda$ denotes a fixed realization of a Hamiltonian.
The ground state, the excited state, and the response operator may all depend on the same $\lambda$.
The measured signal is therefore given by
\begin{equation}
s_\Omega^{(j)} = \int \mathrm{d}\lambda\,p(\lambda)\Tr\left[\rho_{\star, \lambda}X_{\Omega,\lambda}^{(j)}\right]\,.
\end{equation}
In general this does not factorize as $\Tr[\bar{\rho}_\star \bar{X}_\Omega^{(j)}]$ with independently averaged quantities
\begin{equation}
\bar{\rho}_\star = \int \mathrm{d}\lambda\,p(\lambda)\rho_{\star, \lambda}\,, \quad \bar{X}_\Omega^{(j)} = \int \mathrm{d}\lambda\,p(\lambda)X_{\Omega,\lambda}^{(j)}\,.
\end{equation}
Indeed, the difference between the true disorder-averaged signal and the factorized expression is the covariance term
\begin{align}\label{eq:disorder-covariance}
&\int \mathrm{d}\lambda\,p(\lambda)\Tr\left[\rho_{\star, \lambda}X_{\Omega,\lambda}^{(j)}\right]-\Tr\left[\bar{\rho}_\star \bar{X}_\Omega^{(j)}\right]\nonumber\\
&\qquad\qquad =\int \mathrm{d}\lambda\,p(\lambda)\Tr\left[(\rho_{\star, \lambda}-\bar{\rho}_\star)(X_{\Omega,\lambda}^{(j)}-\bar{X}_\Omega^{(j)})\right]\,.
\end{align}
Thus factorization is valid only under additional assumptions, for example if $\rho_{\star, \lambda}$ is independent of $\lambda$, if $X_{\Omega,\lambda}^{(j)}$ is independent of $\lambda$, or if the covariance term in Eq.~\eqref{eq:disorder-covariance} vanishes for the measured windows in the case of an excited state witness construction.
The static-disorder reachable sets are therefore obtained from the same ensemble-resolved map $F_{\Omega,p}^{(n)}$ defined in Eq.~\eqref{eq:ensemble-signal-map}.
The corresponding witness excludes all models in which each disorder realization carries a separable state.
It does not necessarily certify entanglement of the independently averaged state $\bar{\rho}_\star$ unless the factorization through $\bar{\rho}_\star$ and $\bar{X}_\Omega^{(j)}$ is justified.

\section{Spectroscopic entanglement witness\label{app:witness}}

In this section, we prove the witness criterion on the active manifold $\mathcal{H}_\star$ discussed in Sec.~\ref{sec:witness} and provide provide a description for admissible witness operators.
As in Sec.~\ref{subsec:manifold}, the admissible domain contains all density operators supported on the active manifold $\mathcal{D}_\star$ defined in Eq.~\eqref{eq:D-star} and it is assumed that $\mathcal{D}_{\star, \mathrm{sep}} \neq \emptyset$.
Moreover, recall from Sec.~\ref{sec:main} that for a fixed protocol $\Omega$, the effective measurement observables $X_\Omega^{(j)}$ are operators on $\mathcal{H}_\star$, extended by zero to $\mathcal{H}$.
We introduce their real span together with the normalization observable as
\begin{equation}\label{eq:V-span}
\mathcal{V}_\Omega :=\mathrm{span}_{\mathbbm{R}} \left\{ P_\star, X_\Omega^{(1)}, \ldots, X_\Omega^{(n)} \right\}\,.
\end{equation}
The witness statements below are relative to admissible separable states in $\mathcal{D}_{\star,\mathrm{sep}}$ \eqref{eq:D-star-sep}. 

We first recall from Sec.~\ref{sec:witness} the combined response observable and its separable bound,
\begin{align}
W_\Omega(\mathbf{u}) &= \sum_{j=1}^{n}u_j X_\Omega^{(j)}\,,\\
h_{\mathrm{sep}}^{(\Omega, n)}(\mathbf{u}) &= \max_{\sigma\in\mathcal{D}_{\star, \mathrm{sep}}} \Tr[\sigma W_\Omega(\mathbf{u})]\,,
\end{align}
and the shifted operator
\begin{equation}
\widetilde{W}_\Omega(\mathbf{u}) = h_{\mathrm{sep}}^{(\Omega, n)}(\mathbf{u}) P_\star - W_\Omega(\mathbf{u})\,.
\end{equation}
We rephrase the witness operator construction on the separable and full reachable sets $\mathcal{R}_\mathrm{{sep}}^{(\Omega, n)}, \mathcal{R}^{(\Omega, n)}$ in the criterion:
\begin{prop}  \label{prop:witness}
Let $\mathbf{u}\in \mathbbm{R}^n$. For every $\rho \in \mathcal{D}_\star$,
\begin{equation}
F_\Omega^{(n)}(\rho) \notin\mathcal{R}_{\mathrm{sep}}^{(\Omega, n)}\quad\Leftrightarrow\quad \exists\,\mathbf{u}\in\mathbbm{R}^n: \Tr[\rho\widetilde{W}_\Omega(\mathbf{u})]<0\,.
\end{equation}
The map $F_\Omega^{(n)}(\cdot)$ can certify at least one admissible entangled state $\rho\in \mathcal{D}_\star\setminus \mathcal{D}_{\star, \mathrm{sep}}$ if and only if some direction $\mathbf{u}$ satisfies
\begin{equation}\label{eq:h-ineq-app}
h_{\mathrm{sep}}^{(\Omega, n)}(\mathbf{u}) < \lambda_{\max}\left( W_\Omega(\mathbf u)|_{\mathcal{H}_\star} \right)\,.
\end{equation}
Equivalently, $\mathcal{V}_\Omega$ \eqref{eq:V-span} contains an operator with nonnegative expectations on $\mathcal{D}_{\star, \mathrm{sep}}$ and a negative expectation on at least one state in $\mathcal{D}_\star$.
\end{prop}

\begin{proof}
For every admissible state $\rho\in \mathcal{D}_\star$, it holds that 
\begin{equation}
\mathbf{u}\cdot F_\Omega^{(n)}(\rho) = \Tr[\rho W_\Omega(\mathbf{u})]\,.
\end{equation}
Therefore, 
\begin{equation}
h_{\mathrm{sep}}^{(\Omega, n)}(\mathbf{u}) = \max_{\mathbf{s}\in\mathcal{R}_{\mathrm{sep}}^{(\Omega, n)}} \mathbf{u}\cdot\mathbf{s}\,.
\end{equation}
If the map $F_\Omega^{(n)}(\rho)$ lies outside the nonempty compact
convex separable reachable set $\mathcal{R}_\mathrm{sep}^{(\Omega, n)}$, the hyperplane separation
theorem \cite{rockafellar2015} ensures the existence of a direction $\mathbf{u}$ such that
\begin{equation}
\mathbf{u}\cdot F_\Omega^{(n)}(\rho) >  h_{\mathrm{sep}}^{(\Omega, n)}(\mathbf{u})\,.
\end{equation}
Using $\Tr[\rho P_\star] = 1$, this is equivalent to $\Tr[\rho\widetilde{W}_\Omega(\mathbf{u})] < 0$.
To show the converse direction, assume that this expectation is negative and the signal has an admissible separable explanation $\sigma$.
Then, the map $F_\Omega^{(n)}(\rho) = F_\Omega^{(n)}(\sigma)$ implies
\begin{equation}
\Tr[\rho\widetilde{W}_\Omega(\mathbf{u})] =  \Tr[\sigma\widetilde{W}_\Omega(\mathbf{u})] \geq 0\,,
\end{equation}
which contradicts the assumption. 
Since $\mathcal{D}_\star$ contains all density operators on the active manifold,
\begin{equation}
\max_{\rho\in\mathcal{D}_\star} \Tr[\rho W_\Omega(\mathbf{u})] =  \lambda_{\max}\left(W_\Omega(\mathbf{u})|_{\mathcal{H}_\star} \right)\,.
\end{equation}
This can be violated if Eq.~\eqref{eq:h-ineq-app} applies.
The eigenvalue is evaluated on $\mathcal{H}_\star$ as states outside that support are inadmissible.
Finally, any operator in the span $\mathcal{V}_\Omega$ defined in Eq.~\eqref{eq:V-span} can be written
as $\alpha P_\star -  W_\Omega(\mathbf{u})$.
Its expectations are nonnegative on
$\mathcal D_{\star,\mathrm{sep}}$ exactly when
$\alpha\geq h_{\mathrm{sep}}^{(\Omega,n)}(\mathbf u)$.
It has a negative expectation on some admissible state
exactly when
\begin{equation}
\alpha < \lambda_{\max}\left(W_\Omega(\mathbf{u}) |_{\mathcal{H}_\star} \right)\,.
\end{equation}
Such an $\alpha$ exists if and only if Eq.~\eqref{eq:h-ineq-app} holds, which proves the final equivalence.
\end{proof}

Next, we provide a criterion for which joint response operators and thus their expectation values are accessible. More precisely, an operator expectation is accessible from the retained coefficients if any two admissible states with the same signal vector give the same expectation.
For a Hermitian operator $B$ on $\mathcal{H}$, this holds if and only if
\begin{equation}\label{eq:accessible-wit-span}
P_\star B P_\star\in\mathcal{V}_\Omega\,.
\end{equation}
This statement follows from standard tools in operator algebra. 
To prove sufficiency, we write
\begin{equation}
P_\star B P_\star = c P_\star + \sum_j v_j X_\Omega^{(j)}\,.
\end{equation}
Then, for every state $\rho\in\mathcal{D}_\star$,
\begin{equation}
\Tr[\rho B] = c + \mathbf{v}\cdot F_\Omega^{(n)}(\rho)\,,
\end{equation}
which means that the signal determines the expectation value as required. 
In order to show necessity, consider instead the unmeasured operator directions
\begin{align}
\mathcal{N}_\Omega = \Big\{N = N^\dagger : \; & N = P_\star N P_\star,\,\, \Tr[N] = 0,\\\nonumber
& \Tr\left[N X_\Omega^{(j)}\right] = 0 \,\forall j\Big\}\,.
\end{align}
Let $d_\star = \mathrm{dim}\mathcal{H}_\star$ denote the dimension of the active Hilbert space manifold.
For every nonzero operator $N\in\mathcal{N}_\Omega$, we choose $0 < \epsilon < 1/(d_\star \|N\|_\infty)$.
The operators
\begin{equation}
\rho_\pm = \frac{P_\star}{d_\star }\pm\epsilon N
\end{equation}
are admissible states with identical signal vectors.
If the signal determines the expectation of $B$, we have
\begin{equation}
0 = \Tr[B(\rho_+ - \rho_-)] = 2\epsilon\Tr[BN]\,.
\end{equation}
It follows that $P_\star B P_\star$ is orthogonal to every element of $\mathcal{N}_\Omega$.
Within the real space of Hermitian operators supported on
$\mathcal{H}_\star$, we get $\mathcal{N}_\Omega = \mathcal{V}_\Omega^\perp$, where orthogonality is defined with respect to the Hilbert--Schmidt inner product $\langle A,B\rangle = \Tr[AB]$.
This leads to Eq.~\eqref{eq:accessible-wit-span} in finite dimensions as considered in this paper.
In particular, only $P_\star B P_\star$ has to be considered and matrix elements outside this block do not contribute to expectation values on $\mathcal{D}_\star$.
For $\mathcal{H}_\star=\mathcal{H}$, the criterion trivially reduces to the usual condition in terms of the operator span for incomplete measurements \cite{IT06, Spedalieri12}.

\section{Response operator for the excitonic dimer \label{app:dimer-response}}

For completeness, we derive in this appendix the effective response observables for the excitonic dimer example in Sec.~\ref{subsec:excitonic-dimer} following Ref.~\cite{HW13}. More details on the underlying perturbative response theory can be found in Refs.~\cite{Mukamel1995-book, APVM09}. 
We provide the explicit matrix form of the effective response operator under the linewidth model chosen for the illustrations in Sec.~\ref{subsec:excitonic-dimer} and relate the pump-induced correction to the normalized input state. 

\subsection{Input state and probe interaction}

On the full electronic space, the transition dipoles in Cartesian coordinates are given by
\begin{equation}\label{eq:dimer-dipole}
\mu_a^{(+)} = d_{A,a} \sigma_A^+ + d_{B,a} \sigma_B^+ \,,\,\,\,\,\mu_a^{(-)} = (\mu_a^{(+)})^\dagger
\end{equation}
with $ a = x, y, z$, where $\sigma_A^+ = \ket{e_A}\!\bra{g_A}\otimes \mathbbm{1}_B$ and $\sigma_B^+ = \mathbbm{1}_A\otimes \ket{e_B}\!\bra{g_B}$.
The electric-dipole approximation, the interaction with the classical field is described by the Hamiltonian $H_{\mathrm{int}}(t) = -\boldsymbol{\mu}\cdot\mathbf{E}(t)$.
For a linearly polarized probe as considered in this section, $\mathbf{E}(t) = \mathbf{e}[E^{(+)}(t) + E^{(-)}(t)]$, where $\mathbf{e}$ is the unit polarization vector, $E^{(+)}(t) = \mathcal{E}(t)e^{-i\omega_L t}$, and $E^{(-)}(t) = [E^{(+)}(t)]^*$ with laser frequency $\omega_L$.
The field couples to the projected dipole $\mu_{\mathbf{e}} = \mathbf{e}\cdot\boldsymbol{\mu}$.
The components $\mu_{\mathbf{e}}^{(+)}$ and $\mu_{\mathbf{e}}^{(-)} = (\mu_{\mathbf{e}}^{(+)})^\dagger$ raise and lower the electronic excitation number, respectively. Moreover, in the rotating wave approximation (RWA),
\begin{equation}\label{eq:dimer-rwa}
H_{\mathrm{int}}^{\mathrm{RWA}}(t) = - \mu_{\mathbf{e}}^{(+)}E^{(+)}(t) - \mu_{\mathbf{e}}^{(-)}E^{(-)}(t)\,.
\end{equation}

The pump-probe signal records the change in the probe response relative to the sample without the pump pulse.
For pulses that do not overlap, Ref.~\cite{HW13} collect the two pump interactions into an operator, which describes the contribution of second order in the pump field that enters the detected signal, and will be defined in Eq.~\eqref{eq:Delta-rho-app}.
After $E^{(+)}$ has been factored out, the remaining interaction acts through $(i/\hbar)V_{\mathbf{e}}^{(+)}$, where
\begin{equation}
V_{\mathbf{e}}^{(+)}B = [\mu_{\mathbf{e}}^{(+)}, B]\,, \quad \mu_{\mathbf{e}}^{(+)} = \mathbf{e}\cdot \boldsymbol{\mu}^{(+)}\,.
\end{equation}
for a general operator $B$. 
After the probe field has been factored out, propagation without optical fields and dipole detection result in the response
\begin{align}\label{eq:dimer-causal-response}
R_{\mathrm{PP}, a}(t, \Delta\rho) &=\frac{i}{\hbar}\Tr\left[\mu_a^{(-)}G(t)\left(V_a^{(+)}\Delta\rho\right)\right]\,,
\end{align}
where $G(t)$ is the material propagator after the probe interaction.

What is important is that the entanglement test developed in Sec.~\ref{sec:main} concerns the normalized molecular state $\rho\in\mathcal{D}_\star$ at the chosen delay, and not the perturbative correction itself (c.f., Sec.~\ref{subsec:manifold}).
Let $\rho_g = |gg\rangle\langle gg|$ denote the unpumped reference, and define $P_e = \ket{eg}\!\bra{eg} + \ket{ge}\!\bra{ge}$, $\rho_e = P_e\rho P_e$, and the excitation probability $p = \Tr[\rho_e]$.
The part of $\rho$ that is diagonal between excitation sectors and its difference from the ground state reference are
\begin{align}\label{eq:rho-diag-app}
\rho_{\mathrm{diag}} &=\rho_g\rho\rho_g+P_e\rho P_e = (1 - p)\rho_g + \rho_e\,,\\
\Delta\rho &= \rho_{\mathrm{diag}} - \rho_g =\rho_e - p\rho_g\,,\label{eq:Delta-rho-app}
\end{align}
with $\Tr[\Delta\rho] = 0$. 
The term $-p\rho_g$ in Eq.~\eqref{eq:Delta-rho-app} accounts for the loss of ground state absorption as it describes the population removed from the ground state. Moreover, since $\Delta\rho$ is a difference of states, $\Delta\rho$ is not a density operator.

The pump--probe signal is insensitive to coherences between the ground and excited states.
This has the consequence that the response is determined by $\rho_{\mathrm{diag}}$ in Eq.~\eqref{eq:rho-diag-app}. Nevertheless, this does not mean that these coherences need to vanish in the physical state $\rho$.
To see this, denote the operator that represents the isotropic pump--probe response by $\mathcal{P}_{\mathrm{iso}}(\omega)$.
The corresponding response function in the frequency domain can then be written as~\cite{HW13}
\begin{equation}
R_{\mathrm{PP}}^{\mathrm{iso}}(\omega, \Delta\rho) = \Tr[\mathcal{P}_{\mathrm{iso}}(\omega)\Delta\rho]\,.
\end{equation}
We provide the explicit form of $\mathcal{P}_{\mathrm{iso}}(\omega)$ below.
In the model used in Sec.~\ref{subsec:excitonic-dimer}, evolution without optical fields preserves the difference in excitation number. Together with the raising and lowering dipole factors in Eq.~\eqref{eq:dimer-causal-response}, this implies that $[\mathcal{P}_{\mathrm{iso}}(\omega), N] = 0$, where $N = P_e + 2|ee\rangle\langle ee|$ is the total electronic excitation number operator.
Therefore, the third order response has no sensitivity to coherences between different excitation sectors, and as a consequence of that $\rho$ and $\rho_{\mathrm{diag}}$ give the same signal.
Nevertheless, coherences within $\mathcal{H}_e$ remain in the (unnormalized) projected excited state $\rho_e$ and can contribute to the response.
With the response model fixed, the same expression for the response to a weak probe applies to every normalized input in the admissible space of density operators $\mathcal{D}_\star$ defined in Eq.~\eqref{eq:D-star}. 
The following subsections express the differential signal as an expectation value of $\rho$ as required by the construction in Sec.~\ref{sec:main}. In particular, this means that no division by the excitation probability $p = \Tr[\rho_e]$ is needed.

\subsection{Detection}

Next, we relate the response to the correction $\Delta\rho$ defined above to the measured spectrum.
For this, let $R_{\mathrm{PP}}^{\mathrm{iso}}(t,\Delta\rho)$ denote the pump-probe response after the isotropic average. Here, $t$ is the time passed after the probe interaction. Taking the Fourier transform of $R_{\mathrm{PP}}^{\mathrm{iso}}(t,\Delta\rho)$ yields
\begin{equation}\label{eq:dimer-fourier}
R_{\mathrm{PP}}^{\mathrm{iso}}(\omega, \Delta\rho) = \int_0^\infty dt\,e^{i\omega t} R_{\mathrm{PP}}^{\mathrm{iso}}(t, \Delta\rho)\,,
\end{equation}
where $\omega$ is the detected angular frequency.
The correction $\Delta\rho$ is evaluated at the chosen pump-probe delay $T$, which is held fixed in Eq.~\eqref{eq:dimer-fourier}.

For pulses that do not overlap, the induced polarization is a convolution of the probe field with the response to the correction prepared by the pump \cite{Mukamel1995-book, APVM09, HW13}. Then, the component of the third-order molecular polarization selected by the detector $P^{(3)}(\omega, T)$ is approximately
\begin{equation}
P^{(3)}(\omega, T) \approx E_{\mathrm{pr}}(\omega) R_{\mathrm{PP}}\left(\omega, \Delta\rho (T)\right)\,,
\end{equation}
where $E_{\mathrm{pr}}(\omega)$ is the complex spectral amplitude of the probe field. 
The field and polarization transforms use the probe arrival as their time origin as in Ref.~\cite{HW13}.

In self-heterodyne detection, the transmitted probe interferes with the field generated by the pump-induced polarization \cite{Mukamel1995-book}.
Subtracting the pump-off reference removes the probe background.
Neglecting the intensity of the generated field leaves the interference term \cite{HW13}. With the sign convention used here,the signal is
 \begin{align}\label{eq:dimer-detection}
S(\omega, T) &= K(\omega)\mathrm{Im}\left[P^{(3)}(\omega, T)E_{\mathrm{pr}}^*(\omega)\right]\nonumber \\
&  \approx K(\omega)|E_{\mathrm{pr}}(\omega)|^2 \mathrm{Im}R_{\mathrm{PP}} \left(\omega, \Delta \rho(T)\right)\,.
\end{align}
The second line assumes that the prepared state changes negligibly during the probe pulse.
The signal $S(\omega,T)$ describes the pump-induced change in absorption, with positive values corresponding to reduced probe transmission.
The real prefactor $K(\omega)$ converts the absorptive interference term into the recorded signal and includes in particular the optical propagation through the sample.
Furthermore, Eq.~\eqref{eq:dimer-detection} assumes that the field emitted by the pump-induced polarization is much weaker than the transmitted probe. Thus, it accounts for the interference between these fields but neglects the intensity of the emitted field alone~\cite{HW13}. With the given sign convention, a positive signal denotes an increased absorption or, equivalently, a reduced probe transmission.

The dimer example in Sec.~\ref{subsec:excitonic-dimer} serves as an illustration of the construction of the full and separable reachable sets for witnessing exited state entanglement. For this purpose, we take $K(\omega)|E_{\mathrm{pr}}(\omega)|^2$ to be a positive constant over the selected band and absorb it into a common signal unit.
That is an idealization which is independent of the fast probe approximation.
For quantitative comparison with an experiment, instead any frequency dependence of the probe and detection factors should be included in the response before spectral integration.

\subsection{Effective observable}

In the following, we construct the effective observable for the signal needed for the calculation of the full and separable reachable sets.
The dipole operators and material propagator $G(t)$ act on the full four-dimensional electronic space, including the double excited state $\ket{ee}$ and its dual $G^*(t)$ satisfies $\Tr[O G(t)(C)] = \Tr[G^*(t)(O) C]$ for any operators $O$ and $C$ on this space.
Let $\mu_a^{(\pm)}$ denote the dipole components $\mu_{\mathbf{e}}^{(\pm)}$ for polarization along the Cartesian axis $a = x,y,z$.
Then, 
\begin{align}
R_{\mathrm{PP}, a}(t, \Delta\rho) &=  \frac{i}{\hbar}\Tr \left[G^*(t)(\mu_a^{(-)})[\mu_a^{(+)}, \Delta\rho]\right]\\
&= \Tr[\mathcal{P}_a(t)\Delta\rho]\,,\\ 
\mathcal{P}_a(t) &= \frac{i}{\hbar} \left[G^*(t)(\mu_a^{(-)}), \mu_a^{(+)}\right]\,.
\end{align}
Under the assumed magic-angle factorization \cite{HW13}, Fourier transformation leads to 
\begin{equation}\label{eq:Piso_FT-app}
\mathcal{P}_{\mathrm{iso}}(\omega) =  \frac{1}{3}\sum_{a = x,y,z}\int_0^\infty \mathrm{d}t\, e^{i\omega t}\mathcal{P}_a(t)\,.
\end{equation}
The same orientational average is understood for $\Delta\rho$ as defined in Eq.~\eqref{eq:Delta-rho-app}. 

To evaluate Eq.~\eqref{eq:Piso_FT-app}, let $H\ket{m} = \epsilon_m\ket{m}$ and $Q_m = \ket{m}\!\bra{m}$, and pick  energy eigenvectors with definite excitation number.
Moreover, we represent the finite widths of optical transitions through exponential decay of the coherences between energy eigenstates, assign a common decay rate $\kappa > 0$ and use the propagator $G(t) = e^{t\mathcal{L}}$ for
$t\geq 0$, with
\begin{equation}\label{eq:dimer-generator}
\mathcal{L}(C) = -\frac{i}{\hbar}[H, C] + \kappa\left(\sum_m Q_m C Q_m - C\right)\,.
\end{equation}
The dephasing term has Lindblad operators $\sqrt{\kappa} Q_m$, which leaves the populations of the energy eigenstates unchanged, while their coherences decay at rate $\kappa$ in addition to
the coherent evolution generated by $H$.
This decay produces Lorentzian broadening with half-width $\kappa$ in angular frequency units. In particular, Eq.~\eqref{eq:dimer-generator} represents an illustrative choice sufficient for the goal of Sec.~\ref{subsec:excitonic-dimer}.

Moreover, let $\omega_{nm} = (\epsilon_n - \epsilon_m)/\hbar$ denote the transition angular frequencies and define the dipole matrix elements $(\mu_a^{(-)})_{mn} = \bra{m}\mu_a^{(-)}\ket{n}$.
The lowering dipole has no diagonal matrix elements in this basis, which implies that its adjoint propagation reads
\begin{equation}
G^*(t)(\mu_a^{(-)}) = \sum_{m, n}(\mu_a^{(-)})_{mn} e^{-\kappa t - i\omega_{nm}t}\ket{m}\!\bra{n}\,,
\end{equation}
where only transitions with nonzero dipole matrix elements contribute.
With the Fourier convention used in Eq.~\eqref{eq:dimer-fourier}, we get
\begin{equation}
\int_0^\infty \mathrm{d}t\, e^{-\kappa t - i(\omega_{nm} - \omega)t} = \frac{1}{\kappa  +  i(\omega_{nm} - \omega)}\,,
\end{equation}
and substitution into Eq.~\eqref{eq:Piso_FT-app} leads to
\begin{equation}\label{eq:dimer-iso-eval}
\mathcal{P}_{\mathrm{iso}}(\omega) = \frac{i}{3\hbar}\sum_{a, m, n}\frac{(\mu_a^{(-)})_{mn}}{\kappa+i(\omega_{nm}-\omega)}\left[\ket{m}\!\bra{n}, \mu_a^{(+)}\right]\,.
\end{equation}
The absorptive component follows from the Hermitian imaginary part of the response operator. Since $\Delta\rho$ is Hermitian,
\begin{equation}
\mathrm{Im}\left[R_{\mathrm{PP}}^{\mathrm{iso}} (\omega,\Delta\rho)\right] = \Tr\left[\frac{\mathcal{P}_{\mathrm{iso}}(\omega) - \mathcal{P}_{\mathrm{iso}}(\omega)^\dagger}{2i}\Delta\rho\right]\,.
\end{equation}
This eventually leads to $A_{\mathrm{abs}}(\nu)$ in Eq.~\eqref{eq:dimer-Aabs}, with wave number $\nu = \omega/(2\pi c)$.
What is more, for the ground state input, each allowed transition contributes a Lorentzian proportional to
$\kappa/[\kappa^2+(\omega_{ng} - \omega)^2]$, which is weighted by a nonnegative squared transition dipole.
Thus the ground state absorption is nonnegative, and $\kappa$ is the half-width at half-maximum in angular frequency units. However, the differential signal need not have this sign.

To obtain an effective observable on the normalized input domain, we now use $\Delta\rho$ as defined in Eq.~\eqref{eq:Delta-rho-app}. 
As we have explained above, $\rho$ and $\rho_{\mathrm{diag}}$ give the same response. With the common detection scale removed, one obtains
\begin{equation}
S_\Omega(\nu) = \Tr[\Delta\rho A_{\mathrm{abs}}(\nu)]= \Tr[\rho E_\Omega(\nu)]\,,
\end{equation}
where
\begin{align}
E_\Omega(\nu) &= P_\star A_{\mathrm{abs}}(\nu) P_\star - A_g(\nu) P_\star\,,\\
A_g(\nu) &= \Tr[\rho_g A_{\mathrm{abs}}(\nu)]\,.
\end{align}
Since $\Tr[\rho P_\star] = 1$, the reference subtraction can be included in the observable, as in
Eq.~\eqref{eq:dimer-Enu}.
In addition, the term $-p\rho_g$ supplies ground state bleach, while the response to $\rho_e$ includes stimulated emission and absorption into $\ket{ee}$.

For the numerical illustration in Figs.~\ref{fig:dimer-spectra} and \ref{fig:dimer-reachable-sets} in Sec.~\ref{subsec:excitonic-dimer}, we use $\bar\epsilon_m = \epsilon_m/hc,\, \omega=2\pi c\nu,$ and $\kappa=2\pi c\gamma$, where $c$ denotes the speed of light and $\gamma$ is the Lorentzian half-width in wavenumbers.
Next, let $d_0 = |\boldsymbol d_A|$ be the transition dipole magnitude of chromophore $A$ and define $\bar{\mu}_a^{(\pm)} := \mu_a^{(\pm)}/d_0$.
To introduce a fixed response signal unit, we define
\begin{equation}
\mathcal{P}_0 := \frac{d_0^2}{hc\nu_0}
\end{equation}
with $\nu_0=1\,\mathrm{cm}^{-1}$. 
Then, Eq.~\eqref{eq:dimer-iso-eval} becomes 
\begin{align}
\frac{\mathcal{P}_{\mathrm{iso}}(2\pi c\nu)}{\mathcal{P}_0} &= \frac{i\nu_0}{3}\sum_{a, m, n}\frac{(\bar\mu_a^{(-)})_{mn}}{\gamma + i(\bar\epsilon_n - \bar\epsilon_m -  \nu)}\nonumber\\
&\quad \quad \times [|m\rangle\langle n|, \bar\mu_a^{(+)}]\,.
\end{align}
The above definitions specify the units used in Fig.~\ref{fig:dimer-reachable-sets} in Sec.~\ref{subsec:excitonic-dimer}.

\section{Fluorescence observables and (separable) reachable sets \label{app:DBT}}

In this appendix we derive the observables and full and separabe reachable sets used in the fluorescence example of Sec.~\ref{subsec:DBT}.
The preparation model is based on the effective electronic description of  Ref.~\cite{LDWHH24}.

\subsection{Model and state preparation }

Here we introduce the molecular model and the continuous laser drive used to prepare the example input states.

Consider a pair of molecules labelled by $A, B$, and consider bipartite entanglement with respect to the physical partition $A|B$. The reduced basis of the two molecules is $\mathcal{B} = \{\ket{gg},\ket{ge}, \ket{eg}, \ket{ee}\}$.
The lowering operators $\sigma_i, i=A, B$ for the two molecules are $\sigma_A = \ket{g_A}\!\bra{e_A}\otimes \mathbbm{1}_B$ and $\sigma_B = \mathbbm{1}_A \otimes \ket{g_B}\!\bra{e_B}$.
Moreover, we define the observables
\begin{align}\label{eq:DBT-operators}
N &= n_A + n_B\,,  &D& = n_A n_B\,,\nonumber\\
X &= \sigma_A^\dagger\sigma_B + \sigma_B^\dagger\sigma_A\,, &Z& = n_A - n_B\,,\nonumber \\
Y &= -i(\sigma_A^\dagger\sigma_B -  \sigma_B^\dagger\sigma_A)\,.
\end{align}
where $n_i = \sigma_i^\dagger\sigma_i$. 
The operator $N$ counts electronic excitations, $D = \ket{ee}\!\bra{ee}$ projects onto the state with both molecules excited, and $Z$ measures the population imbalance between the two molecules. The expectation values of $X$ and $Y$ give twice the real and imaginary parts of the coherences $\bra{ge}\rho\ket{eg}$, respectively.

Let $\delta =\omega_{\mathrm{L}} - (\omega_A + \omega_B)/2$ denote the detuning of the laser which excites the molecules, where $\omega_L$ is the laser frequency, and $\omega_i, i=A, B$ are the individual transition frequencies of the two molecules $A, B$
In a frame rotating of the laser, the Hamiltonian describing the matter-light interaction is
\begin{equation}\label{eq:DBT-H}
\frac{H_\delta}{\hbar} = -\delta N + \frac{\Delta}{2}Z + J X + \frac{\Omega_R}{2}\sum_{i=A, B}(\sigma_i + \sigma_i^\dagger)\,, 
\end{equation}
where $J$ is the coherent dipole-dipole interaction strength, $\Delta = \omega_A - \omega_B$ denotes the difference between the two molecular transition frequencies, and $\Omega_R$ is the Rabi frequency. 
Following Ref.~\cite{LDWHH24}, we use the approximate the Lindblad generator when the classical light field is turned on
\begin{align}\label{eq:DBT-L}
\mathcal{L}_{\mathrm{on}}^\delta(\rho) &= -\frac{i}{\hbar}[H_\delta,\rho] + \alpha\gamma \mathcal{D}[\sigma_A + \sigma_B]\rho  \nonumber\\
&\quad + (1 - \alpha)\gamma\sum_{i = A, B}\mathcal{D}[\sigma_i]\rho + \gamma_\phi\sum_{i = A, B}\mathcal{D}[n_i]\rho
\end{align}
with dissipator $\mathcal{D}[c]\rho = c\rho c^\dagger - \{c^\dagger c, \rho\}/2$, $\alpha$ is the density of states weighted Franck--Condon factor, $\gamma$ denotes the total decay rate, and $\gamma_\phi$ is the pure dephasing rate. 
For the numerical illustrations in the main text, we use parameters from Ref.~\cite{LDWHH24}, which are $\Delta/(2\pi) = 2.6\,\mathrm{GHz}$, $J/(2\pi) = 1.0\,\mathrm{GHz}$, $\gamma /(2\pi)= 33\,\mathrm{MHz}$, $\gamma_\phi/(2\pi) =10\,\mathrm{MHz}$, $\alpha = 0.11$ and saturation parameter $s = I/I_\mathrm{sat} =  27$ of an uncoupled molecule. The latter is needed to determine the Rabi frequency through the relation $\Omega_R^2 = s\gamma\Gamma_2$, where $\Gamma_2 = (\gamma + \gamma_\phi)/2$ is the rate of decay of the optical coherences of an isolated molecule.

In this model, the state whose entanglement is tested through the signal space membership criterion of Sec.~\ref{sec:main} is the normalized steady state solution of the Lindblad master equation in Eq.~\eqref{eq:DBT-L},
\begin{equation}\label{eq:DBT-ss}
\mathcal{L}_{\mathrm{on}}^\delta(\rho_\delta) = 0
\end{equation}
The solutions of Eq.~\eqref{eq:DBT-ss} are the states whose entanglement is to be tested, i.e., the states marked in Fig.~\ref{fig:DBT-slices} for the three shown examples of $\delta$.
What is important here is that the prepared state depends on the detuning parameter $\delta$. Therefore, spectral windows of a scan over detunings of fluorescence excitation cannot be interpreted as joint expectations of one common input state, in contrast to the probe spectrum at fixed preparation considered in Sec.~\ref{subsec:excitonic-dimer}.
Moreover, the active manifold is the full model Hilbert space spanned by the basis $\mathcal{B}$ specified above, that is $\mathcal{H}_\star = \mathcal{H} = \mathrm{span}(\ket{gg}, \ket{ge}, \ket{eg}, \ket{ee})$.
On the level of the sets $\mathcal{D}_\star$ and $\mathcal{D}_{\star, \mathrm{sep}}$, one does not impose any stationary constraint. The latter only determines the state the experimental/model setup prepares. 

\subsection{Photon detection}

Next, we derive the fluorescence and coincidence observables used to construct the signal map in Sec.~\ref{subsec:DBT}. The input state $\rho$ is the molecular state at time $t = 0$. Here $t=0$ refers to the time where the laser preparation of the steady state ends and the laser drive is assumed be switched off instantaneously. This model assumption is sufficient for our illustrative purposes but should be revisited for comparison to actual experiments. 
The Lindblad generator without the laser drive is denoted by $\mathcal{L}_{\mathrm{off}}$. It contains the dissipators of Eq.~\eqref{eq:DBT-L}, with Hamiltonian $H_{\mathrm{off}}/\hbar = \Delta Z/2 + JX$ in a frame rotating at the mean molecular frequency.
Since the additional term $-\delta N$ in Eq.~\eqref{eq:DBT-H} in the rotating frame of the laser affects only phases between excitation sectors, the detuning $\delta$ modifies the steady state preparation but not the readout observables used for witnessing entanglement.

According to Ref.~\cite{LDWHH24}, the experiment selects fluorescence through vibrational channels. The detection efficiency $\eta$ is assumed to be the same for both molecules. 
Then, the detected jump operators are given by $c_i = \sqrt{\kappa} \sigma_i$, with rate $\kappa = \eta(1-\alpha)\gamma$ (see Eq.~\eqref{eq:DBT-L}).
This leads to the associated Lindblad jump map and photon count rate observable, 
\begin{equation}\label{eq:DBT-detected-jumps}
\mathcal{J}(\rho) = \sum_{i = A, B} c_i\rho c_i^\dagger\,,\quad  C_{\mathrm{det}} = \mathcal{J}^*(\mathbbm{1}_{\mathcal{H}}) = \kappa N\,.
\end{equation}
Emission channels that are not detected are not included in $\mathcal{J}$. Nevertheless, they are contained in $\mathcal{L}_{\mathrm{off}}$, since they also affect the time evolution of the two molecules.

For any input state $\rho\in\mathcal{D}_\star$, Eq.~\eqref{eq:DBT-detected-jumps} allows to construct the effective measurement operator and the adjoint construction of Sec.~\ref{subsec:spectroscopic-map} leads to
\begin{align}
I_\Omega(t; \rho) &= \Tr[\rho E_\Omega(t)]\,,\\
E_\Omega(t) &= e^{t\mathcal{L}_{\mathrm{off}}^*}(C_{\mathrm{det}})\,. 
\end{align}
This recovers Eq.~\eqref{eq:DBT-I} in the main text, where the measured signal is the fluorescence intensity $I_\Omega$ of the experimental protocol. 
The unnormalized coincidence density at equal times at the start of readout is
\begin{equation}\label{eq:DBT-coincidence}
G_\Omega^{(2)}(0, 0) = \sum_{i,j = A,B}\Tr[\rho c_i^\dagger c_j^\dagger c_j c_i] =  2\kappa^2\Tr[D \rho]\,.
\end{equation}
Since $c_i^2=0$, only the two ordered terms with $i\neq j$ contribute, whose sum probes the initial probability of double excitation. Since $G_\Omega^{(2)}(0, 0)$ is a density in two arrival times, it does not yet describe a finite coincidence count. 
The following subsections use these ideal response quantities to
construct the full and separable reachable sets.

\subsection{Spectroscopic map and selected coefficients}

Next, we derive the three expectation values used in Sec.~\ref{subsec:DBT} from the fluorescence signals.
The corresponding effective observables follow from the detected jump map in Eq.~\eqref{eq:DBT-detected-jumps} and the evolution generated by $\mathcal L_{\mathrm{off}}$.
The input state $\rho$ is specified at $t = 0$, when preparation ends and the drive is removed instantaneously, as explained in the previous section.
We use the initial intensity and the slope of its subsequent decay to construct the effective observables, together with the initial coincidence density, as they lead to  simple expressions for the expectations of $N$, $X$, and $D$ in Eq.~\eqref{eq:DBT-operators}. 
The adjoint generator satisfies
\begin{equation}
\mathcal{L}_{\mathrm{off}}^*(N) = -\gamma(N + \alpha X)\,.
\end{equation}
In combination with $C_{\mathrm{det}} = \kappa N$ (see Eq.~\eqref{eq:DBT-detected-jumps}) and Eq.~\eqref{eq:DBT-coincidence} and assuming $\alpha>0$ and $\kappa>0$, this identity allows us to introduce the three quantities
\begin{align}
n &= \Tr[\rho N] = \frac{I_\Omega(0;\rho)}{\kappa}\,, \label{eq:DBT-n}\\
x &= \Tr[\rho X] = -\frac{\dot{I}_\Omega(0^+; \rho)+\gamma I_\Omega(0; \rho)}{\alpha\gamma\kappa}\,,\label{eq:DBT-x}\\
d &= \Tr[\rho D]= \frac{G_\Omega^{(2)}(0, 0; \rho)}{2 \kappa^2}\,,\label{eq:DBT-d}
\end{align}
where $\dot{I}_\Omega(0^+; \rho)$ is the right derivative, which describes the initial temporal variation after removal of the drive, and $\rho \in\mathcal D_\star$. 
Therefore, the three coefficients in Eqs.~\eqref{eq:DBT-n}-\eqref{eq:DBT-d} define the spectroscopic map $F_\Omega^{(3)}(\rho) = (n,x,d)$ in Eq.~\eqref{eq:DBT-map}, whose full and separable reachable sets are derived below.
In particular, the observable family obtained from the fluorescence measurement and its dynamics, without assuming access to arbitrary molecular operators.
The choice of initial quantities is not essential to the framework, and time integrals of the same fluorescence trace would provide alternative coefficients. 

\subsection{Full and separable reachable sets}

We now derive the full and separable reachable sets illustrated in Sec.~\ref{subsec:DBT}. 
For this, we first observe that the coefficients $n, x, d$ in Eqs.~\eqref{eq:DBT-n}-\eqref{eq:DBT-d} define the populations $p_{gg} = 1-n+d$, $p_{ee} = d$, and $p_{ge}+p_{eg} = n - 2d$, where $p_{ij} = \rho_{ij, ij}$.
We abbreviate $c = \bra{ge}\rho\ket{eg}$. Positivity of $\rho$ implies $|c|^2\leq p_{ge} p_{eg}$ and, hence, $|x|\leq n - 2d$.
For a separable state $\rho \in \mathcal{D}_{\star, \mathrm{sep}}$, positivity of the partial transpose also
requires $|c|^2\leq p_{gg} p_{ee}$.
Therefore, the full and separable images of the spectroscopic map follow as
\begin{align}\label{eq:DBT-reachable-app}
\mathcal{R}^{(\Omega, 3)} &= \left\{(n,x,d): d\geq 0,\ 1-n+d \geq 0,|x| \leq n-2d \right\},\nonumber \\
\mathcal{R}_{\mathrm{sep}}^{(\Omega, 3)} &=\left\{(n,x,d)\in\mathcal{R}^{(\Omega,3)}: |x|\leq 2 \sqrt{d(1-n+d)}\right\}\,.
\end{align}
To show that these conditions characterize the full and separable reachable sets exactly, consider the state
\begin{equation}\label{eq:DBT-compl}
\rho(n,x,d) = \begin{pmatrix}
1-n+d & 0 & 0 & 0\\
0 & (n-2d)/2 & x/2 & 0\\
0 & x/2 & (n-2d)/2 & 0\\
0 & 0 & 0 & d
\end{pmatrix}
\end{equation}
in the ordered basis $\mathcal{B}$ introduced earlier.
For every $(n,x,d)\in\mathcal R^{(\Omega,3)}$, the matrix $\rho(n,x,d)$ is a valid density operator and satisfies $F_\Omega^{(3)}(\rho(n,x,d)) = (n,x,d)$.
In addition, the partial transpose of $\rho(n,x,d)$ is positive whenever the additional separable inequality in $\mathcal{R}_{\mathrm{sep}}^{(\Omega, 3)}$ in Eq.~\eqref{eq:DBT-reachable-app} holds. Since the physical system consists of two qubits, this condition is sufficient for separability \cite{HHH96}.
Therefore, Eq.~\eqref{eq:DBT-reachable-app} gives the exact full and separable reachable sets for the coefficients $N, x, d$.

At fixed $n$, the values of $d$ are constrained by $\max(0, n-1) \leq d \leq n/2$.
Moreover, any point on the slice of $\mathcal{R}^{(\Omega, 3)} $ for fixed $n$ has $|x|\leq n-2d$. The corresponding slice of $\mathcal{R}_\mathrm{sep}^{(\Omega, 3)}$ instead satisfies
\begin{equation}
|x| \leq \min\left\{n - 2d, 2 \sqrt{d(1 - n + d)}\right\}\,.
\end{equation}
Each panel of Fig.~\ref{fig:DBT-slices} fixes $n \equiv n_\delta = \Tr[\rho_\delta N]$ and displays the allowed
pairs $(d,x)$ at that value.
This uses that a separable explanation must reproduce the fixed value of $n$ as well as both plotted coordinates. Thus, the panels in Fig.~\ref{fig:DBT-slices} show cross sections of the full and
separable three-dimensional sets. It is important to notice that these are not projections that discard one coefficient. More specifically, the chosen model parameters determine the prepared state and the conversion of fluorescence signals into the expectations of $N,X,D$. Changing the model parameters can move the target point whose entanglement is to be tested and select a different slice of the reachable sets. At the same time, the underlying three-dimensional full and separable reachable sets remain unchanged.

\subsection{Negativity of steady-states}

\begin{figure}[tb]
\includegraphics[width=\linewidth]{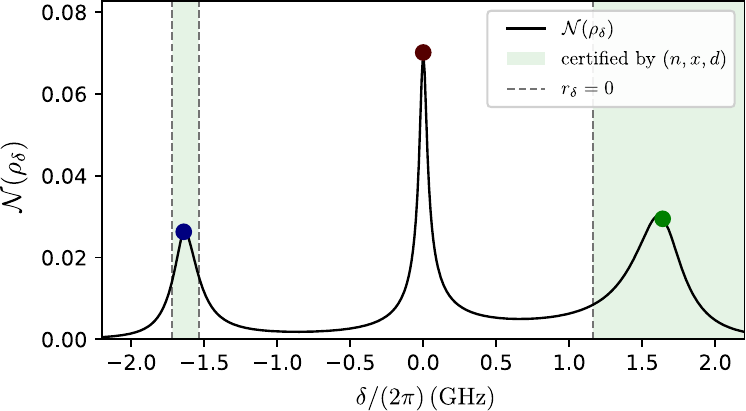}
\caption{Negativity $\mathcal{N}(\rho_\delta)$ of the model steady states for the same parameters and detuning scan as in Fig.~\ref{fig:DBT-residual}. Green areas mark $r_\delta > 0$, where the retained coefficients certify entanglement. Dashed vertical lines indicate zero crossings of the residual. The coloured markers correspond to the three examples shown in Fig.~\ref{fig:DBT-slices}. The negativity is positive at every sampled detuning, including outside the bands that allow for entanglement certification. \label{fig:DBT-negativity}}
\end{figure}

To clarify the interpretation of the residual in Sec.~\ref{subsec:DBT}, we evaluate the negativity of the calculated steady states \cite{VW02}
\begin{align}\label{eq:DBT-negativity}
\mathcal{N}(\rho_\delta) = \frac{\| \rho_\delta^{T_B}\|_1 - 1}{2}\,,
\end{align}
where $T_B$ denotes partial transposition with respect to molecule $B$ and $\|\cdot\|_1$ the trace norm. For the two-qubit system, a state is entangled if and only if $\mathcal{N}(\rho_\delta) > 0$. 
Figure~\ref{fig:DBT-negativity} displays a positive negativity $\mathcal{N}(\rho_\delta)$ at every sampled detuning for the parameter set considered here. This means that the sampled states outside the green certification bands are entangled, although their coefficients $(n_\delta, x_\delta, d_\delta)$ are
reproducible by separable states. This includes the preparation for $\delta=0$, for which the residual is negative, as shown in Fig.~\ref{fig:DBT-residual} in Sec.~\ref{subsec:DBT}. 
The comparison of Fig.~\ref{fig:DBT-residual} and \ref{fig:DBT-negativity} distinguishes the entanglement of the full model state from its certification through the range of the spectroscopic map. 
The residual characterizes the ability to witness entanglement. However, it is not an entanglement measure in contrast to $\mathcal{N}(\rho_\delta)$. 

\subsection{Projection onto two coefficients}

In the following, we illustrate how an entanglement certificate can be lost when the probability for double excitation is omitted from the retained data.
For this comparison, we project the map $F_\Omega^{(3)}(\rho)=(n, x, d)$ onto $F_\Omega^{(2)}(\rho) = (n, x)$. A short calculation shows that the projected sets follow as
\begin{align}\label{eq:DBT-proj}
\mathcal{R}^{(\Omega, 2)} &= \left\{(n,x) : 0\leq n\leq 2,|x|\leq\min(n, 2-n)\right\},\nonumber\\
\mathcal{R}_{\mathrm{sep}}^{(\Omega, 2)} &=\left\{(n,x) : 0\leq n\leq 2, |x|\leq n(2-n)/2\right\}\,.
\end{align}
For fixed $n$, the largest allowed value of $|x|$ follows from Eq.~\eqref{eq:DBT-reachable-app} by choosing the smallest allowed probability of double excitation, which is $d = \max(0,n-1)$. This gives the boundary $|x|=\min(n, 2-n)$ of the full projected reachable set $\mathcal R^{(\Omega,2)}$ above. To obtain $\mathcal{R}_{\mathrm{sep}}^{(\Omega, 2)}$, we first observe that a product state with local excitation probabilities $p_A, p_B$,
\begin{equation}
|x|\leq 2\sqrt{p_A(1 - p_A) p_B(1 - p_B)} \leq \frac{n(2 - n)}{2}\,, 
\end{equation}
where the second inequality is attained for $p_A = p_B = n/2$. A relative phase between the local amplitudes gives every intermediate value of $x$, and concavity preserves the bound under mixtures, which leads to $\mathcal{R}_{\mathrm{sep}}^{(\Omega, 2)}$.

\begin{figure}[t]
\centering
\includegraphics[width=\linewidth]{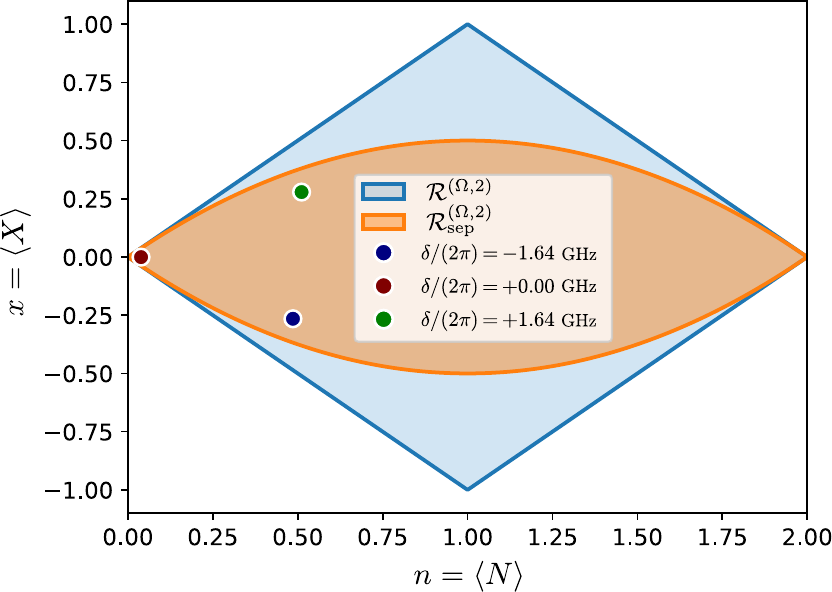}
\caption{Full (blue) and separable (orange) reachable sets projected onto $(n,x)$. Circles denote the three input signals from Fig.~\ref{fig:DBT-slices}.
As a result of the projection and discarding the information about $d$, all three input states admit a separable explanation. However, entanglement in the two resonance inputs can be certified once $d$ is included.\label{fig:DBT-projection-app}}
\end{figure}

We illustrate in Fig.~\ref{fig:DBT-projection-app} the full projected reachable set $\mathcal{R}^{(\Omega, 2)}$ (blue) with its separable subset $\mathcal{R}_{\mathrm{sep}}^{(\Omega, 2)}$ (orange) as defined in Eq.~\eqref{eq:DBT-proj}. The two axes represent the mean excitation number $n$ and the real exchange coherence $x$, while the double excitation probability $d$ is omitted as a result of the projection. The three coloured circles mark the signals of the three preparations used in Fig.~\ref{fig:DBT-slices} in Sec.~\ref{subsec:DBT}. Importantly, all three lie inside $\mathcal{R}_{\mathrm{sep}}^{(\Omega, 2)}$, which implies that the entanglement test via set membership is inconclusive for all three inputs in contrast to Fig.~\ref{fig:DBT-slices}.
In fact, the separable comparison now allows any value of $d$ compatible with $n$ and $x$.
For the two input states with $\delta/ (2\pi) = \pm 1.64\,\mathrm{GHz}$,  a separable state reproduces the plotted coordinates only with a different value of $d$. Once actual value of $d$ is accessible, no separable explanation of the data is possible, as shown by the three-coordinate test in Fig.~\ref{fig:DBT-slices}. What this shows is that omitting the probability of double excitation removes an entanglement certificate despite the input states and physical bipartition are unchanged. In turn, it also demonstrates that enlarging the set of admissible coefficients can allow for witnessing entanglement in states due to the additional information, which has previously been inaccessible.

\section{Two-photon response and reachable sets of the two-orbital model \label{app:two-orbital-TPA}}

In this section, we derive the effective observables used in Sec.~\ref{subsec:RP-TPA} from the electronic Hamiltonian and two-photon transition tensor. The electronic model is taken from Refs.~\cite{Nakano09, Nakano17}, while the optical readout is an additional choice that will be specified below.

\subsection{Model}

We consider two electrons in two localized spatial orbitals. The orbitals $A$ and $B$ are taken to be real and orthonormal. To fix the signs of the fermionic basis states, we order the spin orbitals as $(A\!\uparrow, A\!\downarrow, B\!\uparrow, B\!\downarrow)$ and fix for later the orbital labels $1\leftrightarrow A\!\uparrow, 2\leftrightarrow A\!\downarrow, 3\leftrightarrow B\!\uparrow$ and $4\leftrightarrow B\!\downarrow$.
The one-particle Hilbert space is $\mathfrak{h}= \mathfrak{h}_A \oplus\mathfrak{h}_B$, where $\mathfrak{h}_A\cong \mathfrak{h}_B\cong\CC^2$.
The corresponding fermionic Fock space reads
\begin{equation}
\mathfrak{F} = \bigoplus_{n = 0}^{4}\wedge^n\mathfrak{h}\,, \quad \wedge^0\mathfrak{h}\cong\CC\,.
\end{equation}
We choose its two-electron sector as the active manifold,
\begin{equation}\label{eq:OS-domain}
\mathcal{H}_\star = \wedge^2\mathfrak{h}\,.
\end{equation}
Thus, $\dim(\mathfrak{F}) = 16$ and $\dim(\mathcal{H}_\star) = 6$. With the mode order fixed as stated above, the occupation representation determines $\mathfrak{F}\cong \mathcal{H}_A\otimes\mathcal{H}_B$, where $\mathcal{H}_i = \bigoplus_{n = 0}^{2}\wedge^n\mathfrak{h}_i$ is the four-dimensional local Fock space for $i=A, B$. We test orbital entanglement across this bipartition with both spin and occupation as local degrees of freedom. No additional superselection rules are considered \cite{DMSSZS21, DGS24}. 

For $i<j$, write $|ij\rangle = c_i^\dagger c_j^\dagger|0\rangle$.
A basis of $\mathcal{H}_\star$ is spanned by the singlet and triplet states 
\begin{align}
\ket{C} &= \frac{1}{\sqrt{2}}(\ket{14}  - \ket{23})\,,  \quad 
\ket{I_\pm} = \frac{1}{\sqrt{2}}(\ket{12}\pm \ket{34})\,,\\
\ket{T_+} &= \ket{13}\,, \quad  \ket{T_0}  =  \frac{1}{\sqrt{2}}(\ket{14} +\ket{23})\,,\quad \ket{T_-} = \ket{24}\,.
\end{align}
The label $C$ denotes the covalent singlet with one electron on each orbital. The states $\ket{I_\pm}$ are the symmetric and antisymmetric combinations of the ionic configurations $\ket{\uparrow\downarrow, 0}$ and $\ket{0, \uparrow\downarrow}$. 

Next, consider the spin-independent electronic Hamiltonian
\begin{equation}
H = \sum_{ij} h_{ij} c_i^\dagger c_j + \frac{1}{2}\sum_{ijkl} W_{ij, kl} c_i^\dagger c_j^\dagger c_lc_k\,,
\end{equation}
where $h_{ij}$ and $W_{ij, kl}$ are the matrix elements of the one-particle Hamiltonian and the two-body Coulomb interaction. In the chosen real orbital basis, we adopt the standard notation where $(pq|rs)$ denotes the two-electron Coulomb integral between the orbital products $\phi_p\phi_q$ and $\phi_r\phi_s$, with $p, q, r, s \in\{A, B\}$.
Under the assumed site symmetry, $h_{AA} = h_{BB}$, $(AA|AA)  = (BB|BB)$, and $(AA|AB) = (BB|AB)$.
We then define
\begin{align} 
U &= (AA|AA) - (AA|BB)\,,\quad K =  (AB|AB)\,,\nonumber \\
\tau  &= |h_{AB} + (AA|AB)|\,,
\end{align}
where $U$ is the difference between the on-site and inter-site Coulomb repulsions, $K$ the direct exchange integral, and $\tau$ is an effective transfer magnitude. 

The relative orbital phase is chosen such that $h_{AB} + (AA|AB) = -\tau$. Moreover, define
$H_0 = H|_{\mathcal{H}_\star} - \widetilde{E} \mathbbm{1}_{\mathcal{H}_\star}$ with reference energy $\widetilde{E} = 2h_{AA}+(AA|BB)$. 
The matrix representation of $H_0$ in the ordered basis $(\ket{C},\ket{I_+}, \ket{I_-}, \ket{T_+}, \ket{T_0}, \ket{T_-})$ reads
\begin{equation}\label{eq:OS-H}
H_0 = \begin{pmatrix} K & -2 \tau & 0\\
-2 \tau & U + K & 0 \\
0 & 0 & U - K  \end{pmatrix} \oplus(-K \mathbbm{1}_3)\,,
\end{equation}
where $\mathbbm{1}_3$ is the identity on the triplet sector.
Equal transfer amplitudes from $C$ to the two ionic configurations give the coupling $-2 \tau$ to $I_+$ and cancel for $I_-$. For real orbitals, the matrix element $\bra{12} H\ket{34}$ equals $K$ and leads to the ionic diagonal entries $U+K$ and $U-K$.

To evaluate the two-photon transition tensor, we first express $H_0$ in its energy eigenbasis.
Define $\Lambda = \sqrt{U^2 + 16\tau^2}$ and the mixing coefficients
\begin{align}\label{eq:OS-mixing}
c &= \sqrt{\frac{1 + U/\Lambda}{2}}\,,  \quad s = \sqrt{\frac{1 - U/\Lambda}{2}}\,.
\end{align}
Diagonalization of the singlet block gives
\begin{align}
\ket{g} &= c \ket{C} + s\ket{I_+},\,\,\ket{o} = \ket{I_-},\,\,
\ket{f} = -s\ket{C}  +  c\ket{I_+}. 
\label{eq:OS-eigenstates}
\end{align}
Together with the three triplets $|T_m\rangle$, where $ m  = -1,0,1$, these states form an orthonormal eigenbasis of $H_0$ with eigenenergies
\begin{align}
E_g &= K + \frac{U - \Lambda}{2}\,,\nonumber\\
E_o &= U - K\,,\nonumber\\
E_f &= K + \frac{U + \Lambda}{2}\,,\nonumber\\
E_{T_m} &= -K\,.
\label{eq:OS-eigenenergies}
\end{align}
For $U > 0$, $\tau > 0$, and $0 \leq K < (\Lambda - U)/4$, the lowest singlet is the ground state and $E_g < E_o < E_f$.
To illustrate the witness construction of Sec.~\ref{sec:main}, we choose $\ket{g}$ as the target state, with density operator $P_g = \ket{g}\!\bra{g}$. 
As in Sec.~\ref{subsec:manifold}, the full and separable reachable sets contain the signals of all density operators on the active manifold $\mathcal{H}_\star$ and their subset separable across $A|B$, respectively.

We consider a collinear transition dipole
\begin{align}
\boldsymbol{\mu} &= \mu \,\mathbf{u}\,, \quad \mu = \frac{d_0}{2}(N_A - N_B)\,,
\quad d_0=eR_{AB}\,,
\end{align}
where $\mathbf{u}$ is unit vector, and $R_{AB}$ the effective separation between the centres of the two orbitals labelled by $A$ and $B$. Moreover, the operator $N_i  = \sum_{\sigma=\uparrow, \downarrow} c_{i\sigma}^\dagger c_{i\sigma}$ counts electrons on orbital $i$, and $e > 0$ is the elementary charge. 
The matrix representation of the dipole operator is
\begin{equation}\label{eq:OS-dipole}
\mu = \begin{pmatrix} 0 & a & 0  \\
a  & 0 & b \\
0 & b & 0  \end{pmatrix} \oplus \mathbb{0}_3\,,
\end{equation}
in the ordered basis $(\ket{g}, \ket{o}, \ket{f}, \ket{T_+}, \ket{T_0}, \ket{T_-})$ with $a = d_0s, b = d_0 c$. 
The nonzero matrix elements connect $|g\rangle$ and $|f\rangle$
to the odd singlet $|o\rangle$.
As a consequence, only $|o\rangle$ contributes as an intermediate state to the selected $g\to f$ two-photon transition.

\subsection{Transition tensor}

The $g \leftrightarrow f$ two-photon transition can be isolated by using two weak pulses whose carrier frequencies satisfy $\omega_1 + \omega_2 = \omega_{fg}$, where $\omega_{ji} = (E_j - E_i)/\hbar$ as in Appendix~\ref{app:Ax-Bx}.
We further assume that the magnitudes of the one-photon detunings are much larger than the pulse bandwidths and the corresponding Rabi frequencies. Under these conditions, the optical fields produce only a small admixture of $|o\rangle$ into $|g\rangle$ and $|f\rangle$, which allows for its treatment as a virtual intermediate state in their two-photon transition.

Each measurement starts from the same molecular input state $\rho$. We consider the energy transferred from the optical fields to the molecule during the pulses.
From its perturbative expansion, we select the term proportional to $\mathcal{I}_1\mathcal{I}_2$, where $\mathcal{I}_j$ denotes the intensity scale of pulse $j$ at fixed pulse shape.
Substitution of Eq.~\eqref{eq:OS-dipole} into the two-photon transition tensor in Eq.~\eqref{eq:Tfi_ab} yields
\begin{align}\label{eq:OS-tensor}
T_{\alpha \beta}^{fg}(\omega_1, \omega_2) &= u_\alpha u_\beta\, t_\Omega\,,\\
t_\Omega &= \frac{ab}{\hbar} \left(\frac{1}{\omega_ {og} - \omega_1} + \frac{1}{\omega_{og} - \omega_2}\right)\,,\\
\omega_{og} &=   \frac{U - 4K + \Lambda}{2\hbar}\,.
\end{align}
The two terms in $t_\Omega$ describe absorption of one photon from each field in either order. Within the assumed regime of large one-photon detuning, we set the damping $\gamma_o$ in Eq.~\eqref{eq:Tfi_ab} to zero. For equal photon frequencies on the two-photon resonance, $\omega_1 = \omega_2 = \omega_{fg}/2$, the amplitude reduces to $t_\Omega = 4ab/(U - 4K)$. Both one-photon transitions are resonant for $U = 4K$, such that the dynamics of the intermediate state $\ket{o}$ can not be neglected anymore. Therefore, any system in the vicinity of $U = 4K$ is excluded here. 

To determine the transition amplitude and the scale of the retained signal, the next step is to combine the electronic transition tensor with the amplitudes and temporal profiles of the two optical pulses.
The positive frequency field is
\begin{equation}
\mathbf{E}^{(+)}(t) = \sum_{j = 1}^2\mathbf{e}_j \mathcal{E}_j(t) e^{-i\omega_j t + i\phi_j}\,,
\end{equation}
where $\mathcal{E}_j(t) $ are complex pulse envelopes and the negative frequency part follows as its complex conjugate. At the assumed two-photon resonance, the electronic transition tensor and the pulse envelopes determine the amplitude
\begin{align}\label{eq:OS-pulse-scale}
\mathcal{A}_\Omega &= \frac{t_\Omega}{\hbar} \int_{-\infty}^{\infty}\mathrm{d}t\, \mathcal{E}_1(t)\mathcal{E}_2(t)\,, \quad C_\Omega = \frac{|\mathcal{A}_\Omega|^2}{15}\,.
\end{align}
Within the stated narrow band approximation, $\mathcal{A}_\Omega$  is the dimensionless amplitude for absorption of one photon from each pulse, apart from the polarization factor and an overall phase. The integral accounts for the temporal overlap and complex phases of the pulse envelopes. For the rotational average below, we already include the factor $1/15$ in the definition of $C_\Omega$. We assume $0<|\mathcal{A}_\Omega|^2\ll 1$, which implies that the selected transition has a nonzero amplitude and its probability remains small.

The pulse amplitude is related to population transfer through the effective coupling between $|g\rangle$ and $|f\rangle$.
For a molecule with orientation $R$, the polarization dependence enters through
$b_R = [\mathbf{e}_2\cdot (R\mathbf{u})][\mathbf{e}_1\cdot(R\mathbf{u})]$.
The second order amplitude for the selected $g \to f$ transition is
\begin{align}
a_{fg}^{(2,R)} =  i e^{i(\phi_1 + \phi_2)}b_R \mathcal{A}_\Omega\,,
\end{align}
and the transition probability at the lowest nonzero order in the optical fields follows as
\begin{align}\label{eq:OS-p2R}
p_2^{(R)} := |a_{fg}^{(2,R)}|^2 = |\mathcal{A}_\Omega|^2b_R^2\,.
\end{align}
The reverse process describes stimulated two-photon emission from $|f\rangle$ to $|g\rangle$.
Since the isolated two-level evolution is unitary, it has the same transition probability $p_2^{(R)}$ in Eq.~\eqref{eq:OS-p2R}. 
Initial population of $|o\rangle$ and the three triplet states make no contribution to the selected net energy transfer. 

Next, we combine the absorption and emission probabilities to obtain the signal for an arbitrary input. We average uniformly over the optical phases $\phi_1,\phi_2$ for a fixed molecular preparation. Then, contributions proportional to the initial coherence $\rho_{gf}$ cancel out. As this average is part of the readout process it does not impose restriction on the input state. As both transition directions have probability $p_2^{(R)}$, the final state population is a function of $\rho_{gg}-\rho_{ff}$.
Let $\rho_{\mathrm{out}}$ denote the molecular state after the pulses, for a fixed orientation and optical phases. Relative to the input state $\rho$ before the pulses, define $\Delta\langle P_f\rangle  = \Tr[P_f(\rho_{\mathrm{out}} - \rho)]$, where $P_f = \ket{f}\!\bra{f}$, and $\Delta\langle H_0\rangle = \Tr[H_0(\rho_{\mathrm{out}} - \rho)]$, 
which describe the change in the population of $|f\rangle$ and in molecular energy, respectively.
The selected fourth-order contribution for each density operator $\rho\in\mathcal{D}_\star$ gives
\begin{align}
\overline{\Delta\langle P_f\rangle}^{\,\phi} &= p_2^{(R)}(\rho_{gg} - \rho_{ff})\,, \label{eq:OS-pop-transfer}\\
\overline{\Delta\langle H_0\rangle}^{\,\phi} &= \Lambda p_2^{(R)}(\rho_{gg} - \rho_{ff})\,,\label{eq:OS-transfer}
\end{align}
where the overbar denotes the average over optical phases. Since relaxation is neglected and the fields are assumed to vanish at both endpoints of the protocol, the molecular energy change equals the energy supplied by the fields.
Equation~\eqref{eq:OS-transfer} relates the optical signal to the input populations and provides the basis for the effective observable derived below.

\subsection{Effective observables}

We now express the retained optical signal through the effective observables that define the spectroscopic map of Sec.~\ref{subsec:spectroscopic-map}.
For the selected contribution in Eq.~\eqref{eq:OS-transfer}, division by $\Lambda = E_f-E_g$ gives the dimensionless signal
\begin{align}\label{eq:OS-signal}
S_\Omega^{(R)}(\rho) = \frac{\overline{\Delta\langle H_0\rangle}^{\,\phi}}{\Lambda} = p_2^{(R)}\Tr[\rho (P_g - P_f)]\,. 
\end{align}
To connect Eq.~\eqref{eq:OS-signal} to Appendix~\ref{app:Ax-Bx}, we represent absorption and stimulated emission by the two transition operators
\begin{equation}\label{eq:OS-trans-op}
M_\Omega^{(R, +)} = b_R\mathcal{A}_\Omega\ket{f}\!\bra{g}\,, \quad M_\Omega^{(R, -)} = b_R\mathcal{A}_\Omega^*\ket{g}\!\bra{f}\,.
\end{equation}
We omit overall phases have been omitted as they cancel in the transition probabilities.
The two projectors $P_f$ and $P_g$ select the final level of each process.
Their contributions enter the signal $S_\Omega^{(R)}(\rho)$ with opposite signs. Thus, $S_\Omega^{(R)}(\rho) = \Tr[\rho E_\Omega^{(R)}]$, with
\begin{align}
E_\Omega^{(R)} &= M_\Omega^{(R,+)\dagger} P_f M_\Omega^{(R,+)} - M_\Omega^{(R,-) \dagger} P_g M_\Omega^{(R,-)} \nonumber\\
&= |\mathcal A_\Omega|^2b_R^2 (P_g - P_f)\,.
\end{align}
Each term has the positive form derived in Appendix~\ref{app:Ax-Bx}.
Their difference represents the energy transfer in Eq.~\eqref{eq:OS-transfer} within the approximations explained earlier. 

For an isotropic ensemble of randomly oriented molecules, every molecular orientation has the same input state in the molecular frame. The orientation average then acts on the observable, and the rotational contractions of Appendix~\ref{app:Ax-Bx} lead to 
\begin{align}
\langle b_R^2\rangle_R &= \frac{1+2\cos^2\theta}{15}\,, \quad \cos\theta = \mathbf{e}_1\cdot\mathbf{e}_2\,.
\end{align}
With $C_\Omega = |\mathcal{A}_\Omega|^2/15$ as defined in Eq.~\eqref{eq:OS-pulse-scale}, the averaged response is given by
\begin{align}
E_\Omega(\theta) &= \left\langle E_\Omega^{(R)}\right\rangle_R = C_\Omega (1 + 2\cos^2\theta) (P_g - P_f)\,.
\end{align}
Comparison with Eq.~\eqref{eq:S-TPA} in Sec.~\ref{subsec:spectroscopic-map} identifies the two observables associated with the angular coefficients,
\begin{align}
X_\Omega^{(1)} = C_\Omega (P_g - P_f)\,, \quad X_\Omega^{(2)} = 2C_\Omega(P_g - P_f)\,.
\end{align}
This eventually recovers Eq.~\eqref{eq:OS-response} in Sec.~\ref{subsec:RP-TPA}.

In the notation of Sec.~\ref{subsec:spectroscopic-map}, the coefficients obtained from the angular fit define the spectroscopic map
\begin{align}\label{eq:OS-signal-map}
F_\Omega^{(2)}(\rho) &= \left(\Tr[\rho X_\Omega^{(1)}], \Tr[\rho X_\Omega^{(2)}]\right) \nonumber\\
& = C_\Omega\Tr[\rho (P_g-P_f)](1,2)\,.
\end{align}
Both signal coordinates depend on the same expectation value. The full and separable reachable sets follow from its ranges over $\mathcal{D}_\star$ and $\mathcal{D}_{\star,\mathrm{sep}}$, as shown in the following.  It is worth recalling that $C_\Omega$ is fixed by the molecular model and prescribed pulses and its calibration preserves the state-dependent signal amplitude which is required for the entanglement certificate.

\subsection{Separable reachable set}

Equipped with the effective observables and the spectroscopic map, we are now able to calculate the separable reachable set. Due to Eq.~\eqref{eq:OS-signal-map}, the full and separable reachable sets are determined by the possible values of $\Tr[\rho (P_g-P_f)]$. The spectrum of $Q = P_g - P_f$ is $\{1, -1, 0, 0, 0, 0\}$. This implies that its full expectation range is $[-1,1]$. To obtain the separable range, we first characterize the states that are separable across $A|B$ and supported on the two-electron manifold $\mathcal{H}_\star$.
In a decomposition of an admissible separable state $\sigma$ into product states, every component with nonzero weight must lie in the active manifold $\mathcal{H}_\star$. Consider one such component $|\psi\rangle = |\psi_A\rangle\otimes|\psi_B\rangle$. To determine its possible local occupations, we use that the number fluctuations on the two orbitals are uncorrelated in a product state. Since the total number of electrons is fixed, 
\begin{align}
0 & =\mathrm{Var}_{\psi}(N_A+N_B) =\mathrm{Var}_{\psi_A}(N_A) +\mathrm{Var}_{\psi_B}(N_B)\,.
\end{align}
Clearly, both terms are nonnegative and the possible occupation pairs are $(2,0)$, $(0,2)$, and $(1,1)$.
We can now group the product components by their local occupations. Every admissible separable state takes the form
\begin{align}
\sigma &=w_{20} \ket{\uparrow\downarrow, 0}\!\bra{\uparrow\downarrow, 0}  +w_{02} \ket{0, \uparrow\downarrow}\!\bra{0, \uparrow\downarrow} + w_{11}\sigma_{11}\,,
\label{eq:OS-sep}
\end{align}
where the weights $w_{ij}$ are nonnegative and sum to one. Here $\sigma_{11}$ is a normalized separable state of the two local spins in the sector with one electron on each orbital.
Conversely, every state of the form in Eq.~\eqref{eq:OS-sep} belongs to $\mathcal{D}_{\star, \mathrm{sep}}$. Its expectation of $Q = P_g-P_f$ is the corresponding weighted sum. The separable range of $\Tr[\rho (P_g-P_f)]$ is the convex hull of the values attained by the two ionic configurations and by separable states in the $(1,1)$ sector.

The restriction of $Q$ to the two singlet states $(C, I_+)$ is
\begin{equation}
Q \big|_{\mathrm{span}\{C, I_+\}} = \frac{1}{\Lambda} \begin{pmatrix} U & 4 \tau\\
4 \tau &  -U \end{pmatrix}\,,
\label{eq:OS-Q-block}
\end{equation}
and vanishes on $I_-$ and the three triplet states $\ket{T_m}$.
Moreover, the covalent singlet population for the product of local spin states with Bloch vectors $\mathbf{a}, \mathbf{b}$ is
\begin{align}
\bra{C} \rho_A\otimes\rho_B\ket{C} = \frac{1}{4}(1 -\mathbf{a}\cdot\mathbf{b}) \leq\frac{1}{2}\,.
\end{align}
By linearity, the same upper bound holds for every separable state $\sigma_{11}$ in this sector.
Together with the ionic contributions in Eq.~\eqref{eq:OS-sep}, we get for $U>0$,
\begin{align}
-\frac{U}{2\Lambda} \leq \Tr[\sigma(P_g - P_f)] \leq\frac{U}{2\Lambda}\,,\quad \sigma\in\mathcal{D}_{\star, \mathrm{sep}}\,.
\label{eq:OS-sep-interval}
\end{align}
Either ionic product, $\ket{\uparrow\downarrow, 0}$ or $\ket{0, \uparrow\downarrow}$, attains the lower bound. The upper bound is attained by both $\ket{\uparrow, \downarrow}$ and $ \ket{\downarrow, \uparrow}$.  Convexity of the set of separable states then proves the exact separable image in Eq.~\eqref{eq:OS-ranges}.

\section{Two-photon response and vibrational entanglement for a molecular dimer \label{app:vibronic-complete}}

In this section, we derive in detail the full and separable joint numerical ranges for a molecular dimer whose fragments have local vibronic structure as discussed in Sec.~\ref{subsec:vibronic-dimer}.
For this we first calculate the explicit two-photon transition tensor, the effective observables, and then their separable bounds. The goal is to construct a witness for vibrational entanglement across the physical fragment partition $A|B$ of the dimer in the electronic ground state manifold.

\subsection{Model and active manifold}

The dimer consists of two fragments $A$ and $B$. We first specify their local vibronic structure, which determines the transition energies and vibrational overlaps needed for the two-photon response.
Each fragment $Z \in \{A, B\}$ has an electronic ground state $\ket{g_Z}$, an excited state $\ket{e_Z}$, and one intramolecular vibrational mode. The two electronic states have harmonic potentials with displaced minima and the same vibrational energy spacing $\varepsilon_{v, Z} > 0$.
This is a standard minimal model of local vibronic structure \cite{PKP12}.

Let $\nu = 0, 1, 2, \ldots$ and $m = 0, 1, 2, \ldots$ denote the vibrational quantum numbers on the ground and excited electronic potentials, respectively.
The energy $\varepsilon_{00,Z}$ is the local $0$--$0$ transition energy, defined as the gap between the lowest vibrational levels of the two electronic states.
After subtraction of their common zero-point energy, the local eigenenergies read
\begin{equation}\label{eq:vd-local-energies}
E_Z(g, \nu) = \nu\varepsilon_{v,Z}\,,\quad  E_Z(e, m) = \varepsilon_{00, Z} + m\varepsilon_{v, Z}\,.
\end{equation}
To display the coupling between electronic and vibrational degrees of freedom explicitly, we express both electronic sectors in the oscillator basis centred at the minimum of the ground electronic potential.
In this basis, $b_Z$ and $b_Z^\dagger$ annihilate and create one vibrational quantum. The displacement between the potential minima is described by the Huang--Rhys factor $S_Z\geq 0$ with associated local
reorganization energy $S_Z \varepsilon_{v, Z}$.
We also define $P_{e,Z} = \ket{e_Z}\!\bra{e_Z}$, which projects onto the excited electronic state and acts as the identity on the vibrational space.
Then, the local Hamiltonian is given by
\begin{align}\label{eq:vd-H-local}
H_Z &=\varepsilon_{v,Z}b_Z^\dagger b_Z + (\varepsilon_{00,Z} + S_Z \varepsilon_{v,Z}) P_{e, Z} \nonumber\\
&\quad +\varepsilon_{v,Z} \sqrt{S_Z}\, P_{e,Z}(b_Z + b_Z^\dagger)\,.
\end{align}
Here the energy offset $S_Z\varepsilon_{v,Z}$ ensures that $\varepsilon_{00,Z}$ remains the $0$--$0$ transition energy in Eq.~\eqref{eq:vd-local-energies} and the positive sign of last fixes the convention for the direction of the displacement.
For the optical response of the system, it is convenient to use the vibrational eigenbasis of each electronic potential. Denote these vibrational states by $\ket{\nu; g_Z}$ and $\ket{m; e_Z}$.
Their Franck--Condon overlaps that will be used later are
\begin{equation}\label{eq:FC-overlaps}
f_{m\nu}^{(Z)}  := \langle m; e_Z|\nu; g_Z\rangle_{\mathrm{vib}}\,.
\end{equation}
The subscript emphasizes that this inner product involves only the vibrational factors.

Next, we introduce the admissible input space, which fixes the states over which the separable bounds are evaluated. The input is contained in the electronic ground manifold, with local vibrational support restricted to the two lowest levels of each fragment.
Using $\ket{g_Z,i} := \ket{g_Z}\otimes\ket{i;g_Z}$, we define
\begin{align}\label{eq:vd-input}
\mathcal{H}_{G,Z} &= \mathrm{span}\{\ket{g_Z, 0}, \ket{g_Z, 1}\}\,,\nonumber\\
\mathcal{H}_G  \equiv \mathcal{H}_\star &= \mathcal{H}_{G, A}\otimes \mathcal{H}_{G, B} = \mathrm{span}\{\ket{00}, \ket{01}, \ket{10}, \ket{11}\}\,,
\end{align}
where
\begin{equation}
\ket{ij} := \ket{g_A, i}\otimes\ket{g_B, j}\,,\quad i, j = 0, 1, 2, \ldots\,.
\end{equation}
Entanglement is tested across the physical fragment partition $A|B$. The restriction to the $\mathcal{H}_G$ is an assumption about the input preparation.
For equal vibrational spacings, the states $\ket{20}$ and $\ket{02}$ have the same total energy as $\ket{11}$, which implies that this space is not selected by means of an energy cutoff. 
The vibrational ladders used for the probe dynamics remain complete.
Within the chosen input space, we admit every density operator $\rho_G\in\mathcal{D}(\mathcal {H}_G)$.  This includes states with coherences between different vibrational energies.

Using the local Hamiltonians $H_Z$ from Eq.~\eqref{eq:vd-H-local} the system Hamiltonian without electric field is given by
\begin{equation}
H_0 = H_A\otimes\mathbbm{1}_B + \mathbbm{1}_A \otimes H_B + U P_{e, A}\otimes P_{e, B}, \label{eq:vd-H}
\end{equation}
with projectors
\begin{equation}\label{eq:vd-PeZ}
P_{e,Z}  = \sum_{m = 0}^{\infty}  \ket{e_Z, m}\!\bra{e_Z, m}.
\end{equation}
The eigenenergies of $H_0$ in Eq.~\eqref{eq:vd-H} are required to calculate the two-photon transition tensor in Eq.~\eqref{eq:Tfi_ab}. 
Here $\mathbbm{1}_Z$ is the identity on the full local electronic-vibrational space, and $\ket{e_Z, m} := \ket{e_Z} \otimes \ket{m; e_Z}$ in analogy to $\ket{g_Z,i}$. The summation in Eq.~\eqref{eq:vd-PeZ} includes the complete excited state vibrational space represents the same electronic projector used in Eq.~\eqref{eq:vd-H-local}. The last term in Eq.~\eqref{eq:vd-H} describes exciton--exciton interaction. More specifically, the parameter $U\in \mathbbm{R}$ leads to an interaction energy that shifts every doubly excited level by the same amount independent of the vibrational quantum numbers. The real parameter $U$ should not be confused with a unitary operator. On $\mathcal{H}_G$, the interaction proportional to $U$ vanishes, and $H_0$ reduces to a sum of local vibrational Hamiltonians.
In the minimal model considered here, we do not include a coupling $J$ between the excitonic eigenstates. 
However, the same steps as outlines below can be followed to calculate the full and separable reachable sets for nonzero $J$.  

In this model, we assume the readout determines the population of one doubly excited vibronic level. Let $\lambda_A,\lambda_B\in \{0, 1, 2, \ldots\}$ be the selected vibrational quantum numbers on the excited electronic potentials. The detected state and the corresponding orthogonal projector are given by
\begin{align}\label{eq:vd-detector}
\ket{F} &:=\ket{e_A, \lambda_A} \otimes\ket{e_B, \lambda_B}\,,\quad D_\Omega  =\ket{F}\!\bra{F}\,,
\end{align}
The state $\ket{F}$ is an eigenstate of the Hamiltonian $H_0$ in Eq.~\eqref{eq:vd-H}, with $H_0\ket{F}=E_F\ket{F}$ with eigenenergy
\begin{equation}
E_F = \varepsilon_{00, A} + \lambda_A \varepsilon_{v, A}  +\varepsilon_{00, B} + \lambda_B \varepsilon_{v, B} + U\,.
\end{equation}

To construct the optical response, we now specify the local electronic transition dipole operators.
We use the Condon approximation, which leads to an electronic transition dipole that is independent of the vibrational coordinate.
The transition dipole is assumed to be real, with magnitude $\mu_Z$ and a fixed unit direction $\mathbf{u}_Z$ in the molecular frame. The vibrational eigenfunctions are chosen to be real as well.
Then, 
\begin{align}\label{eq:vd-local-dipole}
\mu_{Z, \alpha}^{(+)}  &= \mu_Z u_{Z, \alpha} \sum_{m, n\geq 0} f_{mn}^{(Z)}\ket{e_Z ,m}\!\bra{g_Z, n}\,, \nonumber\\
\mu_{Z,\alpha} &=\mu_{Z, \alpha}^{(+)}  + \left(\mu_{Z, \alpha}^{(+)}\right)^\dagger\,,  \quad \alpha\in\{x,y,z\}\,.
\end{align}
Here, $\alpha$ labels a Cartesian component in the molecular frame and $\mu_{Z,\alpha}$ denotes the electronic transition-dipole
operator.
The transition dipole of the dimer in this model is
\begin{align}\label{eq:vd-total-dipole}
\mu_\alpha =  \mu_{A, \alpha} \otimes\mathbbm{1}_B + \mathbbm{1}_A \otimes \mu_{B , \alpha}\,.
\end{align}
The leading two-photon amplitude from $\mathcal{H}_G$ to $\ket{F}$ requires one electronic excitation on each fragment. This implies that the electronic transition dipole terms suffice at this order.

\subsection{Two-photon transition tensor}

Next, the sum-over-states transition tensor in Eq.~\eqref{eq:Tfi_ab} will be applied to the dimer Hamiltonian $H_0$ in Eq.~\eqref{eq:vd-H} and the transition dipole in Eq.~\eqref{eq:vd-total-dipole}.
We set $\gamma_n\to 0^+$ in Eq.~\eqref{eq:Tfi_ab}. This limit should be understood under the finite pulse spectral integrals introduced in the next subsection.
Let $\omega$ and $\omega'$ denote angular frequency components of the two fields 1 and 2, respectively.
The corresponding tensor element $T_{\alpha\beta}^{F,ij}(\omega,\omega')$ describes a transition
from $\ket{ij}$ to $\ket{F}$ as explained in Appendix~\ref{app:Ax-Bx}. 
For each input level, the energies and intermediate transition frequencies are
\begin{align}
E_{ij}  &= i\varepsilon_{v, A} + j\varepsilon_{v, B}\,,\quad  \omega_{n, ij} =  \frac{E_n  - E_{ij}}{\hbar}\,,
\end{align}
where $E_n$ is the eigenenergy of the intermediate state $\ket{n}$ under $H_0$ in Eq.~\eqref{eq:vd-H}.
Each of the two dipole interactions must create one electronic excitation.
For a fixed input $\ket{ij}$ and the selected final state $\ket{F}$, there are only two states in the singly excited vibronic manifold $\mathcal H_E$ that can contribute, namely
\begin{align}\label{eq:vd-nA-nB}
\ket{n_A} &=\ket{e_A, \lambda_A; g_B, j}\,,\quad  \ket{n_B} =   \ket{g_A, i; e_B, \lambda_B}\,.
\end{align}
The second interaction acts on the other fragment and leaves the vibrational state of the already excited fragment unchanged. Thus, that vibrational state must coincide with its value in $\ket{F}$.

For a local input basis state $\ket{g_Z, \nu}$, with vibrational quantum number $\nu \in\{0, 1\}$, define
\begin{align}
a_\nu^Z  &:= E_Z(e, \lambda_Z) - E_Z(g, \nu)  \nonumber\\
&= \varepsilon_{00, Z} +  (\lambda_Z - \nu)\varepsilon_{v, Z}\,, \label{eq:vd-anuZ}\\
\Delta^Z_{\lambda_Z, \nu}(\omega)  &:= \frac{a_\nu^Z}{\hbar} - \omega - i0^+\,, \quad Z \in \{A, B\}\,.\label{eq:Delta_X-dimer}
\end{align}
Thus, $E_{n_A}-E_{ij}=a_i^A$ and
$E_{n_B}-E_{ij}=a_j^B$.
Substitution of the local dipole matrix elements gives
\begin{widetext}
\begin{align}\label{eq:Tab-model}
T_{\alpha\beta}^{F,ij}(\omega, \omega') = \frac{\mu_A \mu_B}{\hbar} f_{\lambda_A i}^{(A)}f_{\lambda_B j}^{(B)} \left[u_{B, \alpha} u_{A, \beta}\left(\frac{1}{\Delta^A_{\lambda_A, i}(\omega)} + \frac{1}{\Delta^B_{\lambda_B, j}(\omega')} \right)  + u_{A,\alpha}u_{B,\beta}\left(\frac{1}{\Delta^A_{\lambda_A, i}(\omega')} + \frac{1}{\Delta^B_{\lambda_B, j}(\omega)} \right)\right]\,.
\end{align}
\end{widetext}
The coefficient of $u_{B,\alpha}u_{A,\beta}$ describes excitation of $A$ by field 1 and of $B$ by field 2.
The coefficient of $u_{A,\alpha}u_{B,\beta}$ describes the opposite assignment. 

The interaction term $U P_{e,A}\otimes P_{e,B}$ acts only when both fragments are electronically excited  such that the energy required to reach $\ket{F}$ from an input level $\ket{ij}$ is
\begin{align}\label{eq:vd-final-gap}
E_F - E_{ij} = a_i^A + a_j^B + U\,.
\end{align}
with $a_\nu^Z$ given by Eq.~\eqref{eq:vd-anuZ}. 
For absorption of one photon from each pulse, we have $\hbar(\omega+\omega') = E_F - E_{ij}$. The pulse spectra determine the weights of these frequency pairs in the transition amplitude. Thus, which parts of the two spectra jointly drive the transition is decided by the final energy gap. A change in $U$ shifts this condition and can alter the excitation amplitude for fixed pulses.

\subsection{Pulse amplitudes and transition operator}

The next step is to combine the transition tensor in Eq.~\eqref{eq:Tab-model} with finite pulses to obtain
the second-order amplitude for one absorption from each field. These are standard calculations, which are included here to keep the derivation of the reachable sets self-contained, and we refer to Refs.~\cite{Mukamel1995-book, DDGFS01} as references. 
We then express this amplitude as a transition operator on $\mathcal{H}_G$, which eventually leads to the effective observable used for the witness construction.

We work in the interaction picture with respect to $H_0$ and choose the common pulse center as $t = 0$.
The two reference pulses have Gaussian envelopes,
\begin{align}
\mathcal E_k(t)  &= \mathcal{E}_{0k}  e^{-t^2/(2\tau^2)}e^{-i \varepsilon_k t/\hbar}\,, \quad \varepsilon_k = \hbar\omega_k\,, \nonumber\\
\mathbf{E}_k^{(+)}(t) &= \mathbf{e}_k \mathcal{E}_k(t)e^{i\phi_k}\,, \quad k = 1, 2\,. \label{eq:vd-pulses}
\end{align}
Here $k$ labels the pulses, $\phi_k$ is the carrier phase, and the unit vector $\mathbf{e}_k$ specifies the linear polarization.
In addition, $\mathcal{E}_{0k}\geq 0$ and the common width of the two pulses is characterized by $\tau > 0$. We assume $\omega_k\tau \gg 1$ such that the weight of the Gaussian spectral amplitude at negative frequencies is negligible.
Relaxation and dephasing during excitation and up to the selected population readout are neglected.
To connect the molecular transition tensor in Eq.~\eqref{eq:Tab-model} to these pulses below, we use the Fourier convention
\begin{align}
\widetilde{\mathcal{E}}_k(\omega) &= \int_{-\infty}^{\infty}\mathrm{d}t\, \mathcal{E}_k(t) e^{i\omega t}  =\sqrt{2\pi} \tau\mathcal E_{0k} e^{-\tau^2(\omega - \omega_k)^2/2}\,, \nonumber\\
\mathcal{E}_k(t) &= \int_{-\infty}^{\infty}\frac{d\omega}{2\pi}\, \widetilde{\mathcal{E}}_k(\omega)e^{-i\omega t}\,. \label{eq:vd-pulse-spectrum}
\end{align}
In order to calculate the transition amplitude for the specified pulses, we integrate the molecular tensor over their frequency components.
In second order perturbation theory each contribution is weighted by the spectral amplitudes of the two fields and sums them coherently \cite{DDGFS01}. The contributing frequency pairs satisfy $\omega + \omega'= (E_F - E_{ij})/\hbar$.
We denote the resulting dimensionless Cartesian amplitude by $\mathcal{T}_{\alpha\beta}^{F, ij}$. For one absorption from each field, with the common phase $e^{i(\phi_1 + \phi_2)}$ factored out, this amplitude is given by
\begin{align}
\mathcal{T}_{\alpha\beta}^{F, ij} &= \frac{i}{2\pi\hbar} \int_{-\infty}^{\infty}\mathrm{d}\omega\,\widetilde{\mathcal{E}}_1(\omega)\widetilde{\mathcal{E}}_2(\omega_{F, ij} - \omega) \nonumber\\
&\quad \times T_{\alpha \beta}^{F, ij}(\omega, \omega_{F, ij} - \omega)\,, \label{eq:vd-spectral-convolution}\\
\omega_{F,ij} &:= \frac{E_F - E_{ij}}{\hbar}\,.
\end{align}
Contraction with the polarization vectors at a specified molecular orientation yields the transition amplitude from $\ket{ij}$ to $\ket{F}$. In the time domain, Eq.~\eqref{eq:vd-spectral-convolution} takes the form
\begin{align}\label{eq:vd-finite-SOS}
\mathcal{T} _{\alpha\beta}^{F, ij} &=  - \frac{\mathcal{E}_{01}\mathcal{E}_{02}}{\hbar^2} \sum_n [\bra{F}\mu_\alpha\ket{n}\bra{n}\mu_\beta\ket{ij} \nonumber\\
&\quad\times \mathcal{I}_\tau\left(E_n - E_{ij} - \varepsilon_1, E_F - E_n - \varepsilon_2 \right)
\nonumber\\
&\quad + \bra{F}\mu_\beta \ket{n} \bra{n}\mu_\alpha\ket{ij} \nonumber\\
&\quad\times \mathcal{I}_\tau\left(E_n - E_{ij} - \varepsilon_2, E_F - E_n  - \varepsilon_1\right)]\,.
\end{align}
For each of the two absorption paths (either field 1 or field 2 first), 
\begin{equation}\label{eq:vd-I_tau}
\mathcal{I}_\tau(d_1, d_2) :=\int_{-\infty}^{\infty}\mathrm{d} t_2 \int_{-\infty}^{t_2} \mathrm{d}t_1\, e^{-(t_1^2 + t_2^2)/(2\tau^2)} e^{i(d_1 t_1 + d_2 t_2)/\hbar}\,.
\end{equation}
contains the information about the Gaussian pulse envelopes and the phase factors at the two times of the pulses interacting with the dimer. The arguments $d_1$ and $d_2$ of $\mathcal{I}_\tau$ are the energy detunings of the absorption steps relative to the respective carrier photon energies. 
Abbreviate $T = (t_1+t_2)/2$ and $u = t_2 - t_1\geq 0$. Equation~\eqref{eq:vd-I_tau} splits into a Gaussian integral over $T$ and a half-line integral over $u$, such that
\begin{equation}
\mathcal{I}_\tau(d_1, d_2) = \pi\tau^2 e^{-\tau^2(d_1 + d_2)^2/(4\hbar^2)}  w \left(\frac{\tau(d_2 - d_1)}{2\hbar}\right)\,,
\end{equation}
where $w(z) = e^{-z^2}\mathrm{erfc}(-iz)$. 
For the molecular dimer, Eq.~\eqref{eq:vd-finite-SOS} reduces to $\ket{n_A}$ and $\ket{n_B}$ from Eq.~\eqref{eq:vd-nA-nB}. Thus, 
\begin{equation}\label{eq:vd-T-complete}
\mathcal{T}_{\alpha\beta}^{F, ij}  = u_{B, \alpha} u_{A, \beta} x_{ij} + u_{A, \alpha} u_{B, \beta}  y_{ij}\,,
\end{equation}
with
\begin{align}
x_{ij} &= - \frac{\mu_A \mu_B \mathcal{E}_{01}\mathcal{E}_{02}}{\hbar^2} f_{\lambda_A i}^{(A)} f_{\lambda_B j}^{(B)} \nonumber\\
& \quad \times\Big[ \mathcal{I}_\tau\left(a_i^A - \varepsilon_1,  a_j^B + U - \varepsilon_2\right) \nonumber\\
&\quad + \mathcal{I}_\tau \left(a_j^B - \varepsilon_2, a_i^A + U - \varepsilon_1\right) \Big]\,, \label{eq:vd-x}\\
y_{ij} &= - \frac{\mu_A \mu_B \mathcal{E}_{01} \mathcal{E}_{02}}{\hbar^2} f_{\lambda_A i}^{(A)}f_{\lambda_B j}^{(B)} \nonumber\\
&\quad \times\Big[\mathcal{I}_\tau\left(a_i^A - \varepsilon_2, a_j^B + U - \varepsilon_1\right)
\nonumber\\
&\quad +  \mathcal{I}_\tau\left(a_j^B -\varepsilon_1, a_i^A + U - \varepsilon_2\right)  \Big]\,.\label{eq:vd-y}
\end{align}
Thus, $x_{ij}$ describes the case where field 1 excites fragment $A$ and field 2 excites fragment $B$, and $y_{ij}$ describes the opposite setting. 
The exciton--exciton interaction $U$ is included through the second absorption step, which reaches the doubly excited manifold.
In particular, for any input state $\ket{ij}$ 
\begin{align}
d_1 + d_2  &= E_F - E_{ij} - \varepsilon_1 -  \varepsilon_2 = a_i^A + a_j^B + U - \varepsilon_1 - \varepsilon_2\,.
\end{align}
Thus, each $\mathcal{I}_\tau$ contains the common factor $\exp[-\frac{\tau^2}{4\hbar^2}(E_F - E_{ij} - \varepsilon_1 - \varepsilon_2)^2]$, which illustrates the dependence of $\mathcal{T}_{\alpha\beta}^{F, ij}$ on the energy gap between initial and final states. 

To express $\mathcal{T}_{\alpha\beta}^{F, ij}$ in Eq.~\eqref{eq:vd-T-complete} as an operator on the input space, we introduce the (unnormalized) channel vectors
\begin{equation}\label{eq:vd-channels}
\ket{\chi_1} = \sum_{i,j = 0}^1 x_{ij}^* \ket{ij}\,, \quad \ket{\chi_2} = \sum_{i,j = 0}^1y_{ij}^* \ket{ij}\,.
\end{equation}
The complex conjugation ensures $\langle\chi_1| ij\rangle = x_{ij}$ and $\langle\chi_2 | ij\rangle = y_{ij}$. Then, for a pure input state, its overlap with a channel vector gives the associated amplitude. 
This allows us to introduce the operator-valued analogue of Eq.~\eqref{eq:vd-T-complete} as
\begin{equation}\label{eq:vd-tensor-op}
\mathcal{T}_{\alpha\beta}^{(\Omega)} = u_{B,\alpha} u_{A,\beta}  \ket{F}\!\bra{\chi_1} + u_{A,\alpha}u_{B,\beta}\ket{F}\!\bra{\chi_2}\,. 
\end{equation}
As a consistency check, $\bra{F}\mathcal{T}_{\alpha\beta}^{(\Omega)} \ket{ij} = \mathcal{T}_{\alpha\beta}^{F, ij}$.
The channel vectors in Eq.~\eqref{eq:vd-channels} are determined by the dimer model as well as the pulse parameters. Hence, they apply to all admissible input states $\rho_G\in \mathcal{D}_\star$.

Finally, let $R\in SO(3)$ be the rotation that transforms a vector from the molecular to the laboratory frame. For laboratory frame Cartesian indices $a,b$ and a polarization angle $\mathbf{e}_1\cdot\mathbf{e}_2 = \cos\theta$, the transition operator follows as
\begin{align}\label{eq:vd-M12}
M_{12}^{(R)}  &= \sum_{a, b, \alpha, \beta} e_{2,a}e_{1,b} R_{a\alpha} R_{b\beta} \mathcal{T}_{\alpha\beta}^{(\Omega)} \nonumber\\
&= \ket{F}\left(G_1^{(R)}\bra{\chi_1} + G_2^{(R)}\bra{\chi_2} \right)\,, 
\end{align}
where
\begin{align}\label{eq:vd-G12}
G_1^{(R)}  &= (\mathbf{e}_2\cdot R\mathbf{u}_B)(\mathbf{e}_1 \cdot R \mathbf{u}_A)\,, \nonumber\\
G_2^{(R)} &= (\mathbf{e}_2\cdot R \mathbf{u}_A)(\mathbf{e}_1 \cdot R\mathbf{u}_B)\,.
\end{align}
contain the orientation and polarization dependence of the transition operator.

\subsection{Effective observable}

Here we derive the effective observable $E_\Omega(\theta)$ used to construct the full and separable reachable sets using the dependence of the measured signal on the angle $\theta$ between the two pulses. For this, we first explain how $M_{12}^{(R)}$ derived in Eq.~\eqref{eq:vd-M12} determines the signal. 
More specifically, we construct the effective observable from excitation of $\ket{F}$ by one photon from each pulse. The measured population also contains contributions from two absorptions from a single pulse.
Phase averaging and subtraction of these contributions from a single pulse allow us to separate the contribution used for the reachable sets.

For the two-photon absorption the Tleading contribution to the population of $\ket{F}$ after the probe pulses is fourth order in the amplitudes of the electric fields. 
The complete second-order transition amplitude follows from the time-ordered expansion with the total probe field $\mathbf{E}=\mathbf{E}_1 + \mathbf{E}_2$. Each absorption from pulse $j=1,2$ adds a phase factor $e^{i\phi_j}$, such that to second order
\begin{align}\label{eq:vd-M-full}
M^{(2,R)}(\phi_1, \phi_2) &= e^{2i\phi_1} M_{11}^{(R)} + e^{i(\phi_1+\phi_2) }M_{12}^{(R)} \nonumber\\
& \quad + e^{2i\phi_2 }M_{22}^{(R)}\,.
\end{align}
Here $M_{jj}^{(R)}$ describes the process with two absorptions from a single pulse $j$.
The two mixed terms, with pulse 1 or pulse 2 acting first, yield $M_{12}^{(R)}$ in Eq.~\eqref{eq:vd-M12}.

Moreover, let $Y(\theta, \delta)$ denote the isotropically averaged population of $\ket{F}$, where $\delta = \phi_1 - \phi_2$ refers to the relative carrier phase between the two pulses.
All three amplitudes in Eq.~\eqref{eq:vd-M-full} reach the same final level $\ket{F}$. The cross terms of the three contributions in Eq.~\eqref{eq:vd-M-full} describe interference between the excitation pathways.
For the isotropic ensemble of molecules, we get
\begin{align}\label{eq:vd-Y-theta-phi}
Y(\theta, \delta) &= \int \mathrm{d}R\,\Tr\left[\rho_G M^{(2,R)\dagger} D_\Omega M^{(2,R)}\right]\,,
\end{align}
where $dR$ is the normalized Haar measure on $SO(3)$ with $\int_{SO(3)}dR=1$, and $ D_\Omega = \ket{F}\!\bra{F}$. 
The product of the operators inside the trace in Eq.~\eqref{eq:vd-Y-theta-phi} contains in particular interferences between the three amplitudes in Eq.~\eqref{eq:vd-M-full}. 
The interference term involving $M_{12}^{(R)}$ and either $M_{11}^{(R)}$ or $M_{22}^{(R)}$ comes with a factor of $e^{\pm i\delta}$. The other cross terms between $M_{11}^{(R)}$ and $M_{22}^{(R)}$ carry factors of $e^{\pm2i\delta}$.
It is possible to remove these interference terms by assuming that the relative probe phase can be varied independent of the input preparation and all other pulse parameters. 
One option is to measure the population of $\ket{F}$ for repeated preparations of the same state $\rho_G$ and different phase differences $\delta_\ell = \ell\pi/2$, with $\ell=0,1,2,3$, and calculate the average of the results with equal weights. For those choices of $\delta_\ell$ the interference terms vanish. The phase independent contributions remain, and the four measurements reproduce a uniform average over the relative phase.
Thus, we are left with the individual contributions from $M_{11}^{(R)}$, $M_{12}^{(R)}$, and $M_{22}^{(R)}$ inside the trace in Eq.~\eqref{eq:vd-Y-theta-phi}.

For $i=1,2$, let
\begin{equation}
Y_{ii} = \int \mathrm{d}R\, \Tr\left[\rho_G M_{ii}^{(R)\dagger}D_\Omega M_{ii}^{(R)} \right]
\end{equation}
denote the contributions to $Y(\theta,\delta)$ in Eq.~\eqref{eq:vd-Y-theta-phi} from a single pulse.
For the isotropic ensemble, $Y_{11}$ and $Y_{22}$ add a $\theta$-independent background to the signal after the average over $\delta_\ell$ has been performed, as described above. 
We therefore subtract $Y_{11}, Y_{22}$ to isolate the contribution with one absorption from each pulse, which results in 
\begin{align}
S_\Omega(\theta) &:= \frac{1}{4}\sum_{\ell = 0}^{3} Y(\theta, \ell\pi/2) - Y_{11} - Y_{22}\,.
\label{eq:vd-S-theta}
\end{align}
The extracted signal can be represented as an expectation value of an effective observable $E_\Omega(\theta)$ as (c.f., Eq.~\eqref{eq:S-TPA})
\begin{align}
S_\Omega(\theta) &= \Tr\left[\rho_G E_\Omega(\theta)\right]\,, \nonumber\\
E_\Omega(\theta) &= \int \mathrm{d}R\, M_{12}^{(R)\dagger} D_\Omega M_{12}^{(R)}\,. \label{eq:vd-Etheta}
\end{align}
The operator $E_\Omega(\theta)$ fixes the spectroscopic map and thus allows us to determine the full and separable reachable sets as illustrated in Fig.~\ref{fig:overview} (recall also Sec.~\ref{subsec:spectroscopic-map}). 

\subsection{Parallel local dipoles \label{app:parallel-vd}}

Before we discuss the more general case of non-parallel local dipoles, we show how $E_\Omega(\theta) $ in Eq.~\eqref{eq:vd-Etheta} leads to line segments as reachable sets for parallel dipoles. 
For this, set
\begin{align}
\mu_A &= \mu_B = \mu\,,\,\,\,\,\varepsilon_{00,A} = \varepsilon_{00, B} = \varepsilon_{00}\,,\,\,\,\,\lambda_A = \lambda_B = \lambda\,, \nonumber\\
\varepsilon_{v, A} &= \varepsilon_{v, B} = \varepsilon_v\,, \,\,\,\, f_{mn}^{(A)} = f_{mn}^{(B)} = f_{mn}\,,\,.
\label{eq:vd-identical}
\end{align}
and consider parallel transition dipoles with $\mathbf{u}_A = \mathbf{u}_B = \mathbf{u}$. 
Then, $\mathcal T_{\alpha\beta}^{F,ij}$ in Eq.~\eqref{eq:vd-T-complete} reduces to 
\begin{equation}
\mathcal{T}_{\alpha \beta}^{F, ij} = u_\alpha u_\beta c_{ij}\,,\quad c_{ij} = x_{ij} + y_{ij}\,.
\end{equation}
Moreover, we define
\begin{equation}
\ket{\chi_\parallel}  = \sum_{i,j=0, 1} c_{ij}^*\ket{ij} = \ket{\chi_1} + \ket{\chi_2}\,,
\end{equation}
which leads to 
\begin{equation}
M_{12}^{(R)}  = (\mathbf{e}_2\cdot R\mathbf{u}) (\mathbf{e}_1\cdot R\mathbf{u}) \ket{F}\!\bra{\chi_\parallel}\,.
\end{equation}
Inserting $M_{12}^{(R)}$ into Eq.~\eqref{eq:vd-Etheta} and evaluating the rotational average leads to 
\begin{align}
E_\Omega(\theta) = \frac{1 + 2\cos^2\theta}{15} \ket{\chi_\parallel}\!\bra{\chi_\parallel}
\end{align}
such that the spectroscopic map $F_\Omega^{(2)}$ in Eq.~\eqref{eq:S-TPA} is defined by the two operators
\begin{equation}\label{eq:vd-X12}
X_\Omega^{(1)}=\frac1{15} \ket{\chi_\parallel}\!\bra{\chi_\parallel}\,,\quad  X_\Omega^{(2)} = 2X_\Omega^{(1)}.
\end{equation}
The same effective operators $X_\Omega^{(1)}, X_\Omega^{(2)}$ are obtained for antiparallel dipoles since the overall amplitude sign disappears from the population of the state $\ket{F}$.
Due to the form of $X_\Omega^{(1)}, X_\Omega^{(2)}$ in Eq.~\eqref{eq:vd-X12}, the corresponding signal coefficients $s_\Omega^{(j)} = \Tr[\rho_G X_\Omega^{(j)}]$ are functions of the overlap $p = \bra{\chi_\parallel}\rho_G\ket{\chi_\parallel}$.
Therefore, the full reachable set is given by
\begin{equation}\label{eq:vd-parallel-full}
\mathcal{R}^{(\Omega, 2)}  = \left\{(p/15, 2p/15):  0\leq p\leq\|\chi_\parallel\|^2\right\}\,.
\end{equation}
It remains to determine the separable reachable subset. Let $\Lambda_{\max}$ be the largest Schmidt coefficient of $\ket{\widetilde\chi_\parallel} = \ket{\chi_\parallel}/\|\chi_\parallel\|$.
The largest squared overlap with a product state is $\Lambda_{\max}^2$ \cite{WG03}, and due to convexity of the set of separable admissible states we obtain
\begin{equation}\label{eq:vd-parallel-sep}
\mathcal{R}_{\mathrm{sep}}^{(\Omega, 2)} = \left\{(p/15, 2p/15): 0\leq p\leq\|\chi_\parallel\|^2\Lambda_{\max}^2\right\}\,.
\end{equation}
Thus, strict inclusion of $\mathcal{R}_{\mathrm{sep}}^{(\Omega, 2)} $ in $\mathcal{R}^{(\Omega, 2)}$ holds whenever the normalized channel vector $\ket{\chi_\parallel}/\|\chi_\parallel\|$ is entangled.

This criterion in terms of the Schmidt coefficient is testable directly from the calculated transition amplitudes. In the present model, exchange symmetry yields $y = x^{\mathsf{T}}$ for $x, y$ in Eqs.~\eqref{eq:vd-x} and \eqref{eq:vd-y}. The matrix $C$ of coefficients $c_{ij}$ satisfies
\begin{equation}\label{eq:vd-C-matrix}
C = [c_{ij}] = x + x^{\mathsf{T}} = \begin{pmatrix}
2 x_{00} & x_{01} + x_{10} \\
x_{01} + x_{10} & 2 x_{11} \end{pmatrix}\,.
\end{equation}
The channel vector $\ket{\chi_\parallel}/\|\chi_\parallel\|$ has coefficient matrix $C^*/\|\chi_\parallel\|$ and its Schmidt coefficients are the singular values of $C$ divided by $\|\chi_\parallel\|$. Thus, $\Lambda_{\max} < 1$ if both singular values are nonzero (and $\ket{\chi_\parallel}\neq 0$). Therefore, strict inclusion, $\mathcal{R}_{\mathrm{sep}}^{(\Omega, 2)}\subsetneq \mathcal{R}^{(\Omega, 2)}$, holds whenever
\begin{align}\label{eq:vd-parallel-det}
\mathrm{det}(C) = 4x_{00} x_{11} - (x_{01} + x_{10})^2 \neq 0\,.
\end{align}
Equation~\eqref{eq:vd-parallel-det} illustrates under which conditions the protocol can and cannot witness entanglement. In fact, $\mathrm{det}(C)  \neq 0$ requires that both local input levels in $\ket{ij}$ couple
to the selected excited state vibrational level $\lambda$ and each coefficient $c_{ij}$ contains the Franck--Condon factor $f_{\lambda i}f_{\lambda j}$. Thus, if either $f_{\lambda0}$ or $f_{\lambda1}$ vanishes, the channel vector $\ket{\chi_\parallel}$ is a product vector or zero. As a result the full and separable images coincide in this setting.

\subsection{Nonparallel local dipoles \label{app:vibronic-dimer-2D}}

Here we show how nonparallel local dipoles can lead to two-dimensional images of the spectroscopic map.
To simplify the calculations, we again assume identical fragments $A$ and $B$, as specified in Eq.~\eqref{eq:vd-identical}. The cosine of the angle between their local dipoles, $\eta = \mathbf{u}_A\cdot\mathbf{u}_B$, is now allowed to differ from $\pm 1$. We derive the resulting spectroscopic map and determine the conditions under which its image is two-dimensional, as well as the conditions that allow for witnessing entanglement. 

\subsubsection{Rotational average and coefficient operators}

We first derive the effective observable and determine the signal coefficients used to construct the spectroscopic map as in Eq.~\eqref{eq:rot-invariants}. 
Let $\mathbf{a} = R\mathbf{u}_A$ and $\mathbf{b} = R\mathbf{u}_B$. The combination of rotational invariance and symmetry under interchange of the two Cartesian indices associated with each repeated vector leads to
\begin{align}\label{eq:vd-fourth-moment}
\langle a_i a_j b_k b_l\rangle_R  =\frac{2 - \eta^2}{15}\delta_{ij}\delta_{kl} + \frac{3\eta^2 - 1}{30}(\delta_{ik}\delta_{jl} + \delta_{il}\delta_{jk})\,.
\end{align}
Using Eq.~\eqref{eq:vd-G12} and $\eta = \mathbf{u}_A\cdot\mathbf{u}_B$, we arrive at the rotational factors (see also \cite{AT77})
\begin{align}\label{eq:vd-rotation-factors}
\left\langle \left(G_1^{(R)}\right)^2 \right\rangle_R &= \left\langle \left(G_2^{(R)}\right)^2\right \rangle_R = \alpha_\eta(\theta)\nonumber\\
& =\frac{2-\eta^2 + (3\eta^2 - 1)\cos^2\theta}{15},
\nonumber\\
\left\langle G_1^{(R)}G_2^{(R)}\right\rangle_R &=\beta_\eta(\theta) = \frac{3\eta^2 - 1 + (\eta^2 + 3)\cos^2 \theta}{30}\,. 
\end{align}
Recall the definition of the channel vectors in Eq.~\eqref{eq:vd-channels}. We then introduce Hermitian operators
\begin{align}\label{eq:vd-Sigma-Xi}
\Sigma_\Omega &= \ket{\chi_1}\!\bra{\chi_1} + \ket{\chi_2}\!\bra{\chi_2}\,, \nonumber\\
\Xi_\Omega  &=  \ket{\chi_1}\!\bra{\chi_2} + \ket{\chi_2}\!\bra{\chi_1}\,.
\end{align}
Then, the effective response operator follows from Eq.~\eqref{eq:vd-rotation-factors} as
\begin{align}\label{eq:vd-effect-Sigma}
E_\Omega(\theta)  &= \alpha_\eta(\theta)\Sigma_\Omega  + \beta_\eta(\theta)\Xi_\Omega  \nonumber\\
&= X_\Omega^{(1)} + X_\Omega^{(2)}\cos^2\theta\,. 
\end{align}
The two operators that define the corresponding spectroscopic map $F_\Omega^{(2)}$ are given by
\begin{align}
X_\Omega^{(1)} &= \frac{2 -  \eta^2}{15}\Sigma_\Omega  + \frac{3\eta^2 - 1}{30}  \Xi_\Omega\,, \nonumber\\
X_\Omega^{(2)} &= \frac{3\eta^2 - 1}{15}\Sigma_\Omega + \frac{\eta^2 + 3}{30}\Xi_\Omega\,.
\end{align}
As explained in Sec.~\ref{subsec:spectroscopic-map} and illustrated in Sec.~\ref{subsec:vibronic-dimer}, the signal coefficients $s_\Omega^{(j)} = [\rho_G X_\Omega^{(j)}]$ can be obtained from an angular fit by performing a scan of the measured signal over the angle $\theta$. 

As a consistency check, we observe that the same result is obtained from the three Cartesian contractions evaluated using the transition tensor $\mathcal{T}^{(\Omega)}$, that is 
\begin{align}\label{eq:vd-invariants}
\mathcal{I}_{F, \Omega} &:= \sum_{\alpha, \beta}  \mathcal{T}_{\alpha\alpha}^{(\Omega)\dagger} D_\Omega\mathcal{T}_{\beta \beta}^{(\Omega)} =  \eta^2(\Sigma_\Omega + \Xi_\Omega)\,, \nonumber\\
\mathcal{I}_{G, \Omega} &:= \sum_{\alpha, \beta} \mathcal{T}_{\alpha \beta}^{(\Omega)\dagger} D_\Omega \mathcal{T}_{\alpha \beta}^{(\Omega)} = \Sigma_\Omega + \eta^2 \Xi_\Omega\,,  \nonumber\\
\mathcal{I}_{H,\Omega} &:= \sum_{\alpha, \beta} \mathcal{T}_{\alpha\beta}^{(\Omega)\dagger} D_\Omega\mathcal T_{\beta\alpha}^{(\Omega)} = \eta^2\Sigma_\Omega + \Xi_\Omega\,.
\end{align}
Indeed, using Eq.~\eqref{eq:vd-invariants} we get
\begin{align}
X_\Omega^{(1)} &= \frac{-\mathcal{I}_{F,\Omega} + 4\mathcal{I}_{G, \Omega} - \mathcal{I}_{H, \Omega}}{30}\,, \nonumber\\
X_\Omega^{(2)} &= \frac{3\mathcal{I}_{F, \Omega} - 2\mathcal{I}_{G, \Omega} + 3 \mathcal{I}_{H,\Omega}}{30}\,.
\end{align}
In the case of collinear dipoles as in Appendix~\ref{app:parallel-vd}, all three contractions reduce to $\ket{\chi_\parallel}\!\bra{\chi_\parallel}$ with the correct prefactor, which recovers Eq.~\eqref{eq:vd-X12}. 

\subsubsection{Channel symmetry and reachable sets}

In the following, we discuss the conditions under which the two-photon absorption protocol with possibly nonparallel local dipoles is able to witness entanglement. The general line of argument is the same as in Appendix~\ref{app:parallel-vd}. 

The matrix representing the coefficient map from $(\Sigma_\Omega, \Xi_\Omega)$ to $(X_\Omega^{(1)}, X_\Omega^{(2)})$ has determinant
\begin{align}\label{eq:vd-coeff-det}
\mathrm{det}\begin{pmatrix} (2 - \eta^2)/15 & (3\eta^2 - 1)/30\\
(3\eta^2 - 1)/15 & (\eta^2 + 3)/30 \end{pmatrix}  = \frac{(1 -  \eta^2)(1 + 2\eta^2)}{90}\,.
\end{align}
The coefficient matrix is invertible for $|\eta| < 1$. What remains to determine are the conditions under which $\Sigma_\Omega$ and $\Xi_\Omega$ are independent on the active input space. This matters because independent $\Sigma_\Omega, \Xi_\Omega$ will lead to genuinely two-dimensional reachable sets, thus increasing the likelihood of witnessing entanglement. 

Let $P$ swap the two local vibrational factors. Identical monomers and a symmetric final level imply $y = x^{\mathsf{T}}$ and $\ket{\chi_2} = P\ket{\chi_1}$. Moreover, define
\begin{align}
\ket{\chi_s} &= \frac{\ket{\chi_1} + \ket{\chi_2}}{2}\nonumber\\
&= x_{00}^*\ket{00} + \frac{(x_{01} + x_{10})^*}{2}(\ket{01} + \ket{10}) + x_{11}^*\ket{11}\,,\label{eq:chi-s}\\
\ket{\chi_a} &= \frac{\ket{\chi_1} - \ket{\chi_2}}{2} =\frac{(x_{01} - x_{10})^*}{2}(\ket{01}- \ket{10})	\label{eq:chi-a}
\end{align}
with norms 
\begin{align}
N_s &= \|\chi_s\|^2 = |x_{00}|^2 + |x_{11}|^2 + \frac{|x_{01} + x_{10}|^2}{2}\,,\\
N_a &= \|\chi_a\|^2 = \frac{|x_{01}-x_{10}|^2}{2}\,.
\end{align}
By construction, the above two vectors satisfy $\langle\chi_s|\chi_a\rangle = 0$. 
Furthermore, $x_{01}\neq x_{10}$ implies that the antisymmetric channel $\ket{\chi_a}$ \eqref{eq:chi-a} is nonzero. 
Using $\ket{\chi_a}, \ket{\chi_s}$, the operators $\Sigma_\Omega, \Xi_\Omega$ in Eq.~\eqref{eq:vd-Sigma-Xi} simplify to
\begin{align}\label{eq:vd-Sigma-Xi-chi}
\Sigma_\Omega &= 2\left(\ket{\chi_s}\!\bra{\chi_s} + \ket{\chi_a}\!\bra{\chi_a}\right)\,, \nonumber\\
\Xi_\Omega &= 2\left(\ket{\chi_s}\!\bra{\chi_s}  - \ket{\chi_a}\!\bra{\chi_a}\right)\,.
\end{align}
Therefore, $\Sigma_\Omega$ and $\Xi_\Omega$ are linearly independent if both channel states $\ket{\chi_a}, \ket{\chi_s}$ are nonzero.
Since $[\Sigma_\Omega, \Xi_\Omega]=0$, which can be seen from the orthogonality $\langle \chi_s|\chi_a\rangle = 0$, also the operators $X_\Omega^{(1)}$ and $X_\Omega^{(2)}$ commute. Substitution of Eq.~\eqref{eq:vd-Sigma-Xi-chi} in Eq.~\eqref{eq:vd-effect-Sigma} yields
\begin{align}\label{eq:vd-effect}
E_\Omega(\theta) &= \frac{3 + \eta^2 + (1 + 7\eta^2)\cos^2\theta}{15}\ket{\chi_s}\!\bra{\chi_s}\nonumber\\
&\quad  + \frac{(1 - \eta^2)\sin^2\theta}{3}\ket{\chi_a}\!\bra{\chi_a}\,.
\end{align}

For $N_s, N_a > 0$, let $P_s = \ket{\chi_s}\!\bra{\chi_s}/N_s$ and $P_a = \ket{\chi_a}\!\bra{\chi_a}/N_a$ denote rank-one orthogonal projectors. For $|\eta| < 1$, the joint eigenspaces of  $(X_\Omega^{(1)}, X_\Omega^{(2)})$ are the symmetric and antisymmetric channel directions $\mathrm{span}\{\ket{\chi_s}\}$ and  $\mathrm{span}\{\ket{\chi_a}\}$, and their two-dimensional orthogonal complement $\mathrm{span}\{\ket{\chi_s},\ket{\chi_a}\}^{\perp}$ within $\mathcal{H}_G$. The three distinct joint eigenvalue pairs are
\begin{align}\label{eq:vd-joint-eigenvalues}
\bd\lambda_s &:= \frac{N_s}{15}(3 + \eta^2, 1 + 7\eta^2) =: (x_s, y_s)\,, \nonumber\\
\bd\lambda_a &:= \frac{N_a(1 - \eta^2)}3 (1, -1) =: (x_a, -x_a)\,, \nonumber\\
\bd\lambda_0  &:= (0, 0)\,.
\end{align}
Let $p_s =\Tr[\rho_GP_s], \,\, p_a = \Tr[\rho_G P_a]$ with $p_s,p_a\geq 0$ and $p_s + p_a\leq 1$.
Then, for any input density operator $\rho_G\in \mathcal{D}_\star$, 
\begin{equation}\label{eq:vd-linear-map}
\mathbf{s} := \left(s_\Omega^{(1)}, s_\Omega^{(2)}\right) = p_s\bd\lambda_s + p_a\bd\lambda_a\,. 
\end{equation}
All reachable signal vectors $\mathbf{s}\in \RR^2$ are attained by a mixture supported on the three joint eigenspaces. Therefore, the full reachable set is given by the convex hull of $\bd\lambda_a, \bd\lambda_s$ and $\bd\lambda_0$, that is
\begin{equation}\label{eq:vd-full-region}
\mathcal{R}^{(\Omega, 2)} = \mathrm{conv}\{\bd\lambda_0, \bd\lambda_s, \bd\lambda_a\}\,.
\end{equation}

In the following we explain under which conditions the full reachable set $\mathcal{R}^{(\Omega, 2)} $ is two-dimensional and when it reduces to a lower-dimensional set. The three vertices of $\mathcal{R}^{(\Omega, 2)}$ in Eq.~\eqref{eq:vd-full-region} determine its dimension, and the set is a two-dimensional triangle if $\bd\lambda_s$ and $\bd\lambda_c$ are linearly independent. We can see from the determinant
\begin{equation}
\mathrm{det}\begin{pmatrix} x_s & x_a\\ y_s & -x_a \end{pmatrix} = -\frac{4N_s N_a(1 - \eta^2)(1 + 2\eta^2)}{45}\,.
\end{equation}
that $\mathcal{R}^{(\Omega, 2)}$ is two-dimensional if the following three conditions apply:
\begin{equation}\label{eq:vd-2D-condition}
|\eta| < 1\,,\quad N_s  > 0\,,\quad N_a > 0\,. 
\end{equation}
If only one channel vertex $\bd\lambda_s$ or $\bd\lambda_s$ is nonzero, the image $\mathcal{R}^{(\Omega, 2)}$ a line segment, and the set reduces to $\{(0,0)\}$ for $\bd\lambda_s=\bd\lambda_c=\bd\lambda_0$.
As we show by a simple argument in the next subsection, the same conditions in Eq.~\eqref{eq:vd-2D-condition} ensure that the separable reachable set is two-dimensional.

\subsubsection{Separable reachable set \label{app:vd-separable-set}}

It remains to derive the image of the admissible separable states $\mathcal{D}_{\star, \mathrm{sep}}$ under the spectroscopic map $F_\Omega^{(2)}$ specified by the operators $X_\Omega^{(1)}, X_\Omega^{(2)}$. 
In the following, we first discuss the dimensionality of the separable reachable set, the result of which has been anticipated above. An analytic expression for the separable reachable set is derived afterwards under the assumption of identical fragments (recall Eq.~\eqref{eq:vd-identical}). This analytic solution further explains how the measurement channels determine which signals can be assigned to entangled states. 

The two-qubit PPT criterion \cite{Peres96,HHH96} implies that the maximally mixed state $\mathbbm{1}_4/4$ is an interior point of the set of separable states.
Under Eq.~\eqref{eq:vd-2D-condition}, the spectroscopic map has rank two. Therefore, the image of this neighbourhood has a nonempty interior in the space of retained signal coefficients. 
Since $\mathcal{R}_{\mathrm{sep}}^{(\Omega,2)} \subseteq\mathcal{R}^{(\Omega,2)}$, it follows immediately that Eq.~\eqref{eq:vd-2D-condition} provides necessary and sufficient conditions for both reachable sets to be two-dimensional.

For the model of identical fragments, we derive the exact expression for $\mathcal{R}_\mathrm{sep}^{(\Omega, 2)}$. As shown in Eq.~\eqref{eq:vd-linear-map}, the signal depends on both $p_s$ and $p_a$.
Therefore, we first determine the set of pairs $(p_s, p_a)$ that are compatible with admissible separable states $\rho_G\in \mathcal{D}_{\star, \mathrm{sep}}$. 
Equation~\eqref{eq:vd-linear-map} connects these pairs $(p_s, p_a)$ to the measured signal coefficients. 
We first express symmetric channel $\ket{\chi_s}$ in a common Schmidt basis for the two fragments.
The coefficient matrix $(C_s)_{ij} = \langle ij|\chi_s\rangle$ satisfies $C_s = C_s^{\mathsf{T}}$.
Moreover, the concurrence of the normalized state $\ket{\widetilde{\chi}_s} = \ket{\chi_s}/\sqrt{N_s}$ is given by
\begin{equation}\label{eq:vd-concurrence}
c =\frac{2\left\vert\mathrm{det}(C_s)\right\vert}{N_s} \in[0,1]\,.
\end{equation}
We denote a common local Schmidt basis for $Z=A,B$ by $\{\ket{\bar{0}}_Z,\ket{\bar{1}}_Z\}$. The states $\ket{\bar{0}}_Z,\ket{\bar{1}}_Z$ are orthonormal linear combinations of $\ket{g_Z, 0}$ and $\ket{g_Z, 1}$, with the same coefficients on both fragments. The bar distinguishes the Schmidt basis from the original vibrational energy basis.
In this Schmidt basis, the normalized symmetric channel reads
\begin{equation}
\ket{\widetilde{\chi}_s}  = \sqrt{r} \,\ket{\bar{0}}_A\otimes\ket{\bar{0}}_B + \sqrt{1-r} \, \ket{\bar{1}}_A\otimes\ket{\bar{1}}_B\,.
\end{equation}
Here $r= (1+\sqrt{1-c^2})/2\in[1/2,1]$ is the larger squared Schmidt coefficient.

Next, we bound $p_s$ at fixed $p_a$. To this end, consider a normalized product input state $\ket{w}_A\otimes\ket{v}_B$. In the common local Schmidt basis introduced above,
\begin{align}
\ket{w}_A &= w_0 \ket{\bar{0}}_A + w_1 \ket{\bar{1}}_A\,, \nonumber\\
\ket{v}_B &= v_0 \ket{\bar{0}}_B + v_1 \ket{\bar{1}}_B\,.
\end{align}
We denote their corresponding coefficient vectors by $\mathbf{w} = (w_0, w_1)^{\mathsf{T}}$ and $\mathbf{v}= (v_0, v_1)^{\mathsf{T}}$, with $\mathbf{w}^\dagger \mathbf{w} = \mathbf{v}^\dagger\mathbf{v} = 1$.
Since $P_a = (\mathbbm{1}_4 - P)/2$, where $P$ swaps the two local factors, the channel overlaps are
\begin{align}
p_a &= \frac{1 - |\mathbf{w}^\dagger \mathbf{v}|^2}{2} \in[0, 1/2]\,, \\
p_s &= \left\vert \sqrt{r}\, w_0 v_0  +\sqrt{1-r}\, w_1 v_1 \right\vert^2\,.
\end{align}
To simplify the notation, define $a = |w_0 v_0|$ and $b = |w_1 v_1|$.
Normalization of the local states and the Cauchy--Schwarz inequality imply $a + b\leq 1$.
In addition, the reverse triangle inequality leads to
\begin{equation}
|a - b| \leq  |\mathbf{w}^\dagger \mathbf{v}| = \sqrt{1 - 2p_a}\,.
\end{equation}
These two inequalities constrain the possible values of $p_s$. 
The  equalities $a+b=1$ and $|a-b| = \sqrt{1-2p_a}$ can be attained simultaneously. 
To see this, choose the product state $\ket{w}_A\otimes\ket{v}_B$ with
\begin{align}
\ket{w}_A &= \cos\vartheta \ket{\bar{0}}_A + i\sin\vartheta\ket{\bar{1}}_A\,,
\nonumber\\ 
\ket{v}_B &= \cos\vartheta\ket{\bar{0}}_B - i\sin\vartheta\ket{\bar{1}}_B\,,
\end{align}
where $0\leq \vartheta \leq \pi/4$. The overlaps of the state $\ket{w}_A\otimes\ket{v}_B$ with the two channels $\ket{\chi_s}$, $\ket{\chi_a}$ follows as
\begin{align}
p_a &= \frac{\sin^2(2 \vartheta)}{2}\,, \\
p_s &= \left(\sqrt r\,\cos^2\vartheta + \sqrt{1-r}\,\sin^2\vartheta \right)^2 =: f_c(p_a)\,.\label{eq:vd-fc}
\end{align}
For an arbitrary product input state, the inequalities from the Cauchy-Schwarz and inverse triangle inequality together with $r\geq 1/2$ give
\begin{align}
\sqrt{p_s} &\leq \sqrt{r} a + \sqrt{1-r} b \nonumber\\
&= \frac{\sqrt{r} + \sqrt{1 - r}}{2} (a + b) + \frac{\sqrt{r} - \sqrt{1 - r}}{2} (a-b) \nonumber\\
& \leq \frac{\sqrt{r} + \sqrt{1 - r}}{2} + \frac{\sqrt{r} - \sqrt{1 - r}}{2}\sqrt{1 - 2p_a}\,.
\end{align}
Taking the square of this expression leads to
\begin{align}
p_s \leq f_c(p_a) = \frac{1 - (1 - c)p_a + \sqrt{(1 - c^2)(1 - 2p_a)}}{2}\,.
\end{align}
The function $f_c(p_a)$ is concave on $[0,1/2]$. Since both channel overlaps are linear in the density operator, this bound is extended to all separable states by Jensen's inequality \cite{Jensen1906}. 
The product states constructed above lead to equality for every $p_a\in [0, 1/2]$.  As $\vartheta$ varies over the interval $0\leq \vartheta \leq \pi/4$, these product states attain all possible largest values of $p_s$ for fixed $p_a$.

The minimum $p_s = 0$ is attainable for every $p_a\in [0,1/2]$ through mixtures of two product states, as shown next. The state $\ket{\bar{0}}_A\otimes\ket{\bar{1}}_B$ has $(p_s, p_a) = (0, 1/2)$.
The other endpoint, $(p_s, p_a)=(0, 0)$, is attained by $\ket{w_{\mathrm{d}}}_A\otimes\ket{w_{\mathrm{d}}}_B$, where
\begin{equation}\label{eq:vd-dark-product}
\ket{w_{\mathrm{d}}}_Z  = \frac{(1-r)^{1/4}\ket{\bar{0}}_Z + i r^{1/4}\ket{\bar{1}}_Z}{\sqrt{\sqrt{1-r} + \sqrt{r}}}\,,\quad Z\in\{A,B\}\,.
\end{equation}
The product state $\ket{w_{\mathrm{d}}}_A\otimes\ket{w_{\mathrm{d}}}_B$ is invariant under the swap operator $P$ and, thus, $p_a=0$. Its overlap with $\ket{\widetilde{\chi_s}}$ vanishes as well because the coefficients in Eq.~\eqref{eq:vd-dark-product} ensure that the contributions from $\ket{\bar{0}}_A \otimes \ket{\bar{0}}_B$ and $\ket{\bar{1}}_A \otimes \ket{\bar{1}}_B$ cancel. Therefore, we get $p_s = 0$ for $\ket{w_{\mathrm{d}}}_A\otimes\ket{w_{\mathrm{d}}}_B$.
This implies that for any value of $p_a\in[0, 1/2]$, a mixture with weight $2p_a$ on $\ket{\bar{0}}_A \otimes \ket{\bar{1}}_B$ and weight $1 - 2p_a$ on $\ket{w_{\mathrm{d}}}_A \otimes \ket{w_{\mathrm{d}}}_B$ has $(0, p_a)$. Moreover, any mixture of this separable state with the product state $\ket{w_A}\otimes\ket{ v_B}$ constructed above at the same $p_a$, preserves $p_a$. These mixture allow to obtain any value of $p_s$ between zero and the value attained by that product state.

Therefore, the set of tuples $(p_s, p_a)$ that is compatible with a separable state in $\mathcal{D}_{\star, \mathrm{sep}}$ is given by 
\begin{equation}\label{eq:vd-sep-region}
\mathcal{P}_{\mathrm{sep}} = \left\{(p_s,p_a) : 0\leq p_a\leq\frac{1}{2},\, 0\leq p_s\leq f_c(p_a)\right\}\,.
\end{equation}
where $f_c(p_a)$ is defined in Eq.~\eqref{eq:vd-fc}. In particular, the $\mathcal{P}_{\mathrm{sep}}$ depends only on the concurrence $c$ of the normalized symmetric channel $\ket{\widetilde{\chi}_s}$ (recall Eq.~\eqref{eq:vd-concurrence}). 
In combination with Eq.~\eqref{eq:vd-linear-map} this eventually leads to the separable reachable set
\begin{equation}\label{eq:vd-Rsep}
\mathcal{R}_{\mathrm{sep}}^{(\Omega,2)} = \left\{ p_s\bd\lambda_s + p_a\bd\lambda_a : (p_s,p_a)\in \mathcal{P}_{\mathrm{sep}} \right\}\,.
\end{equation}

The vertex $\bd\lambda_a$ lies outside the separable reachable set $\mathcal{R}_{\mathrm{sep}}^{(\Omega, 2)}$ whenever the conditions in Eq.~\eqref{eq:vd-2D-condition} ensure a two-dimensional reachable set. The signal point $\mathbf{s} = \bd\lambda_a$ requires $(p_s,p_a) = (0,1)$, which violates the separable bound $p_a\leq 1/2$. Hence, it follows that $\mathcal{R}_{\mathrm{sep}}^{(\Omega,2)} \subsetneq\mathcal{R}^{(\Omega,2)}$, and the measurement protocol allows for witnessing entanglement in this parameter regime.  Moreover, for maximal concurrence, that is $c = 1$, the set $\mathcal{P}_{\mathrm{sep}}$ is the square $[0,1/2]^2$, whose image in signal space is a parallelogram. In contrast, for $c < 1$, the $p_s = f_c(p_a)$ is curved. This is the regime which we illustrate in Fig.~\ref{fig:TPA-dimer} in Sec.~\ref{subsec:vibronic-dimer}. At $p_a = 0$, the maximum value $f_c(0) = r$ that is attainable with a separable state, equals the largest squared Schmidt coefficient of $\ket{\widetilde{\chi}_s}$. Thus, it reduces to the bound obtained in Appendix~\ref{app:parallel-vd}.

\bibliography{Refs2}

\end{document}